\documentclass{egpubl}
\usepackage{eg2026}
 
\ArxivPaper

\CGFccby
\usepackage[T1]{fontenc}
\usepackage{dfadobe}  

\usepackage{cite}  % comment out for biblatex with backend=biber 
\BibtexOrBiblatex
\electronicVersion
\PrintedOrElectronic

\ifpdf \usepackage[pdftex]{graphicx} \pdfcompresslevel=9
\else \usepackage[dvips]{graphicx} \fi

\usepackage{egweblnk} 
\PassOptionsToPackage{table}{xcolor}

\def\cg{\cellcolor{green!40}}
\def\cy{\cellcolor{yellow!40}}
\def\cred{\cellcolor{red!40}}

\usepackage{booktabs} % For formal tables
\usepackage[normalem]{ulem}
\usepackage{pgfplots}
\usepackage{tikz}
\usepackage{microtype}
\usepackage{newtxtext} % newtxmath
\pgfplotsset{compat=1.17}
\usepackage{graphicx} %Loading the package
\usepackage[colorinlistoftodos,prependcaption,textsize=tiny]{todonotes}
\usepackage{multicol}
\usepackage{multirow}
\usepackage{diagbox}
\usepackage{placeins}
\usepackage{amsfonts}

\definecolor{OIorange}{RGB}{230,159,0}
\definecolor{OIskyblue}{RGB}{86,180,233}
\definecolor{OIgreen}{RGB}{0,158,115}
\definecolor{OIyellow}{RGB}{240,228,66}
\definecolor{OIblue}{RGB}{0,114,178}
\definecolor{OIvermillion}{RGB}{213,94,0}
\definecolor{OIpurple}{RGB}{204,121,167}

\newcommand{\beginsupplement}{%
        \setcounter{table}{0}
        \renewcommand{\thetable}{S.T\arabic{table}}%
        \setcounter{figure}{0}
        \renewcommand{\thefigure}{S.F\arabic{figure}}%
        \renewcommand{\thesection}{S\arabic{section}}%     
     }
\newcommand{\vect}[1]{\boldsymbol{#1}}

\definecolor{agreen}{rgb}{.2,.65,.2}

\def \modifycolor{black}

\usepackage{arydshln}

\usepackage{graphicx}
\usepackage{amsmath}
\usepackage{booktabs}

\definecolor{jorange}{rgb}{8,.5,.0}

\newcommand\nir{near-infra red}

\title[Multi-Spectral Gaussian Splatting]%
      {Multi-Spectral Gaussian Splatting with Neural Color Representation}

\author[Lukas Meyer \emph{et al.}]{\parbox{\textwidth}{\centering Lukas Meyer\textsuperscript{1}\orcid{0000-0003-3849-7094}, Josef Grün\textsuperscript{1}, Maximilian Weiherer\textsuperscript{1}\orcid{0009-0001-4615-4684}, Bernhard Egger\textsuperscript{1}\orcid{0000-0002-4736-2397}, Marc Stamminger\textsuperscript{1}\orcid{0000-0001-8699-3442} and Linus Franke\textsuperscript{1,2,*} \orcid{0000-0001-8180-0963} \\ \vspace{0.3cm} $^1$ Visual Computing Erlangen, Friedrich-Alexander-Universität Erlangen-Nürnberg-F\"urth, Germany\\$^2$ Inria, Université Côte d’Azur, France}}

\begin{document}

\teaser{
\centering
\includegraphics[width=1\linewidth]{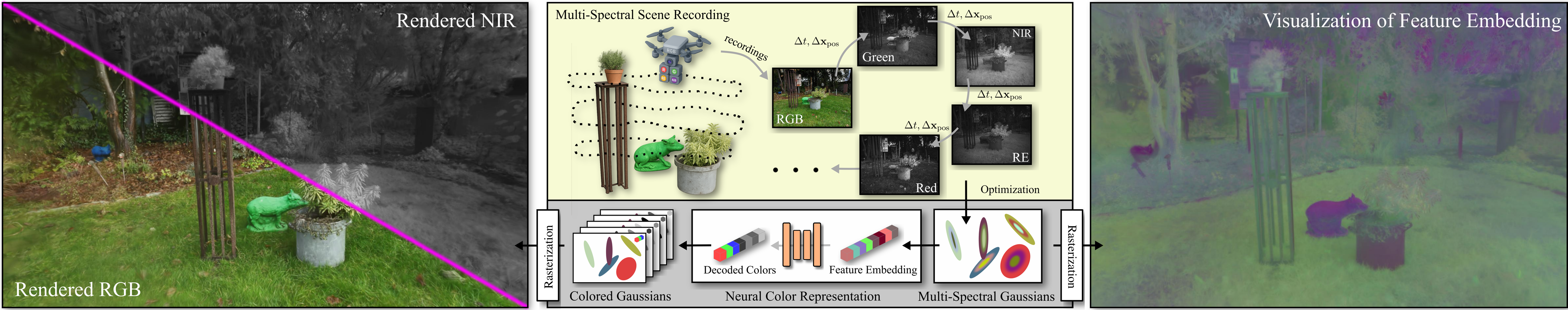}
\caption{Our approach integrates multi-spectral images taken from different capture positions and cameras (middle top) into a single 3D Gaussian-based~\cite{kerbl3Dgaussians} representation. Therefore, we introduce a neural color model to jointly optimize and store all spectral bands. To render individual spectral bands, we use a shallow MLP to decompose the feature embedding into its constituent bands (middle bottom). The left image shows a rendering of RGB and \nir, while the right image visualizes the feature embedding. Notice how regions with similar spectral characteristics coalesce in the learned feature space.}
\label{fig:teaser}
\vspace{-1mm}
}

\maketitle

\begin{abstract}%
3D Gaussian Splatting (3DGS)~\cite{kerbl3Dgaussians} has transformed novel-view synthesis from RGB images, yet remains restricted to the visible spectrum. Many applications, including agricultural monitoring, rely on multi-spectral imaging, where spectral camera alignment and scalability pose major challenges.

We present \textsc{MS-Splatting}---a multi-spectral 3DGS framework enabling unified multi-view consistent reconstruction and rendering across both visible and invisible spectra.
Our key component is a neural color representation that encodes per-primitive features shared across spectral bands, decoded through a shallow multi-layer perceptron into spectrum-specific radiance. By leveraging inter-band correlations, this formulation enhances detail while reducing memory consumption compared to independent band modeling via per-channel modeling with spherical harmonics.

Our method enables accurate parallax-free novel-view vegetation index rendering for plant monitoring and enhances RGB novel view synthesis quality by exploiting details revealed through multi-spectral bands.
Our evaluation demonstrates that MS-Splatting exceeds the current leading methods in both categories. In addition, we introduce a multi-spectral dataset from aerial captures covering outdoor environments, specifically designed for evaluating these applications.
We will release our code and dataset to facilitate further research. \textcolor{\modifycolor}{The project page is located at: \url{https://meyerls.github.io/ms_splatting}}

%
% End generated code
%
%\vspace{0.1cm}
\end{abstract}

\begingroup \renewcommand\thefootnote{}\footnotetext{ \textsuperscript{*} Corresponding author\\
Email: \texttt{\{firstname.lastname\}@fau.de}  \\
%Visual Computing Erlangen, Friedrich-Alexander-Universität Erlangen-Nürnberg, Cauerstraße 11, 91058 Erlangen, Germany
}        
\endgroup

\maketitle

\section{Introduction}
\label{sec:intro}

The rise of recent (neural) rendering techniques such as Neural Radiance Fields (NeRFs)~\cite{mildenhall2020nerf} and 3D Gaussian Splatting (3DGS)~\cite{kerbl3Dgaussians}, which are optimized from captured images in the \textit{visible} (RGB) light spectrum, has greatly enhanced interaction and processing of 3D scenes. 
Specifically, 3DGS provides the state-of-the-art technique in this field by reconstructing the radiance field of the scene through a set of anisotropic 3D Gaussian primitives, allowing for efficient rendering and optimization~\cite{mallick2024taming,meuleman2025onthefly} and scaling to large scenes~\cite{kerbl2024hierarchical,lin2024vastgaussian}.

On the other hand, \textit{multi-spectral} cameras, used to obtain images across different spectral bands (e.g., \nir~or specific narrow bands of visible light) in expansive outdoor environments, have become prevalent and are available in both handheld and aerial formats; however, usually as independent camera sensors without calibration or synchronized shutters. 
Such cameras enable the acquisition of detailed surface characteristics~\cite{4043437, albanwan2024imagefusionremotesensing} and material properties~\cite{namin2012classification} by observing the isolated or combined emitted multi-spectral surface radiance of different spectral bands.
The increasing affordability of these cameras has also enabled widespread adoption in agriculture and forestry, particularly for plant monitoring using \textit{Vegetation Indices}~\cite{Huang2020}.
Most processing pipelines involving these indices, however, demand precise co-registration of spectral channels, which is challenging due to parallax introduced by motion capture, wind effects, and variations in shutter timing between spectral cameras. 

In this paper, we integrate multi-spectral capturings with radiance field rendering techniques into a unified 3DGS reconstruction model, allowing us to both increase novel view synthesis (NVS) quality in visible (RGB) light spectra, as well as to use NVS to precisely render aligned multi-spectral images, which form the base for downstream tasks such as agricultural monitoring.
Furthermore, different from previous methodologies in the field, we take care not to introduce scaling constraints, e.g., positional dependencies common in NeRF or other neural field frameworks, which limit application in large-scale scenarios such as drone-scanned agriculture fields. 

\begin{figure}
    \definecolor{newblock}{HTML}{F100FF} 
    \centering

    \includegraphics[width=1\linewidth]{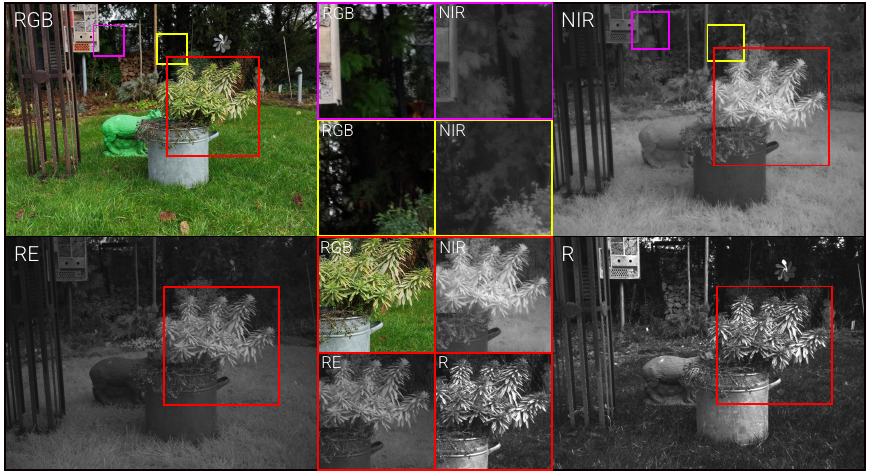}

    \caption{Visualization of different spectral images, which allow for quality enhancement and compression for radiance fields. Near-infra red~(NIR) reveals details not visible in RGB (\textcolor{black}{magenta} and \textcolor{black}{yellow} box) which allows for finer reconstruction. Light reflectance characteristics between bands (RE: red-edge; R: narrow 32nm red band) are shared (\textcolor{black}{red} box), which allows for compression.
    }
    \label{fig:cross_talk_and_compression_visuals}
\end{figure}

Since we have to merge multi-spectral information from multiple independent cameras, we face the challenges of reconstructing the trajectories in a joint space, merging the single bands into a joint representation, and efficiently querying and scaling this information to the requirements of agricultural applications.

Our key insight is that we can exploit the correlations in the interaction of different light spectra with scene surfaces for efficient reconstruction and rendering.
The correlations are twofold: Firstly, different spectra reveal different details of materials and surfaces, resolving ambiguities in the optimization and also sharpening details.
Secondly, nearby wavelengths of light are reflected and absorbed similarly by most surfaces, which allows to compress information. 
A visualization of these effects can be seen in Fig.~\ref{fig:cross_talk_and_compression_visuals}

Consequently, instead of modeling all spectra in isolation, e.g., with separate sets of spherical harmonics per band~\cite{thermalgaussian, gruen2025towards}, we introduce a unified multi-spectral color representation for 3DGS.
We refer to this representation as \textit{neural color representation}, for which we split the radiance emittance model into two parts: (1) a neural feature vector per primitive, which acts as the surface characteristic convolved with the multi-spectral irradiance\footnote{This is because we do not model light transport but optimize the radiance field. 3DGS and similar particle-based radiance field methods represent the scene as a set of emitting primitives, each with a view-dependent radiance.}, and (2) a global function (approximated with an MLP) for generating emittance that calculates per-primitive radiance based on the outgoing direction and the demanded light spectrum.

This formulation efficiently exploits multi-spectral correlations and allows the per-primitive representation to be concise; however, it requires the underlying primitives to be grouped by spectral characteristics.
To further facilitate this, we incorporate a multi-spectral aware densification formulation, allowing for each primitive in the reconstruction to achieve consistent surface properties.
With our pipeline, we are able to increase RGB rendering quality by up to 1 dB in PSNR compared to a common 3DGS model~\cite{kerbl3Dgaussians} through cross-spectral information exchange, commonly called \textit{spectral cross-talk}.
Furthermore, we reduce memory consumption by 88\% compared to 3DGS extended to the multi-spectral domain~\cite{gruen2025towards}.

To evaluate our approach for agricultural applications, we introduce a new drone-captured multi-spectral dataset and showcase the effectiveness of our method in large-area capturing and computing vegetation indices.
For the latter, we exploit the fact that rendering novel views from our multi-spectral model allows synthesizing a shared optical sensor not present in real-world capturing devices, something that is not leveraged in previous approaches.

In summary, our key contributions are as follows:
\begin{itemize}
    \item We introduce \textsc{MS-Splatting}, a multi-spectral neural scene representation based on 3DGS which seamlessly incorporates multiple different visible and invisible spectra of light capturings. 
    \item We propose a novel \textit{neural color representation} that efficiently encodes multiple spectral bands into a joint representation shared across different spectra.
    \item We are the first to compute vegetation indices using novel view synthesis, allowing detailed parallax-free plant monitoring for agricultural applications.
\end{itemize}
%We will release both the code and the multi-spectral dataset.
Our code and the multi-spectral dataset can be found on our project page: \url{https://meyerls.github.io/ms_splatting}

\section{Related Work}

\subsection{Novel View Synthesis and Radiance Fields}

Methods for novel view synthesis historically build on image-based rendering~\cite{shum2000review,shum2008image}, whereby images are warped onto a geometry proxy obtained through active depth measuring or Structure-from-Motion~\cite{snavely2006photo,schonberger2016structure} and multi-view stereo~\cite{goesele2007multi,schoenberger2016mvs}.
Recent advances in neural rendering~\cite{tewari2020state,tewari2022advances} allow direct entangling of 3D reconstruction and rendering through differentiable rendering. 
Neural Radiance Fields (NeRFs)~\cite{mildenhall2020nerf} optimize a 3D volume abstracted in an MLP, sparking a field with many subsequent quality and speed increases~\cite{zhang2020nerf++,barron2022mipnerf360,barron2021mipnerf,mueller2022instant,barron2023zip}. 
These methods abstract the complete appearance in an MLP or feature grid; however, scaling this methodology to large-scale areas is computationally intensive, both in optimization and rendering~\cite{turki2022mega}.

Recently, point-based approaches proved to be a great alternative in the speed-to-quality tradeoff, first represented through point sampling or splat rendering with neural filtering~\cite{aliev2020neural}. 
Thereby, high-dimensional per-point features are optimized for each point of a geometry proxy to capture the local appearance and decoded with a CNN~\cite{adop, trips, kopanas2021perviewopt} or an MLP~\cite{kopanas2022neural}. %inovis, vet, rakhimov2022npbgpp,
We draw inspiration from these approaches, as we optimize high-dimensional appearance features per primitive to capture multi-spectral appearances and radiance; however, we specifically use this as a scheme to unify appearance channels. 

\subsection{3D Gaussian Splatting}

\textit{3D Gaussian Splatting} (3DGS)~\cite{kerbl3Dgaussians} revolutionized point-based radiance field rendering by representing points as 3D Gaussian distributions $G(\mathbf{x}) = e^{-\frac{1}{2}(\mathbf{x} - \boldsymbol{\mu})^{T} \boldsymbol{\Sigma}^{-1} (\mathbf{x} - \boldsymbol{\mu})}$, with position $\boldsymbol{\mu}$ and covariance $\boldsymbol{\Sigma}$, and omitting neural components entirely.
Rendering is done with alpha blending $\mathbf{C} = \sum_{i \in N} \mathbf{c}_{i}\alpha_{i}T_i $, with $T_i = \prod_{j=1}^{i-1} (1-\alpha_{j})$ for each pixel with all contributing Gaussians $N$ with colors $c$.
Notably, colors are hereby stored in spherical harmonics with three degrees.

Subsequent works improved performance~\cite{radl2024stopthepop,hahlbohm2025htgs,yu2024mip,kheradmand20243d,niemeyer2024radsplat}, compression~\cite{3DGSzip2024,Niedermayr_2024_CVPR}, scalability~\cite{kerbl2024hierarchical,lin2024vastgaussian}, and dynamic scenes~\cite{luiten2023dynamic, wu20244dgs}; however, the original algorithm remains remarkably consistent in reconstruction quality.
We refer to recent surveys for more details~\cite{chen2024survey,dalal2024gaussian,fei20243d,wu2024recent}.

Neural components in radiance fields have been excessively used by \textit{in-the-wild} scenarios~\cite{martinbrualla2020nerfw} adapted to 3DGS, which we take inspiration from.
WildGaussians~\cite{kulhanek2024wildgaussians} use per-Gaussian optimizable embeddings during training, combined with per-image embeddings to harmonize different illuminations in the dataset. The embeddings serve as affine color transformations, allowing control of the illumination, with Wild-GS~\cite{xu2024wildgs} and GS-W~\cite{zhang2024GSW} following similar methodologies.
Scaffold-GS~\cite{tao2024scaffoldgs} and Octree-GS~\cite{ren2024octree} use neural components to spawn Gaussians on-the-fly, decoding the anchor Gaussians' neural attributes to all parameters, including color.
Compression for Gaussian representations~\cite{hac2024,3DGSzip2024} similarly uses neural condensation; however, not due to physical properties or in relation to spectral imaging.

Also not related to multi-spectral rendering, \textit{FeatSplat}~\cite{martins2024featuresplattingbetternovel} is methodologically closest to our approach. They learn a feature vector per Gaussian that encodes both color and semantic information. 
%However, rather than alpha-blending splat colors, they splat these feature vectors onto the image pixels, where they are later decoded by a multilayer perceptron (MLP) to predict both the pixel's color and semantic label. 
The important difference to our method is that they first alpha-blend the feature channels into one feature image, and then decode this image in \textit{screen space} to RGB colors---an idea also explored in temporal Gaussian processing~\cite{Li_STG_2024_CVPR}.
For the multi-spectral reconstruction domain, this blending of neural surface characteristics diminishes the effect of spectral correlations between surface materials and limits spectral cross-talk.

Compared to the aforementioned approaches, we introduce a per-Gaussian neural encoding and decoding, which, to the best of our knowledge, is unexplored in the 3DGS literature.
We note, however, that this color model shines in a multi-spectral or multi-model scenario and shows little benefit when employed in the RGB domain only, as also shown in our evaluation.

\begin{figure*}[t]
    \centering
    \includegraphics[width=1\linewidth]{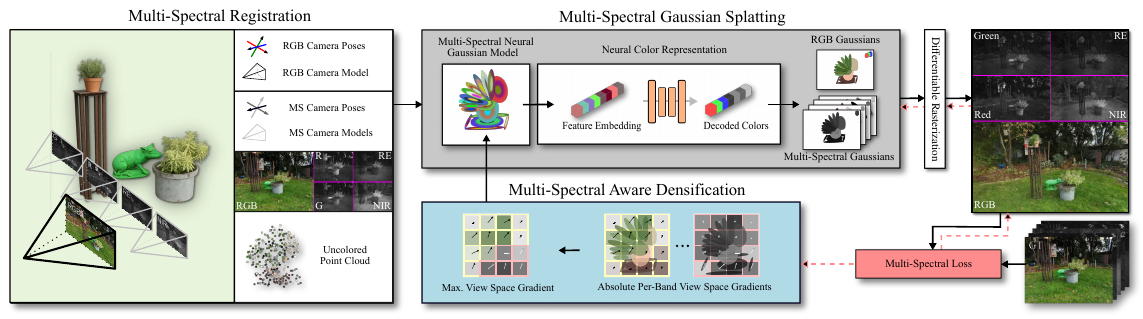}
    %\vspace{-0.6cm}
    \caption{Overview of the \textsc{MS-Splatting} pipeline. After initial Structure-from-Motion registration across all channels simultaneously, we initialize a multi-spectral neural Gaussian model shared for all spectral channels. Thereby, colors of all spectral bands are encoded in a per-Gaussian feature vector and decoded with a tiny MLP. During optimization, the shared Gaussian model and per Gaussian features are optimized by drawing a random view and spectral band under a multi-spectral loss formulation.}
    \label{fig:ms_splatting_pipeline}
    \vspace{-3mm}
\end{figure*}

\subsection{Multi-Modal Radiance Fields}

Multi-modal strategies integrate various sensing modalities, including passive reflectance (multi-spectral), passive emission (thermal), and active illumination (LiDAR or radar). Conversely, single-modal methods utilize only one sensing modality. Multi-spectral and hyper-spectral imaging are encompassed within this category, varying in spectral resolution: multi-spectral sensors collect approximately 4–10 distinct channels, whereas hyper-spectral cameras acquire hundreds of narrower, continuous bands, up to about 600 channels covering 400 nm to 2500 nm~\cite{Adam2010}. This work primarily concentrates on passive multi-spectral reflectance.

\paragraph*{Thermal Imaging.} Although our method does not specifically target it, we assess thermal imaging, which detects emissions using infrared light, owing to its recent surge in popularity.
ThermalNeRF~\cite{lin2024thermalnerf} and Xu et al.~\cite{Jiacongthermalnerf} use thermal data to enhance low-light reconstruction, with the latter also improving RGB through thermal cues. 
Ye et al.~\cite{ye2024thermalnerfneuralradiancefields} focus on thermal input training alone, while \"Ozer et al.~\cite{mert} assess various NeRF architectures to find optimal multi-modal configurations across thermal, \nir, and depth. 
Both \textit{ThermoNeRF}~\cite{hassan2024thermonerf} and Thermal-NeRF~\cite{ye2024thermalnerfneuralradiancefields} utilize dual neural fields for RGB and thermal synthesis. 
In the 3DGS realm, Thermal3D-GS~\cite{chen2024thermal3dgs} pioneered thermal-only splatting. 
\textit{ThermalGaussian}~\cite{thermalgaussian} builds on this by combining RGB and thermal Gaussians. They employ spherical harmonics for separate color modeling.

\paragraph*{Multi- and Hyperspectral Imaging.}
Recent advances have explored the integration of multi- and hyperspectral imagery into neural scene representations. X-NeRF~\cite{poggi2022xnerf} introduces a cross-modal neural field that jointly models RGB, multi-spectral, and \nir~bands, learning camera extrinsics during optimization rather than relying on pre-calibration. 
Similarly, Spec-NeRF~\cite{spec-nerf} captures separate visible-spectrum bands while simultaneously estimating each camera’s spectral sensitivity functions to enable accurate reconstruction. 
SpectralNeRF~\cite{spectralnerf} takes a different approach by training on eleven discrete spectral channels to predict per-band spectral maps, which are then fused into high-quality RGB images using a U-Net–style architecture. 
Extending to hyper-spectral data, Hyperspectral NeRF~\cite{Chen24arxiv_HS-NeRF} embeds continuous wavelength information directly into its radiance field, allowing for detailed, per-wavelength radiance predictions across more than 128 channels.

HyperGS~\cite{HyperGS} abstracted and reduced the spectral high-dimensional hyperspectral data domain to a latent space using a pre-trained autoencoder.
During training, the Gaussians optimize latent spectral signatures, which a view-conditioned MLP maps to view-dependent spectral and opacity features that are then decoded back into full spectra at rendering time.
Importantly, this method targets only object-centric scans in the hyperspectral domain, and it uses a positional neural field for color embeddings.
For our domain, their methodology unfortunately does not suffice as we target \textit{unbounded} scenes, and multi-spectral cameras lack the high spectral resolution of hyperspectral imaging.
Closest to our approach is Grün et al.~\cite{gruen2025towards}, who represent multi-spectral data in Gaussian Splatting as separate sets of spherical harmonics (similar to \textit{ThermalGaussian}~\cite{thermalgaussian} for thermal data), but optimized jointly after an initial RGB-only reconstruction.
This, however, limits cross-talking effects and has high memory consumption due to the large number of SH parameters.

In contrast to all these methods, our approach is the first to use a physically inspired \textit{neural color representation} to jointly optimize spectral bands. 
As we show in the results, this approach leads to superior performance in the reconstruction of multi-spectral scenes and increases quality in RGB renderings through allowing spectral cross-talk.

\section{Overview}
The input to our method comes from a set of $n$ cameras, $C_j$.
Each of these cameras has its own intrinsic parameters and covers one or multiple (spectral) bands.
We aim our method to be versatile; as such, we assume no calibration or rig between the different cameras.
With this, recordings can include independent trajectories, resulting in an input of $n$ sets, $\mathcal{I}=\{\mathcal{I}_1,\mathcal{I}_2,\dots,\mathcal{I}_n\}$, where $\mathcal{I}_j$ contains images captured from camera $C_j$.
For our recorded dataset (see Sec.~\ref{ssec:ms_dataset}) we use $n=5$ cameras: an RGB camera, and four multi-spectral cameras capturing red (R), green (G), red edge (RE), and \nir~(NIR). 
Hereby, the RGB camera captures the three wide bands, while the multi-spectral cameras capture narrower bands of the light (see Sec.~\ref{ssec:ms_dataset} and the supplemental).
In this capturing, the cameras mounted on a drone exhibit an irregular capturing pattern due to winds, drone motion and unsynchronized shutters, and are treated as independent shuffled sets.

Based on the input images, we optimize a 3D Gaussian-based scene representation with our \textit{neural color representation} which allows for novel view synthesis across multiple spectral bands. 
Our framework operates in two stages: In the first step (Sec. \ref{subsec:ms_camera}), we compute camera poses and a sparse point cloud from $\mathcal{I}$ using classical Structure-from-Motion (SfM)~\cite{schoenberger2016mvs} \textit{without} assuming shared camera parameters across multiple sensors. 
Then, in the second stage (Sec. \ref{subsec:ms_splatting}), we use previously computed poses and point cloud to initialize and train the unified \textit{neural color representation} within our multi-spectral Gaussian Splatting model, which, eventually, yields a compact multi-spectral scene representation.

\section{Multi-Spectral Camera Calibration} % dont like this section's name.
\label{subsec:ms_camera}

Existing multi-spectral and multi-modal scene representations \cite{mert, thermalgaussian, lin2024thermalnerf} usually assume \textit{shared} camera parameters (intrinsics \textit{and} extrinsics) across multiple channels, and hence need some form of cross-modal camera calibration to superimpose images captured using different sensors.
Once aligned, camera parameters are obtained by applying SfM only to the RGB images.
Although simple, this workflow requires either a fixed setup and special hardware for camera calibration, or the use of error-prone methods to precisely align RGB and multi-spectral images after they have been captured.

In this work, we neither assume shared camera intrinsics nor extrinsics across multiple bands, allowing for images to be captured from different viewpoints and/or points in time, assuming the overall lighting conditions did not change drastically, e.g., no change from daytime to nighttime. 
Specifically, we found that processing all images in $\mathcal{I}$ with SfM \textit{at once} leads to satisfying results if we use separate, per-camera intrinsics.
This works because multi-spectral and RGB images show similar high-frequency geometric details when having a comparable spatial resolution \cite{710832}, as is the case for most commercially available multi-spectral cameras.
Thus, standard SfM's feature detection (e.g., using SIFT~\cite{lowe1999object}) and matching is well applicable to match RGB and multi-spectral features \textit{across spectral bands}.
Please see Fig.~\ref{fig:sift_feature_matching} in the supplemental for an illustration. 

The drawback of this initialization approach is that typical 3DGS initialization cannot be implemented. 
This occurs because the SfM point cloud, which consists of 2D features elevated and refined in 3D, is not separated across different bands. 
Consequently, initially coloring the point cloud becomes challenging since some feature points lack discernible RGB samples from the images, e.g., as feature points are similar in spectral bands, but wildly different in RGB. Standard procedures~\cite{schoenberger2016mvs} involve reusing spectral band colors for those points, resulting in a mixed colored point cloud.
We solve this issue by discarding colorization and replacing it with a brief color warm-up phase at the start of training (see Secs.~\ref{ssec:initalization} and \ref{ssec:qualitative_multispectral}).

\section{Multi-Spectral Gaussian Splatting}
\label{subsec:ms_splatting}

The aim of our method is to optimize all available spectral channels into a single representation, which is both scalable and allows cross-spectral correlations to be exploited.
Compared to grid- or neural field-informed approaches~\cite{poggi2022xnerf,HyperGS}, we opt for a solely primitive-based representation, which is the base for our \textit{shared geometry}. 
This particle-based representation automatically extends to unbounded scenarios and is extendable with spatial partitioning~\cite{kerbl2024hierarchical}.
We extend this shared geometry formulation with a single \textit{neural color representation} instead of per-band spherical harmonics, effectively enabling information to be shared \textit{across} multiple channels.
Fig.~\ref{fig:ms_splatting_pipeline} provides an overview of our pipeline.

Our neural color-based model (Sec.~\ref{ssec:neural_color}) is first initialized (Sec.~\ref{ssec:initalization}) and then optimized using a differentiably Gaussian renderer~\cite{kerbl3Dgaussians,ye2025gsplat} with a randomly drawn spectral image $\mathcal{I}_j$ and our multi-spectral loss function including feature regularization (Sec.~\ref{ssec:loss}).
To ensure sufficiently detailed reconstruction, we employ a multi-spectral aware densification strategy (Sec.~\ref{ssec:msdensify}).

\subsection{Neural Color Representation}\label{ssec:neural_color}
Our key insight is that RGB and multi-spectral channels profit from exchanging information between each other (spectral cross-talking) instead of being treated separately.
To implement this finding, we replace traditional spherical harmonics with an optimizable, per primitive feature vector $\vect{f}_i\in\mathbb{R}^d$, which stores the \textit{view-dependent} multi-spectral radiance shared over the full vector. 

In comparison to separate spherical harmonics~\cite{gruen2025towards, thermalgaussian} using a high per-Gaussian number of parameters (usually 16 per modality/band), this necessitates compression across spectral bands and promotes contributions from various bands to refine details like appearance and edges, aiding in resolving ambiguities during reconstruction. 
This is further supported by our use of a low dimensionality ($d=8$ for all bands jointly) as discussed in Sec.~\ref{ssec:ablation}.

This feature vector is decoded before rendering (thus in world space), resulting in the full-spectrum color vector $\hat{\mathbf{c}}_i\in\mathbb{R}^B$, where $B_j$ represents the channels count per camera. In our captured dataset (Sec.~\ref{ssec:ms_dataset}), this equates to $B=7$, comprising three RGB and four multi-spectral channels.
For decoding, we use a shallow MLP $\Phi$:
\begin{equation}
	\hat{\mathbf{c}}_i =\Phi({f}_i \oplus {s};{\Theta}),
\end{equation}
where $\vect{s}\in\mathbb{S}^2$ is the normalized outgoing viewing direction, in spherical coordinates. 
$\Theta$ are the learnable parameters of the MLP and $\oplus$ denotes vector concatenation.

Importantly, $\Phi$ is \textit{not} a neural field, as it does not depend on the Gaussians' positions.
Compared to NeRF~\cite{poggi2022xnerf} or Gaussian Splatting sampling a neural field~\cite{HyperGS}, our formulation does \textit{not} store multi-spectral radiance in the MLP weights $\Theta$, but uses the MLP as an emittance function. 
This is similar in spirit to a BRDF, which emits light based on outgoing direction, surface parameters, and incoming light.
Hereby, for us, surface parameters are combined with the incoming light in the feature vector $\vect{f}_i$.
This is an important distinction, as it separates the required size of $\Phi$ from the radiance model, thus allowing large unbound scenes only based on primitives.

For $\Phi$, we use a shallow 3-layer MLP with one hidden layer of dimension 32. 
We apply exponential linear unit (ELU) activations~\cite{Clevert2015FastAA} to all layers except the output, which uses a sigmoid activation to constrain outputs between [0,\,1]. 
No encodings~\cite{mueller2022instant,tancik2020fourfeat} are used.

To differentiable render the individual colors of each spectral band, we follow the default 3DGS rasterization pipeline~\cite{kerbl3Dgaussians}:
The pixel color is obtained by alpha-blending the decoded color band $\hat{\mathbf{c}}_i$ of each contributing projected primitive.

\subsection{Initialization}\label{ssec:initalization}

We initialize Gaussian primitives following Kerbl et al.~\cite{kerbl3Dgaussians}, with positions from the SfM point cloud and scale through kNN-distance.
Gaussian Splatting benefits also from good color initialization~\cite{kerbl3Dgaussians}, which is not complete in the multi-spectral calibration scenario (see Sec.~\ref{subsec:ms_camera}) due to unreliable point colors in the multi-spectral SfM point cloud.

To resolve this and stabilize reconstruction, we use a \textit{warm-up initialization} phase.
Point features $\vect{f}_i$ are initialized from a zero-mean normal distribution with a standard deviation of 0.2, and MLP weights $\Theta$ are initialized Kaiming uniformly. 
For the first 500 iterations, we freeze the covariance, mean, scaling, and opacity to solely let the feature vectors and MLP parameters optimize.
%by only training the feature vectors and MLP parameters.
This effectively recovers per-channel color values and successfully prevents our model from large, color-induced gradients which would cause the initial primitive positions to jump around. 

We use this warm-up initialization phase in addition to the default warm-up used by Kerbl et al.~\cite{kerbl3Dgaussians}, whereby earlier training iterations use lower-resolution images.

\subsection{Loss Function and Training Regime}\label{ssec:loss}
We employ the following loss function to train our model:
\begin{equation}
    \mathcal{L}_{\text{MS}}=\mathcal{L}_{\text{3DGS}}(\hat{I},I)+\lambda_{\text{norm}}\sum_{i=1}^S\left( \|\vect{f}_i\|_2 - 1 \right)^2,
    \label{eq:loss}
\end{equation}

where $I$ denotes a randomly chosen image from the set $\mathcal{I}_j$ of all images captured using the $j$-th camera, $\hat{I}$ is the corresponding rendered image, \textcolor{\modifycolor}{and $S$ is the total number of primitives in the scene}.
$\mathcal{L}_\text{3DGS}$ is the standard 3DGS loss, 
%In each iteration, we first randomly select a  from our dataset $\mathcal{I}$, and then choose at random a number of channels for which we finally compute the standard 3DGS loss,
\begin{equation}
	\mathcal{L}_{\text{3DGS}}(\hat{I},I) = (1-\lambda)\mathcal{L}_1(\hat{I},I) + \lambda \mathcal{L}_{\text{D-SSIM}}(\hat{I},I).
\end{equation}
%where $\hat{I}^{(b)}$ denotes the $b$-th channel of the rendered and $I_k^{(b)}$ the $b$-th channel of the ground truth image.
%Moreover, $[B]:=\{1,2,\dots,B\}$.
The second term in Eq. (\ref{eq:loss}) regularizes the per-Gaussian feature vectors $\vect{f}_i$. 
%\bernhard{hier wird noch eq2 erklärt aber sehr spät, früher?}
We back-propagate this loss through our model to update the parameters of the MLP $\Theta$, feature vectors $\vect{f}_i$, as well as the conventional 3DGS parameters $\{\vect{\mu}, \vect{\Sigma}, \vect{\sigma}\}$.

When sampling the rendering camera $C_j$, we observe that RGB guides reconstruction quality over the whole image. Therefore, we determined empirically that the sampling weight for RGB cameras should be increased to four times that of multi-spectral cameras, so that, after four multi-spectral images, an RGB image is used.

We use $\lambda=0.2$ and $\lambda_{\text{norm}}=0.1$ in all our tests, and evaluate our loss function in Eq. (\ref{eq:loss}) with a single randomly chosen image per iteration.  
For an ablation of the regularization, see the supplemental. %Tab~\ref{tab:supp_norm_ablation}
% this is the most general form of our loss. we only sample one I_k and one b per iteration!

\iffalse

As a loss function we adapt the original version of 3DGS 
\begin{equation}
	\mathcal{L}_{\textit{3DGS}} = (1-\lambda)\mathcal{L}_1 + \lambda \mathcal{L}_{\textit{SSIM}}
\end{equation}
and extend it to all available channels trough:
\begin{equation}
	\mathcal{L}_{MS} = \sum_{S \in \mathcal{S}} \lambda_s \mathcal{L}_{\textit{3DGS}}.
\end{equation}
$\mathcal{S}$ are the indices for the spectral bands/frequencies for color, red, green, red edge and \nir~and 
$\lambda_s = \frac{N_{I_s}}{N}$ defines the ratio between the number of images for on channel to all $N$ images

As the feature embedding $\mathbf{f}$ is also a learnable parameter we normalize the feature vector trough
\begin{equation}
    \mathcal{L}_{\text{norm}} = \sum_{i} \left( \|F_i\|_2 - 1 \right)^2
\end{equation}
where $\mathbf{F} \in \mathbb{R}^{N\times d}$ are the per Gaussian feature embeddings.
Finally the loss function build up to
\begin{equation}
	\mathcal{L} =  \mathcal{L}_{MS} + \lambda_{\textit{norm}}\mathcal{L}_{\text{norm}}
\end{equation}

\fi

\subsection{Multi-Spectral Aware Densification}\label{ssec:msdensify}

We utilize the Gaussian densification strategy first introduced by \cite{kerbl3Dgaussians} and later refined by \cite{ye2024absgs}, but adapt it to better respond to multi-spectral input data. In every iteration, for every rendered Gaussian $G_i$, we calculate the homodirectional view-space positional gradient
\begin{equation}
\small
\hat{g}_i^{(j)} = \left( \sum_{k=1}^{m} \left| \frac{\partial L_k}{\partial \mu_{i,x}^{(2d)}} \right|, \sum_{k=1}^{m} \left| \frac{\partial L_k}{\partial \mu_{i,y}^{(2d)}} \right| \right),
\end{equation}
where $j\in[n]:=\{1,2,\dots,n\}$, $\mu_{i}^{(2d)}$ is the view-space position of $G_i$, $m$ is the number of pixels rendered for the Gaussian, and $L_k$ is the loss function in Eq. (\ref{eq:loss}) computed for the $k$-th pixel. We choose a densification interval of 300 iterations, during which the calculated $\hat{g}_i^{(j)}$ are accumulated. In the next densification step, if the criterion
\begin{equation}
\small
\max_{j \in [n]} \left( \frac{\sum \left \lVert \hat{g}_i^{(j)} \right \rVert }{n^{(j)}} \right) > \tau_{grad}
\end{equation}
is met for a Gaussian, we split or clone it according to the approach laid out by \cite{kerbl3Dgaussians}, where $n^{(j)}$ is the number of view-space gradients accumulated in the densification interval for the current band. We choose $\tau_{grad}=0.0008$ in our experiments. 
 
Note that we separately accumulate view-space gradients for each spectral band, and then select the maximum average gradient for the densification criterion. Since high view-space gradients correlate with under- or over-reconstructed areas of the scene \cite{kerbl3Dgaussians}, this approach recognizes when a region is properly modeled for one spectral band, but is insufficiently reconstructed for another, and introduces new Gaussians to the area accordingly. Therefore, our modification to the densification strategy allows for proper reconstruction of high-frequency details that are only visible in one or a few of the captured spectral bands.
See Sec.~\ref{ssec:ablation} for an a evaluation compared to the default densification strategy.

\iffalse
\subsubsection{Network Implementation}
Our spectral feature decoding network is a small MLP with three layers: an input layer, a hidden layer, and an output layer. The hidden layer has a width of 32 units. The output layer has dimension 7, corresponding to one value per channel (RGB: 3; R: 1; G: 1; RE: 1; NIR: 1). We apply exponential linear unit (ELU) activations~\cite{Clevert2015FastAA} to all layers except the output, which uses a sigmoid activation to constrain outputs to the range [0,\,1]. The network input consists solely of the feature embedding concatenated with the normalized viewing direction, without any positional encoding.
\fi

\section{\textsc{MS}-Splatting Dataset}
\label{ssec:ms_dataset}

We recorded our own dataset -- the \textit{MS-Splatting Dataset} -- as as no suitable outdoor multi-view dataset with multi-spectral imagery was available. 
Voynov et al.~\cite{10205168} provide an indoor multi-sensor dataset containing RGB, \nir, and depth data; Poggi et al.~\cite{poggi2022xnerf} offer an unposed, forward-facing indoor dataset with RGB, multi-spectral, and \nir~channels (16 scenes, 10 bands), although they estimate poses by learning relative sensor transformations, which lie outside the scope of our work. Finally, Hyperspectral NeRF~\cite{Chen24arxiv_HS-NeRF} captured an indoor hyperspectral dataset of eight scenes with 128 bands, which is not yet publicly released.

%\subsection{Our Dataset}

Our dataset consists out of seven outdoor scenes recorded with a DJI Mavic 3M drone \cite{dji_mavic_3m}. This drone features a dual-camera system: one RGB sensor and four multi-spectral cameras that cover the red (R), green (G), red-edge (RE), and \nir~ (NIR) bands. While the green and red bands overlap with the RGB spectrum, their narrow capture bandwidth (32 nm) provides higher spectral precision. We note that spectral bands can only be captured sequentially---due to hardware restrictions---which can introduce spatial misalignment, especially in windy scenarios and due to the drone’s stabilization system. Fig.~\ref{fig:motivation_image2} in the supplemental illustrates a typical misalignment caused by this sequential capture.
Further details on the drone's imaging system, spectral band characteristics, and an overview of the dataset's evaluation images can also be found in the supplemental.

The dataset includes 360° panoramas of a garden, a dormant cherry tree, an apple orchard during bud swelling, and a lake with a construction trailer; 180° views of a house with solar panels; forward-facing sequences of an apple orchard at harvest; and a top-down survey of a golf course. An overview of the datasets can be seen in the evaluation images in the supplements. The dataset comprises out of 81-136 images per channel in total out of 405-680 images. RGB images are captured with a focal length of 24 mm, while each multi-spectral camera has a focal length of 25 mm. For training we split each dataset in 90\% training and 10\% evaluation data. We also release the indices of the evaluation data for all datasets.  The dataset of default and warped is cropped to $1800$px$\times1350$px.

\section{Evaluation}

We compare our method against current state-of-the-art approaches to confirm the effectiveness of our \textit{neural color representation} and to show that our framework can be readily extended to data from additional modalities, such as thermal. Moreover, in Sec.~\ref{sec:agricultre_application}, we demonstrate how \textsc{MS-Splatting} can be used to compute vegetation indices for agriculture applications, and in the supplemental how to leverage our learned feature embedding to perform multi-spectral clustering to visualize areas with similar spectral information.

Our model builds upon \textit{gsplat} \cite{ye2025gsplat} and is implemented within the \textit{Nerfstudio} \cite{nerfstudio} framework. 
Unless otherwise noted, we trained our model for 120,000 iterations using the Adam optimizer with a learning rate of 0.005, as described in the previous section.

\begin{table*}[] 
\scriptsize
\definecolor{cellgreen}{RGB}{247,203,153}
%\fontsize{4pt}{4pt}\selectfont
\centering
\caption{Results over all seven scenes of our captured dataset. See also Fig.~\ref{fig:mainpaper_comp_images_ms_single_tree_2} as well as Figs.~\ref{fig:comp_images_ms_solar},\ref{fig:comp_images_ms_bauwagen}, \ref{fig:comp_images_ms_LWG_vegetation}, \ref{fig:comp_images_ms_single_tree_1}, \ref{fig:comp_images_ms_golf} and \ref{fig:comp_images_ms_garden} in the supplemental for visual comparison. The colors indicate \colorbox{green!40}{best} and \colorbox{yellow!40}{second best}. Independent Cameras (IC) indicates if a method is able to fuse multiple independent cameras into the same model.
For methods supporting IC, we use the original dataset, for the other we train with warped multi-spectral images.
%Depending on the method, we train with warped multi-spectral images (W), or the original dataset (O) with independent camera positions. ThermalGaus. (W) uses the original implementation by \cite{thermalgaussian}, whereas ThermalGaus. (O) 
$\dagger$ indicates our re-implementation described in the supplementary (Sec. \ref{sec:sup_tg_reimplementation}).}
\setlength{\tabcolsep}{3pt}
\resizebox{\textwidth}{!}{
\begin{tabular}{ l | c | c c c c c c | c c c c c c | c | c | c | c   }
\toprule
& & \multicolumn{6}{c|}{PSNR $(\uparrow)$} & \multicolumn{6}{c|}{SSIM $(\uparrow)$} & \multicolumn{1}{c|}{LPIPS$(\downarrow)$} & \multicolumn{1}{c|}{SAM$(\downarrow)$} & \multicolumn{1}{c|}{SCM$(\uparrow)$} & \multicolumn{1}{c}{SID$(\downarrow)$}\\\midrule
 & IC & All & RGB & G & R & RE & NIR & All & RGB & G & R & RE & NIR  & RGB & All-MS & All-MS & All-MS \\
\midrule
ThermalGaus.~\cite{thermalgaussian} & $\times$ & 23.54 & 19.81 & 23.98 & 25.66 & 23.78 & 24.49 & 0.700 & 0.565 & 0.710 & 0.763 & 0.717 & 0.743 & 0.371                                   & -     & -     & -     \\
ThermoNeRF~\cite{hassan2024thermonerf} & $\times$ & 18.91 & 15.39 & 19.48 & 20.62 & 18.78 & 20.26 & 0.500 & 0.255 & 0.542 & 0.616 & 0.514 & 0.575 & 0.723                                & -     & -     & -     \\
3DGS~\cite{kerbl3Dgaussians} & \checkmark & 23.62 & 20.86 & 22.92 & 25.23 & 23.33 & 25.75 & 0.721 & 0.623 & 0.667 & 0.769 & 0.730 & 0.815 & \cg 0.283                                     & 0.146 & 0.812 & 0.044 \\
ThermalGaussian $ \dagger$ & \checkmark & 24.18 & 20.17 & 24.04 & 26.09 & 24.14 & 26.50 & 0.718 & 0.564 & 0.695 & 0.788 & 0.732 & 0.814 & 0.387                                         & 0.127 & 0.843 & 0.035 \\
TIMS~\cite{gruen2025towards}& \checkmark & 24.99 & \cg 21.29 & 24.62 & 26.61 & 25.07 & 27.42 & 0.755 & \cg 0.647 & 0.719 & 0.810 & 0.766 & 0.834 & \cg 0.283                             & 0.116 & 0.860 & 0.031 \\
MultiSpec-FeatSplat & \checkmark & \cy 25.20 & 20.75 & \cy 25.12 & \cy 27.49 & \cy 25.13 & \cy 27.57 & \cg 0.766 & \cy 0.641 & \cg 0.733 & \cg 0.828 & \cg 0.781 & \cy 0.848 & \cy 0.301  & \cellcolor{yellow!40}0.114 & \cellcolor{yellow!40}0.879 & \cellcolor{yellow!40}0.026 \\
Ours & \checkmark & \cg 25.65 & \cy 21.17 & \cg 25.42 & \cg 27.79 & \cg 25.79 & \cg 28.13 & \cy 0.763 & 0.633 & \cy 0.729 & \cy 0.827 & \cy 0.780 & \cg 0.849 & 0.306                    & \cellcolor{green!40}0.109 & \cellcolor{green!40}0.888 & \cellcolor{green!40}0.024\\
\bottomrule
\end{tabular}
}    \vspace{-0.3cm}
\label{tab:large_comparison}
\end{table*}

% why qualitative comparison? it's both, quant and qual. I'll rename this section...
\subsection{Comparison on Multi‐Spectral Data}
\label{ssec:qualitative_multispectral}

\paragraph*{Methods.}
Multi-spectral radiance field methods rarely provide their code open-source; as such, we are limited in methods to fairly compare with.
We compare against 3D Gaussian Splatting (\textit{3DGS})~\cite{kerbl3Dgaussians}, trained on each spectral band individually, resulting in individual models.
Furthermore, we compare against the joint-optimized setup of Grün et al.~\cite{gruen2025towards} who use 3D Gaussians with shared Spherical Harmonics for spectral color channels (denoted as \textit{TIMS}).
A comparison to other multi-spectral methods was, due to non-availability of source code, not possible.

We further compare our method against two open‐source thermal baselines: \textit{ThermalGaussian} \cite{thermalgaussian} and \textit{ThermoNeRF} \cite{hassan2024thermonerf}.
As they normally use only RGB and thermal data, we trained \textit{ThermalGaussian} and \textit{ThermoNeRF} four times, pairing RGB with one additional multi-spectral band per run. 
%For metric evaluation, we averaged RGB over all these runs.

Furthermore, we re-implemented \textit{ThermalGaussian}’s “One Multi Modal Gaussian” variant within our framework, denoted with $\dagger$.
However, our implementation extends each Gaussian with extra spherical harmonics to handle multiple (i.e., not just one as in the original implementation) spectral bands simultaneously.
More implementation details can be found in the supplemental.

Finally, we also compare \textit{MultiSpec-FeatSplat}, which is our multi-spectral adaptation of the approach of FeatSplat~\cite{martins2024featuresplattingbetternovel}. In this method, the Gaussian feature vectors are first rendered to a 2D image, before the MLP is applied in screen-space, producing a colored image. The MLP is also provided with the respective camera position in the scene. We utilize the same feature dimension, MLP size, training strategy, and training duration as in our method.

\paragraph*{Datasets.}
We compare on our captured dataset (Sec.~\ref{ssec:ms_dataset}) as well as the X-NeRF dataset~\cite{poggi2022xnerf}. 
Note that due to the unavailability of the source code and evaluation protocol, direct comparison to X-NeRF was not possible.
For our dataset, we hold out every 10th image for test, while we use every 6th image for X-NeRF (see supplemental Tab.~\ref{tab:xnerf_dataset_indices}), aligning with the number of test images stated~\cite{poggi2022xnerf}. 
Importantly, the test images may differ from those selected by X-NeRF, as those specifics are unknown.

To evaluate methods such as \textit{ThermalGaussian}~\cite{thermalgaussian} and \textit{ThermoNeRF}~\cite{hassan2024thermonerf} which assume shared camera parameters across all spectral bands and are therefore not directly applicable to our dataset, we first superimpose RGB and multi-spectral images by estimating pairwise homographies between each of them (i.e., between RGB + R, RGB + G, RGB + RE, and so on). We then warp multi-spectral images into the RGB frames. Due to focal-length differences, however, the warped multi-spectral images exhibit black borders; we, therefore, crop all images to $1800$px$\times1350$px to effectively remove these margins. 
We trained our model and the other methods on the original, unaligned data, but cropped images to the same regions to ensure a fair comparison. 

For the other Gaussian Splatting-based methods, we also employ the warmup strategy described in Sec.~\ref{ssec:initalization} optimizing the SH base color; otherwise, instabilities occur.

\paragraph*{Metrics.}

To evaluate our results, we relied on both color, perceptual and spectral metrics on the test images. On the one hand, we utilized standard metrics such as PSNR, SSIM, and LPIPS. While LPIPS is a learned perceptual metric trained on RGB images, it is not very expressive in the settings of multi-spectral imagery, but it is included for the sake of completeness in the supplemental. 
On the other hand, we evaluated the multi-spectral images in terms of their spectral accuracy. Therefore, we employed the spectral-similarity metrics Spectral Angle Mapper (SAM)~\cite{Kruse_SAM}, Spectral Correlation Mapper (SCM)~\cite{carvalho_SCM}, and Spectral Information Divergence (SID)~\cite{Chang_SID}. A detailed explanation of these metrics is provided in the supplementary material.

\begin{table}[]

\caption[]{\label{tab:comparisonxnerf}X-NeRF dataset~\cite{poggi2022xnerf}, results averaged on all scenes on RGB, multi-spectral (MS) and infra-red (IR). In this dataset we have 10 multi-spectral bands and 1 band for IR.}
%GROUPED: Average across all XNerF datasets. Ours: 240.000 iterations, f=16, 3x RGB sampling. Gruen et al: 240.000 iterations, 30.000 warmup. 3DGS/TG-O: RGB 30.000, IR 30.000, MS combined (ThermalGaussian-Ours) 200.000}
%
\scriptsize
\centering
\setlength{\tabcolsep}{3pt}
\resizebox{1.01\linewidth}{!}{
\begin{tabular}{ l | c c c c | c c c c | c | c | c | c  }
\toprule
& \multicolumn{4}{c|}{PSNR $(\uparrow)$} & \multicolumn{4}{c|}{SSIM $(\uparrow)$} & \multicolumn{1}{c|}{LPIPS$(\downarrow)$} & \multicolumn{1}{c|}{SAM$(\downarrow)$} & \multicolumn{1}{c|}{SCM$(\uparrow)$} & \multicolumn{1}{c}{SID$(\downarrow)$}\\
\midrule
 & All & RGB & MS & IR & All & RGB & MS & IR & RGB & All-MS & All-MS & All-MS\\
\midrule
%3DGS & \cy 31.66 & \cg 35.36 & \cy 31.34$\dagger$ & \cg 31.14 & \cy 0.919 & \cg 0.948 & \cy 0.916$\dagger$ & \cy 0.922 & \cg 0.243& \colorbox{green!40}{0.063$\dagger$} & \colorbox{yellow!40}{0.878$\dagger$} & \colorbox{yellow!40}{0.021$\dagger$} \\
3DGS & 30.99 & \cg 35.36 & 30.54 & \cg 31.14 & 0.896 & \cg 0.948 & 0.888 & \cy 0.922 & \cg 0.243                   & 0.091 & 0.816 & 0.049 \\
TG.$ \dagger$ & \cy 31.47 & 32.87 & \cy 31.51 & 29.63 & \cy 0.916 & 0.929 & \cy 0.914 & 0.918 & 0.311              & \colorbox{yellow!40}{0.065} & \colorbox{yellow!40}{0.869} & \colorbox{yellow!40}{0.021} \\
TIMS & 29.07 & 31.45 & 29.02 & 27.20 & 0.889 & 0.932 & 0.884 & 0.895 & 0.288                                       & 0.069 & 0.858 & \colorbox{yellow!40}{0.021} \\
Ours & \cg 32.02 & \cy 32.88 & \cg 32.15 & \cy 29.93 & \cg 0.930 & \cy 0.940 & \cg 0.929 & \cg 0.926 & \cy 0.280    & \colorbox{green!40}{0.063} & \colorbox{green!40}{0.879} & \colorbox{green!40}{0.016} \\
\bottomrule
\end{tabular}
}
\label{tab:x_nerf_evaluation_small}
\end{table}

\paragraph*{Results.}
Tab.~\ref{tab:large_comparison} presents the results, averaged over all scenes of our dataset. 
RGB metrics for \textit{ThermalGaussian} and \textit{ThermoNeRF} were computed by averaging respective results over the four runs. 
As seen, our approach improves PSNR by over 1.4~dB, increasing SSIM by 6~\%, and reducing LPIPS by 23~\% compared to the previous state-of-the-art \textit{ThermalGaussian}, and also outperforms \textit{TIMS}~\cite{gruen2025towards} as well as our screen-space splat implementation.
Notably, even on RGB data alone, our method outperforms \textit{3DGS} in PSNR and SSIM, demonstrating the cross-spectral benefit of incorporating additional bands. 
Furthermore, our method provides state-of-the-art results on all three spectral metrics, showcasing exceptional spectral reconstruction quality. 

Beyond improvements in perceptual metrics, we also demonstrate enhancements in spectral-similarity metrics across all compared methods. 
Compared to the state-of-the-art \textit{ThermalGaussian}, we reduce the spectral angle (SAM) by 17\%, increase the spectral correlation (SCM) by 5\%, and shrink the spectral information divergence (SID) by 50\%.
This highlights the stronger joint representational power of our approach on RGB and multi-spectral data. A per dataset listing of the spectral similarity metrics can be found in the supplemental in Tab.~\ref{tab:spectral_sim_dataset}.
In addition, we provide radar (spider) charts -- analogous to spectral response curves of hyperspectral cameras -- showing raw values across RGB and multi-spectral pixels in Fig.~\ref{fig:value_spectral_spider} and per-band errors in Fig.~\ref{fig:error_spectral_spider}.

Visual comparisons for all spectral channels are shown in Fig.~\ref{fig:mainpaper_comp_images_ms_single_tree_2}. 
Hereby, our method is able to reconstruct the fine leaves of the fruit treetop in all bands (top, blue crop), as well as provides great background reconstructions (bottom, red crop).
Further visual comparisons are in the supplemental in Fig.~\ref{fig:comp_images_ms_solar}, ~\ref{fig:comp_images_ms_bauwagen}, ~\ref{fig:comp_images_ms_LWG_vegetation} \ref{fig:comp_images_ms_single_tree_1}, \ref{fig:comp_images_ms_golf} and \ref{fig:comp_images_ms_garden}. Per-scene results are provided in the supplements.  

In Tab.~\ref{tab:x_nerf_evaluation_small}, we evaluate \textit{3DGS}, \textit{ThermalGaussian}$ {\dagger}$, \textit{TIMS}, and our method on all scenes of the X-NeRF dataset~\cite{poggi2022xnerf}. 
For \textit{3DGS}, we trained each spectral band independently, including the multi-spectral channels. More information on the training can be found in the supplemental Sec.~\ref{sec:x_nerf_training}.
Overall, the results demonstrate that our method achieves the best average PSNR and SSIM across RGB, multi-spectral, and infrared (IR) data. 
Compared to the \textit{3DGS} baseline, we improve the PSNR on multi-spectral data by 1.7~dB. 
  The slightly better results on RGB and IR can be attributed to the noisy registration of RGB with multi-spectral and IR images, since the multi-spectral images with a size of $510$px$\times240$px are relatively small, IR has a low resolution with  $1024$px$\times1024$px with a large field of view and as we treat the 10 multi-spectral bands and the single IR band as separate camera models.
For the multi-spectral images, we further show that our method achieves the best scores on spectral-similarity metrics, confirming once more the representational power of our proposed \textit{neural color representation}. 
Additional tables with per-scene spectral-similarity results as well as per-band evaluations are provided in the supplementary material.

\begin{figure*}[htbp]
    \centering
    \includegraphics[width=\textwidth]{images/appendix/ms-fruit-part.pdf} \\
    \vspace{10pt}
    \includegraphics[width=\textwidth, trim={0 0 0 4.5cm},clip]{images/appendix/ms-single-tree-2.pdf} \\
    \caption{Visual comparison on the \textsc{FRUIT TREES} (top) and \textsc{SINGLE TREE} (bottom) scenes with all available spectral-bands. The used \textit{ThermalGaussian} method in this comparison is our multi-spectral re-implementation.}
%    \caption{Visual comparison on the SINGLE TREE scene with all available spectral-bands. The used ThermalGaussian method in this comparison is our multi-spectral re-implementation.}
    \label{fig:mainpaper_comp_images_ms_single_tree_2}
\end{figure*}

\subsection{Comparison on Thermal Data}

\begin{figure*}[h]
    \centering
    \includegraphics[width=1\textwidth]{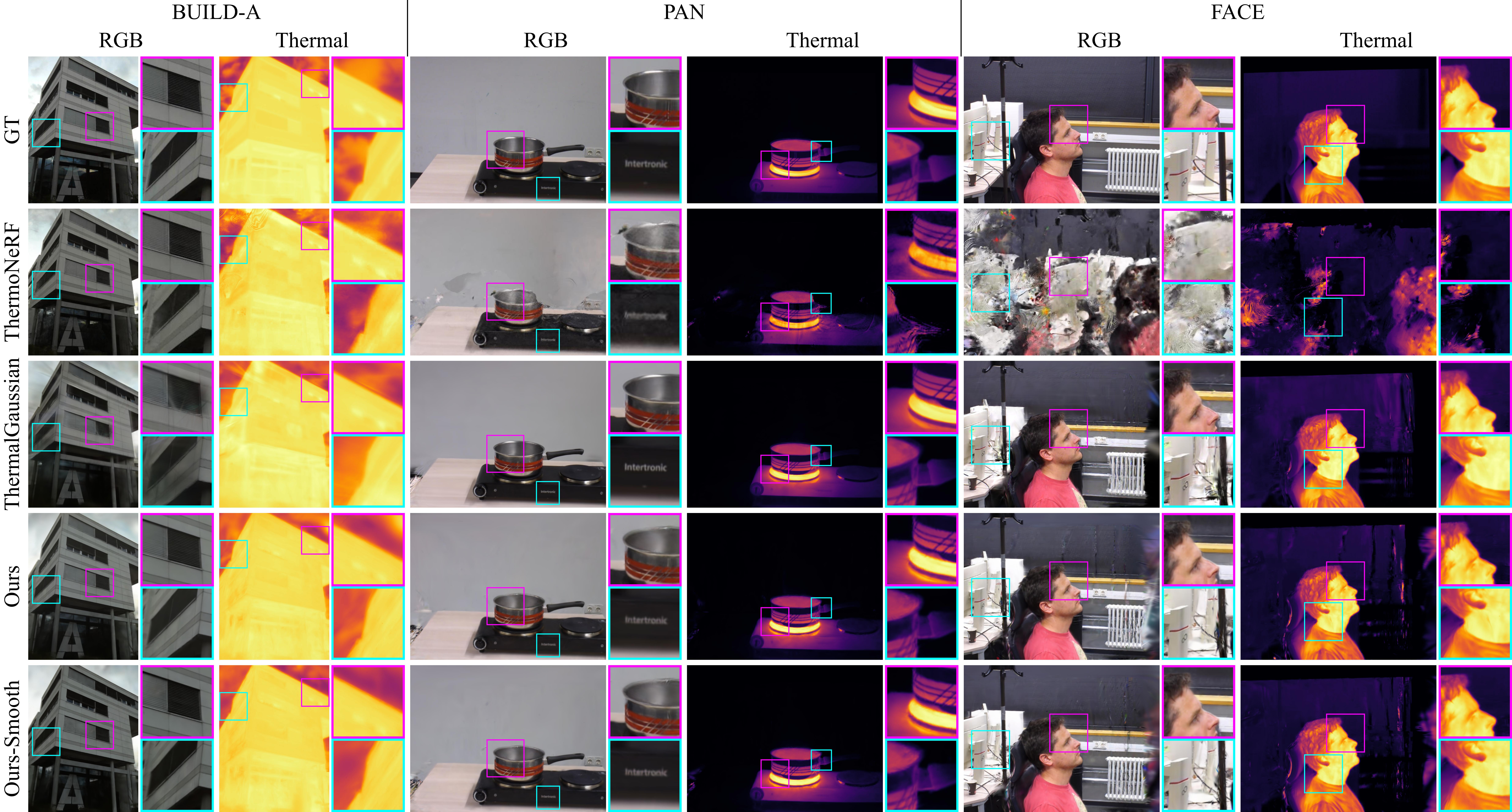}
    \caption{Visual comparison of the thermal scenes BUILD-A, PAN and FACE on \textit{ThermalGaussian}, \textit{ThermoNeRF} and ours. The used \textit{ThermalGaussian} method is the original implementation by ~\cite{thermalgaussian}. Ours-Smooth uses the smoothness loss proposed by~\cite{thermalgaussian}. 
    %Additional implementation information can  be found in the supplement.
    }
    \label{fig:eval_thermal_imags}\vspace{-3mm}
\end{figure*}

We also compared the original implementations of ThermalGaussian~\cite{thermalgaussian} and \textit{ThermoNeRF}~\cite{hassan2024thermonerf} against our method on thermal data.
To do so, we selected three scenes from the ThermalMix dataset~\cite{mert} and three scenes from the dataset used in \textit{ThermoNeRF}~\cite{hassan2024thermonerf} (details can be found in the supplementary). 
Here, the datasets do provide calibrated pairs of thermal and RGB images, and we used the provided camera parameters for all methods.
We evaluated our approach using the same configuration as in Sec.~\ref{ssec:qualitative_multispectral}. 
Additionally, we tested the influence of the smoothness loss $\mathcal{L}_{\mathrm{smooth}}$ introduced by \cite{thermalgaussian} on our method. 
More information can be found in the supplements.

The averaged results across all scenes are presented in Tab.~\ref{tab:thermal_averages}, with per-scene details provided in the supplementary material. 
Visual comparisons can be found in Fig.~\ref{fig:eval_thermal_imags} and show sharper RGB reconstructions (e.g., \textsc{PAN}, blue crop) and less noisy thermal results (\textsc{FACE}, red crop and background).
As presented, our approach outperforms both \textit{ThermalGaussian} and \textit{ThermoNeRF}, with the exception of comparable quality in thermal PSNR. 
This demonstrates that \textsc{MS-Splatting} generalizes to other modalities without further adaptation. 
Please note that thermal LPIPS was evaluated on grayscale images; thus, its expressiveness is limited.

\begin{table}[h]
%\vspace{-5mm}
\caption{Average performance across six thermal scenes from the ThermalMix dataset~\cite{mert} and the dataset from Thermo-NeRF~\cite{hassan2024thermonerf}. Details can be found in the supplemental.}
\vspace{0.2cm}
\centering
\scriptsize
\begin{tabular}{l | cc| cc| cc}
\toprule
            & \multicolumn{2}{c|}{PSNR $(\uparrow)$} 
            & \multicolumn{2}{c|}{SSIM $(\uparrow)$} 
            & \multicolumn{2}{c}{LPIPS $(\downarrow)$} \\ 
Method                  & RGB   & T  & RGB    & T &  RGB   & T \\ \midrule
ThermalGaussian               & 24.91 & \cg 31.34 & 0.824 & 0.931 & 0.226 & 0.107 \\ 
ThermoNeRF                    & 16.54 & 22.45 & 0.546 & 0.765 & 0.376 & 0.295 \\
Ours                          & \cy 26.31 & 31.18 & \cy 0.883 & \cy 0.934 & \cy 0.166 &\cy 0.104 \\
Ours + $\mathcal{L}_{\mathrm{smooth}}$ & \cg 26.66 & \cy 31.23 & \cg 0.892 & \cg 0.935 & \cg 0.157 &  \cg 0.095 \\
\bottomrule
\end{tabular}
\label{tab:thermal_averages}
\end{table}

\subsection{Spectral Cross-Talking Evaluation}

To further investigate multi-spectral rendering in our method and the effects of spectral correlations and spectral crosstalk, we conducted the following study on our captured dataset.
We systematically extend RGB with successively added different spectral bands in every permutation.
The results of this study can be found in Tab.~\ref{tab:spectralcrosstalk}.

Only using RGB with our \textit{neural color representation} decreases quality by 0.28 dB, suggesting that neurally compressing from \textit{3DGS} full Spherical Harmonics (48 floats) to our neural feature vector ($d=8$ floats) introduces slight quality loss. 
To achieve optimal RGB rendering, combining it with a narrow band red channel (R) and \nir~(NIR) yields superior results, outperforming baseline \textit{3DGS} by 0.76 dB and the neural variant by over 1 dB. 
Interestingly, the NIR and R channels directly formulate the Normalized Vegetation Index (refer to Sec.~\ref{sec:agricultre_application}), providing insights into plant health. This implies our approach benefits from knowledge of plant material characteristics for enhanced RGB NVS.

For all bands, the overall quality drops slightly across channels compared to the ideal configurations. 
While this suggests too strong compression, increasing feature vector sizes do benefit results (see Sec.~\ref{ssec:ablation}).
All in all, this configuration still shows a great tradeoff between multi-spectral reconstruction quality and ease of use.

\begin{table}[h]
\caption[]{\label{tab:spectralcrosstalk}Spectral cross-talking effects with RGB, green (G), red (R), red-edge (RE), and \nir~(NIR) on \textsc{MS-Splatting}. The colors indicate the \colorbox{green!40}{best} and \colorbox{red!40}{worst} results. }
\setlength\tabcolsep{2pt}
\resizebox{1.0\linewidth}{!}{
\begin{tabular}{ l | c c c c c | c c c c c | c  }
\toprule
& \multicolumn{5}{c|}{PSNR $(\uparrow)$} & \multicolumn{5}{c|}{SSIM $(\uparrow)$} & LPIPS $(\downarrow)$\\
\midrule
& RGB & G & R & RE & NIR & RGB & G & R & RE & NIR & RGB \\
\midrule
RGB & \cred 20.58 & - & - & - & - & \cred 0.611 & - & - & - & - & \cred 0.316 \\
RGB + G & 21.24 & 25.13 & - & - & - & 0.641 & \cred 0.720 & - & - & - & 0.283 \\
RGB + R & 21.51 & - & 27.77 & - & - & 0.649 & - & 0.827 & - & - & \cg 0.273 \\
RGB + RE & 21.45 & - & - & 25.65 & - & 0.650 & - & - & 0.777 & - & 0.276\\
RGB + NIR & 21.47 & - & - & - & 28.14 & 0.644 & - & - & - & \cred 0.845 & 0.287 \\
RGB + G + R & 21.38 & \cred 25.11 & 27.66 & - & - & 0.640 & 0.726 & \cred 0.826 & - & - & 0.289 \\
RGB + G + RE & 21.30 & 25.35 & - & 25.59 & - & 0.643 & 0.727 & - & \cred 0.776 & - & 0.290 \\
RGB + G + NIR & 21.48 & \cg 25.48 & - & - & 28.17 & 0.645 & 0.730 & - & - & 0.847 & 0.289 \\
RGB + R + RE & 21.52 & - & 27.75 & 25.73 & - & 0.650 & - & 0.829 & 0.781 & - & 0.283\\
RGB + R + NIR & \cg 21.62 & - & \cg 27.84 & - & 28.16 & \cg 0.651 & - & \cg 0.832 & - & \cg 0.850 & 0.284 \\
RGB + RE + NIR & 21.34 & - & - & \cg 25.86 & \cg 28.23 & 0.645 & - & - & 0.782 & 0.847 & 0.291 \\
RGB + G + R + RE & 21.27 & 25.18 & \cred 27.54 & \cred 25.46 & - & 0.642 & 0.729 & 0.828 & 0.779 & - & 0.293 \\
RGB + G + R + NIR & 21.30 & 25.27 & 27.67 & - & 27.88 & 0.642 & \cg 0.731 & 0.830 & - & 0.849 & 0.292 \\
RGB + G + RE + NIR & 21.24 & 25.30 & - & 25.79 & 28.06 & 0.642 & 0.729 & - & \cg 0.783 & \cg 0.850 & 0.298 \\
RGB + R + RE + NIR & 21.33 & - & 27.59 & 25.48 & \cred 27.76 & 0.644 & - & 0.829 & \cg 0.783 & \cg 0.850 & 0.293 \\
\text{All Bands} & 21.17 & 25.42 & 27.79 & 25.79 & 28.13 & 0.633 & 0.729 & 0.827 & 0.780 & 0.849 & 0.306 \\
\bottomrule
\end{tabular}
}%

\end{table}

\vspace{1cm}

\subsection{Memory \& Training Time}

When considering real-time visualization on edge devices, such as remote controls in the agricultural domain (e.g., immersive NDVI exploration, see Sec.~\ref{sec:agricultre_application}), the efficiency and compression capabilities of our method are of particular importance. 
To this end, we analyzed the memory consumption of the evaluated methods on two datasets (\textsc{Lake} and \textsc{Garden}) after training. 

Compared to \textit{3DGS}, our method \textsc{MS-Splatting} requires only 22\% of the original storage. 
Relative to \textit{TIMS}, we achieved an 88\% reduction on the \textsc{Lake} scene, decreasing memory usage from 2.7\,GB to just 326\,MB. 
This remarkable efficiency results from our \textit{neural color representation}, which jointly encodes cross-spectral information and proves more effective than spherical harmonics. 
By contrast, the immense storage demand of \textit{TIMS} and \textit{ThermalGaussian}$ \dagger$ arises from the large number of spherical harmonic coefficients (16 floats per channel/band per Gaussian). 
An overview of the required storage and the number of primitives per method is provided in the supplements in Tab.~\ref{tab:memory_splats_lake_garden}.

In addition to reduced memory consumption, our method also achieves faster training. 
Averaged over all datasets, \textsc{MS-Splatting} reaches a training time of $02{:}30{:}23$, which is 16\% faster than \textit{ThermalGaussian}$ \dagger$ ($02{:}59{:}16$) and 38\% faster than \textit{TIMS} ($04{:}01{:}35$). 
This speedup originates from the optimization of our compact MLP and feature vectors, in contrast to the substantially larger number of spherical harmonic coefficients.

\subsection{Sparse Captures}
We conducted an experiment in which we lowered the amount of non-RGB spectral images for reconstruction to evaluate performance under the assumption that RGB captures are cheaper and more easily available.
As seen in Tab.~\ref{tab:less_spectral_images} in the supplemental, even with an 80\% reduction in images, the overall reconstruction quality of spectral channels only drops by about 2-3 dB, which still enables passable renderings.

\subsection{Ablation}
\label{ssec:ablation}

In this section, we evaluate the impact of our key design choices. Specifically, we assess the new densification strategy and the effect of the dimension of the feature embedding and network size. Additional ablations can be found in the supplements.

\paragraph*{Densification Strategy}
\label{sssec:densification_strategy}

To evaluate our new multi-spectral densification strategy, we compared the standard {3DGS} against our \textit{Multi-Spectral Aware Densification} strategy in Tab.~\ref{tab:densification_ablation}.
From the results shown, we conclude that our new densification strategy slightly improves our overall performance and peaks at a densification interval of 300. 
The larger interval compared to {3DGS} can be attributed to the average being accumulated over multiple steps to compute the gradient correctly.

\begin{table}[]
\begin{minipage}[t][][]{0.48\linewidth}
\caption[]{\label{tab:densification_ablation}Densification strategy with alternating densification intervals. 3DGS (\textit{Default}) uses an interval of 100 iterations.}
\footnotesize
\centering
\setlength\tabcolsep{2pt}
\resizebox{1.02\columnwidth}{!}{
\begin{tabular}{ l | c c c }
\toprule
Split Interval & PSNR& SSIM & LPIPS\\
\midrule
\textit{Default}-100  & 25.02 & 0.748 & 0.310 \\
\textit{Default}-200  & 25.02  & 0.748 & 0.317\\
\textit{Default}-300  & 25.08  & 0.748 & 0.319 \\
\textit{Default}-400  & 24.93 & 0.749 & 0.320\\
\textit{Default}-500  & 24.86 & 0.747  & 0.323\\
Ours - 100      & 24.93 & 0.750 & \cg 0.267 \\
Ours - 200      & 25.08 & 0.754 & 0.271 \\
Ours - 400      & 25.09 & 0.757 & 0.275 \\
Ours - 300      & \cg 25.16 & \cg 0.758 & 0.271 \\
Ours - 500      & 24.85 & 0.756 & 0.281 \\
\bottomrule
\end{tabular}}
\end{minipage}\hfill
\begin{minipage}[t][][]{0.48\linewidth}
    \caption[]{\label{tab:network_ablation}Impact of MLP size (L: Layer depth, W: Layer width) on rendering quality. Results are averaged over all datasets.}%

\footnotesize
\centering
\setlength\tabcolsep{2pt}
\resizebox{1.02\columnwidth}{!}{
\begin{tabular}{ l | c c c  }
\toprule
Hidden Size & PSNR &SSIM & LPIPS\\
\midrule
$L=0$, $W=16$ & 24.89  & 0.747 & 0.312 \\
$L=0$, $W=32$ & 24.91  & 0.746 & 0.314 \\
$L=0$, $W=64$ & 24.96  & 0.747 & 0.312 \\
$L=0$, $W=128$ & 24.97  & 0.747 & 0.311 \\
$L=1$, $W=16$ & 24.59  & 0.742 & 0.317 \\
$L=1$, $W=32$ & 25.02  & 0.748 & \cg 0.310 \\
$L=1$, $W=64$ & 25.03  & 0.748 & 0.311 \\
$L=1$, $W=128$ & \cg 25.04  &\cg 0.749 & \cg 0.310 \\
$L=2$, $W=16$ & 24.97  & 0.747 & 0.312 \\
$L=2$, $W=32$ & 24.74  & 0.746 & 0.314 \\
$L=2$, $W=64$ & 24.81  & 0.746 & 0.315 \\
$L=2$, $W=128$ &  23.85 & 0.721 & 0.373 \\
\bottomrule
\end{tabular}
}\end{minipage}
\vspace{-3mm}
\end{table}

% wieoso heisst das feature dimension? hier wird mehr getestet, nicht nur dim...
\paragraph*{Feature Dimension}
\label{sssec:feature_dim}

To isolate the effect of the feature embedding dimension, we swept this parameter while keeping the MLP architecture fixed (one hidden layer of width 32). Fig.~\ref{fig:feature_ablation} shows the relative change in PSNR and LPIPS (SSIM follows a similar trend with smaller fluctuations) as functions of the embedding size \(d\). The results indicate that a shallow MLP with \(d=8\) yields the best overall performance. Notably, even \(d=2\) provides sufficient capacity to encode substantial information across all input bands.
Please note that even though we use seven distinct bands in our dataset, representing the view-dependent band radiance can require more than seven parameters, similar to \textit{3DGS} using 48 parameters for view-dependent RGB.

\begin{figure}[h]

\definecolor{MS_RGB}{HTML}{666666}      % RGB composite (red accent, camera standard)
\definecolor{MS_G}{HTML}{4DAF4A}        % Green
\definecolor{MS_R}{HTML}{D73027}        % Red (slightly darker than RGB composite)
\definecolor{MS_RE}{HTML}{FF7F00}       % Red Edge (orange transition zone)
\definecolor{MS_NIR}{HTML}{762A83}      % Near Infrared (deep purple, "beyond red")

    \center
    \hspace{-0.8\linewidth}
    \begin{tikzpicture}
    \begin{axis}[
      font=\tiny,
      width=0.55\linewidth,
      height=5cm,
      xlabel={Feature Dim $d$},
      ylabel={Relative PSNR Change (\%)},
      xtick={2,4,8,16,32,64},
      ytick={-0.5,0,1,2,3,4},
      ymin=-0.6, ymax=4.2,
      xmode=log, log basis x=2,
      grid=both, grid style={gray!30, thin},
      axis background style={fill=gray!5},
      major grid style={gray!40},
      minor tick num=1,
      legend style={
        at={(0.02,0.98)}, anchor=north west,
        draw=gray!50, fill=white, font=\small
      },
      ylabel shift=-0.3cm,
      mark options={scale=1.2},
      thick
    ]
    % RGB
    \addplot[MS_RGB, mark=*, dashed, line width=0.15mm] coordinates {
        (2, 0.0)   (4, 1.4)   (8, 3.2)
        (16,2.8)   (32,2.4)   (64,2.1)
    };

    % G
    \addplot[MS_G, mark=square*, dotted, line width=0.15mm] coordinates {
        (2, 0.0)   (4,-0.1)   (8,1.7)
        (16,1.5)   (32,1.2)   (64,0.8)
    };

    % R
    \addplot[MS_R, mark=triangle*, solid, line width=0.15mm] coordinates {
        (2, 0.0)   (4,0.3)    (8,2.1)
        (16,1.9)   (32,1.9)   (64,1.5)
    };

    % RE
    \addplot[MS_RE, mark=diamond*, dash dot, line width=0.15mm] coordinates {
        (2, 0.0)   (4,-0.4)   (8,1.8)
        (16,1.8)   (32,1.2)   (64,1.0)
    };

    % NIR
    \addplot[MS_NIR, mark=pentagon*, densely dashed, line width=0.15mm] coordinates {
        (2, 0.0)   (4,0.1)    (8,3.0)
        (16,2.7)   (32,1.9)   (64,1.8)
    };

    \end{axis}

    \hspace{0.5\linewidth}

    \begin{axis}[
        font=\tiny,
        width=0.55\linewidth,
        height=5cm,
        xlabel={Feature Dim $d$},
        ylabel={Relative LPIPS Change (\%)},
        xtick={2,4,8,16,32,64},
        ytick={-10,-8,-6,-4,-2,0},
        ymin=-10.5, ymax=0.2,
        xmode=log, log basis x=2,
        grid=both, grid style={gray!30, thin},
        axis background style={fill=gray!5},
        major grid style={gray!40},
        minor tick num=1,
        ytick pos=left,
        legend style={
          at={(0.75,1.2)}, anchor=north east,
          draw=gray!50, fill=white, font=\tiny
        },
        ylabel shift=-0.3cm,
        legend columns=-1
        mark options={scale=1.2},
        thick
        ]
    % RGB
    \addplot[MS_RGB, mark=*,, dashed, line width=0.15mm] coordinates {
        (2,  0.0)   (4, -5.2)   (8, -9.1)
        (16, -8.6)  (32, -8.1)  (64, -7.6)
    };
    \addlegendentry{RGB}

    % G
    \addplot[MS_G, mark=square*, dotted, line width=0.15mm] coordinates {
        (2,  0.0)   (4, -3.1)   (8, -5.3)
        (16, -4.8)  (32, -4.8)  (64, -4.5)
    };
    \addlegendentry{G}

    % R
    \addplot[MS_R, mark=triangle*, solid, line width=0.15mm] coordinates {
        (2,  0.0)   (4, -4.0)   (8, -6.6)
        (16, -6.3)  (32, -6.3)  (64, -6.0)
    };
    \addlegendentry{R}

    % RE
    \addplot[MS_RE, mark=diamond*, dash dot, line width=0.15mm] coordinates {
        (2,  0.0)   (4, -2.8)   (8, -5.0)
        (16, -4.1)  (32, -4.4)  (64, -4.1)
    };
    \addlegendentry{RE}

    % NIR
    \addplot[MS_NIR, mark=pentagon*, densely dashed, line width=0.15mm] coordinates {
        (2,  0.0)   (4, -3.1)   (8, -5.6)
        (16, -4.9)  (32, -4.5)  (64, -4.9)
    };
    \addlegendentry{NIR}

    \end{axis}
    \end{tikzpicture}\vspace{-2mm}
    \caption{Relative changes in PSNR and LPIPS for the lightweight MLP and with varying feature‐embedding dimension $d$\textcolor{\modifycolor}{, reported in percentage ($\%$). The first feature dimension, $d=2^1$, is used as the reference for both plots. All reported changes are relative to this baseline.}}
  \label{fig:feature_ablation}

\end{figure}

\paragraph*{Network Size}
\label{sssec:supp_network_size}

To assess the impact of network capacity, we evaluated our pipeline using MLPs of various sizes. Specifically, we varied the number of hidden layers (0, 1, or 2) and the width of each layer (from 16 to 128 neurons). Tab.~\ref{tab:network_ablation} summarizes the results, averaged over all datasets. Overall, a single hidden layer yields the best trade-off across configurations. Extremely shallow networks (no hidden layers) or very deep, wide networks (two layers of width 128) struggle to process to accurate color representations, resulting in washed-out and dull renderings.

%\begin{table}[h]
%\end{table}

\section{Agriculture Applications}
\label{sec:agricultre_application}

For \textsc{MS-Splatting}, we regard the agricultural domain as the most promising field of application. Novel view synthesis has recently been adopted in the agricultural domain, with applications ranging from horticulture~\cite{FruitNeRF, FruitNeRFpp, AgriNeRF}, to greenhouse monitoring~\cite{pagnerf, TomatoNeRF, nerf_for_tomato}, in-field crop observation~\cite{PanicleNeRF, hq_3D_phenotyping, 3dgs_leaf_size}, and laboratory-based plant phenotyping~\cite{splanting, PeanutNeRF}.

\paragraph*{Vegetation Indices.}
We place particular emphasis on computing various Vegetation Indices (VI). VIs are a spectral imaging transformation designed to quantify vegetation type, plant mass, and health in a scene \cite{vegetation_indices}.
When light penetrates a surface, it is partly reflected, absorbed, or transmitted depending on the material and wavelength.
Soil typically absorbs and reflects visible wavelengths while only little light is being transmitted, whereas vegetation absorbs most viable light but strongly reflects and transmits \nir~wavelengths.
This pronounced NIR reflectance arises from the internal leaf's structure, which efficiently scatters light~\cite{vegetation_indices}. 

Consequently, many vegetation indices, such as the normalized difference vegetation index (NDVI), the green NDVI (GNDVI), and the soil-adjusted vegetation index (SAVI), all exploit the strong NIR reflectance of vegetation. In the simplest form, NDVI is computed on a per-pixel basis as
\begin{equation}
\mathrm{NDVI} = \frac{\mathrm{NIR} - R}{\mathrm{NIR} + R}.
\label{eq:ndvi}
\end{equation}
This formula implicitly assumes that the red and NIR sensors are perfectly co-registered or that the scene distance is large relative to the sensor baseline, as is typically the case for top-down drone with high altitude or satellite imagery. However, in near-field scenarios (orchards, greenhouses, phenotyping rigs), separate multi-spectral sensors often capture images from slightly different viewpoints. Without precise image registration, even small misalignment (for example, due to wind or vehicle motion) can introduce significant artifacts.

\paragraph*{Evaluation.}
Mutual information (MI)–based registration is deemed the baseline, since MI is well known for aligning multi-modal imagery without assuming any specific intensity relationships between channels~\cite{ms-registration-mi, ms-registration, mm-registration-mi}. Details are provided in the supplementary material, however, results as shown in the left panel of Fig.~\ref{fig:motivation_image2} of the supplement are unsatisfying.

\begin{figure}
\centering
        \includegraphics[width=\linewidth]{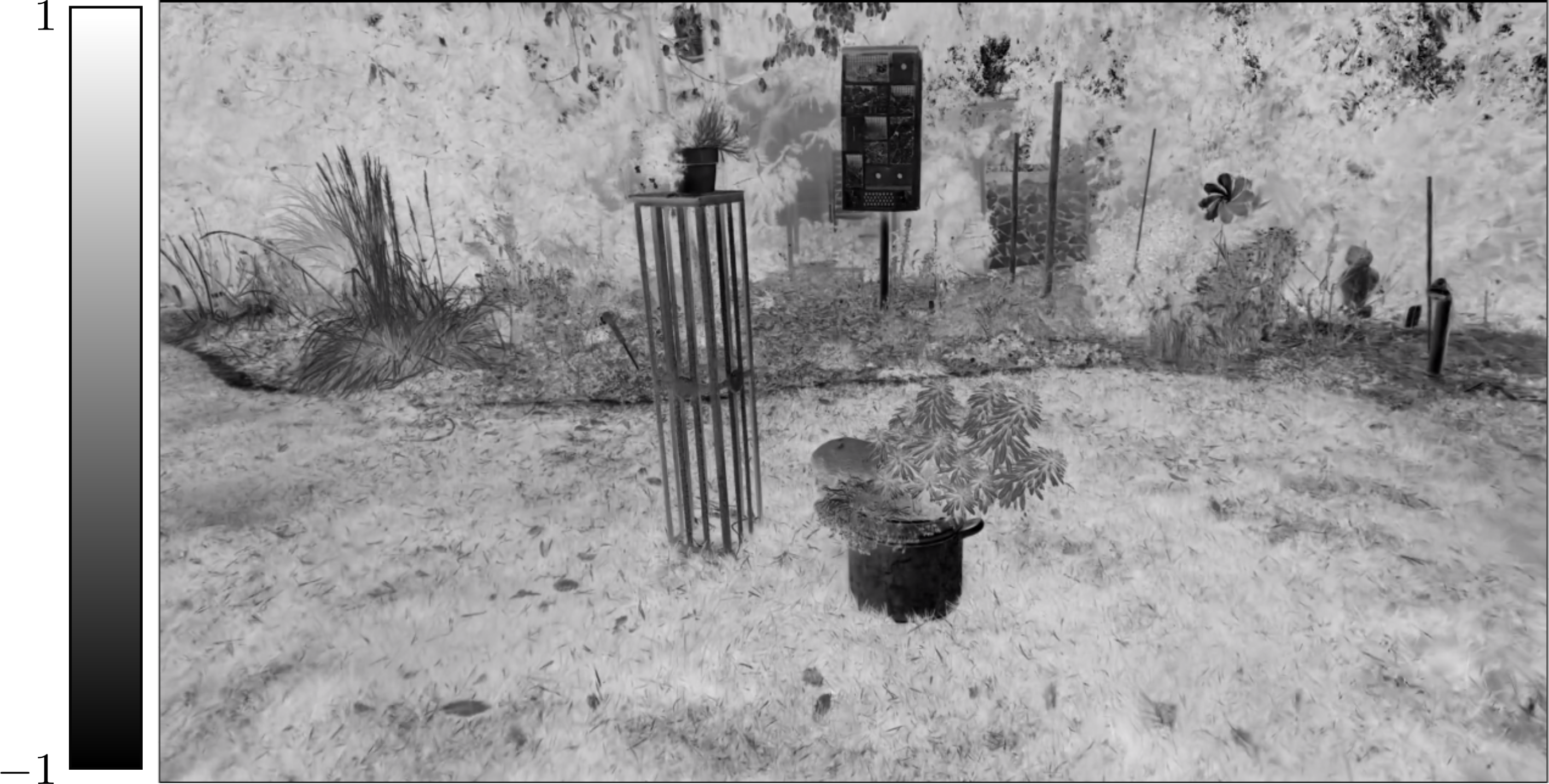}
\caption{NDVI rendered by synthesizing the red and \nir~channels via \textsc{MS-Splatting} from a novel perspective. Values between $-1$ and $0$ indicate inanimate object, $0$ to $0.33$ diseased, $0.33$ to $0.66$ moderately healthy and $0.66$ to $1$ very healthy plants.} % (Right) Spectral-feature clustering based on the learned per-splat embeddings, with clusters reflecting similar spectral properties.
\label{fig:application_figure_ndvi}
\end{figure}

To avoid these registration challenges entirely, we exploit our \textsc{MS-Splatting} model for NDVI (and other VI) rendering. 
%While novel view synthesis has not been used previously to compute vegetation indices, a related work trains directly on precomputed NDVI from laboratory hyperspectral scans~\cite{oefelein2023projektstudie}. 
%In contrast, our method jointly learns a neural scene representation across all spectral bands, allowing us to render any VI “on the fly” by simply synthesizing the red and NIR images and combining them via Eq.~(\ref{eq:ndvi}). 
An example NDVI output from \textsc{MS-Splatting} is shown in Fig.~\ref{fig:application_figure_ndvi}.
For qualitative evaluation, NDVI through NVS relies on the NVS quality of \nir~and narrow-band red, for which the evaluation of Tabs.~\ref{tab:large_comparison} and \ref{tab:comparisonxnerf} showcase state of the art results, outperforming related works by more than 0.5 dB. 
We also note that for NDVI rendering, only reconstructing the model from RGB, R and NIR show even slightly better results (see Tab.~\ref{tab:spectralcrosstalk}).

The supplement included several additional experiments:
As we test in Sec.~\ref{sec:vi_eval_long_baseline} our representation is also able to achieve great results in a synthetic setup.
%Lastly, in Figs.~\ref{fig:error_spectral_spider} and \ref{fig:value_spectral_spider}, we visualize reconstruction quality spider plots. 

\if false

If it comes to near-field applications such as in orchards, greenhouses, or optical plant phenotyping vegetation index computation with multi-spectral cameras becomes a challenging task. Due to separate sensors recording from different positions at varying time steps a image registration is necessary
Without the use of image registration it would neglect any temporal and spatial misalignment, e.g. due to external influences and lead to a significant image misalignment, which can be seen in Fig.~\ref{fig:motivation_image2} on the left. 
As a baseline we used mutual information (MI) as an multi-spectral image registration technique. It is well known to work on multi-modal data \cite{ms-registration-mi, ms-registration, mm-registration-mi} and makes no assumption regarding the intensity relations between multi-modal data. More details can be found at the supplements.

To tackle this problem to render the NDVI (or any other VI) we utilized our \textsc{MS-Splatting} model. Since Novel view synthesis has not been used yet to compute VIs. The only comparable work is from \cite{oefelein2023projektstudie} which directly trains on precomputed NDVI data from a hyperspectral camera in a laboratory setup.
Through our approach we can render multiple VIs according to the available spectral-bands with no need of cumbersome image registration. 
To render e.g. the NDVI, we simply render the same image for both NIR and red modalities and combine them according to Eq.~\ref{eq:ndvi}. A visualization of the NDVI rendered by \textsc{MS-Splatting} is depicted in Fig.~\ref{fig:application_figure_ndvi} on the left.

\todo[inline]{Highlight the benefits! Near-field. etc. }

handheld devices or small drones enable ground-based collection of multi-spectral data, providing more detailed insights into different plant layers.

Vegetation indices can be generated from various data sources with different spatial resolutions. Satellite imagery, such as data from Landsat \cite{landsat}, enables large-scale differentiation between soil and vegetation while also tracking temporal variations. Advanced satellite platforms like Sentinel-2 and Landsat-8 provide imagery with spatial resolutions ranging from  10 and 30 m$^2$ per pixel and offer a revisit cycle of approximately seven days \cite{agricolus2025}. However, the long-term acquisition of satellite data can become cost-prohibitive, limiting its feasibility for continuous monitoring \cite{VIc1353691}.

Farmers who prefer not to rely on external service providers can use multi-spectral drones to map their land and compute relevant vegetation indices using commercial software \cite{pix4dfields, webodm}. A standard surveying approach involves capturing an area from a bird’s-eye view. However, for indices like the Normalized Difference Vegetation Index (NDVI), the corresponding wavelengths do not penetrate leaves, making the index applicable only to the canopy cover.

For near-field applications such as orchards, greenhouses, or optical plant phenotyping, it is beneficial to analyze not only the canopy layer but also the underlying vegetation structure. In such cases, handheld devices or small drones enable ground-based collection of multi-spectral data, providing more detailed insights into different plant layers.

A common limitation of commercial multi-spectral cameras is their reliance on multiple distinct sensors for capturing individual spectral bands. For example, the \emph{DJI Mavic 3 Multispectral} (DJI 3M) \cite{dji_mavic_3m} is equipped with five separate camera sensors. This design presents challenges, including differences in sensor positioning, which can complicate the accurate computation of vegetation indices in near-field scenarios. Additionally, because spectral bands are captured sequentially, there is a risk of temporal misalignment, especially in dynamic environments. A misalignment is visualized in Fig.~\ref{fig:motivation_image2} and demonstrate the erroneous overlay of two multi-spectral bands. This misalignment is due to the sequential recording of the data and is caused by a dynamic environment and the drones internal flight stabilization mechanism.

\fi

\section{Limitations and Future Work}
\textcolor{\modifycolor}{One limitation of our work is that, despite strong quantitative performance, the reconstructed RGB images can occasionally appear dull. We attribute this effect to the fact that both the per-primitive feature embeddings and the neural network continue to adapt until the end of training, allowing subtle color shifts caused by spectral cross-talk to persist. This behavior arises because extremely shallow networks tend to underfit color, whereas very deep and wide networks may require substantially more training iterations to achieve accurate color reproduction. Incorporating a multispectral-aware color loss or an additional RGB color-regularization term could mitigate these effects and help better preserve the original color appearance.
In the same context, it would be interesting to investigate whether all rasterized Gaussian's are relevant for all spectral bands, or whether certain Gaussian's can be ignored for specific bands (e.g., NIR, which is typically less textured). This could be achieved by additionally optimizing a per-band opacity term~\cite{3dgs_with_nerf_sigma} that selectively amplifies or suppresses individual Gaussian's for particular spectral channels.}

\textcolor{\modifycolor}{Additionally in the pipeline of SfM, feature matching across multi-spectral images should be investigated further. While the matching is effective and reliable when images are captured from approximately the same viewpoint and with similar (high) spatial resolution, it remains to be investigated how strongly varying image resolutions and significantly different camera poses affect reconstruction quality.}

%A more in-depth study is needed to understand how the feature-embedding design affects color fidelity.

For future work, it would be valuable to evaluate this framework on truly high-dimensional hyperspectral data. Additionally, following the example of Hyperspectral NeRF~\cite{Chen24arxiv_HS-NeRF}, learning a continuous per-wavelength representation could enable interpolation between wavelengths and further improve spectral reconstruction.
In agricultural settings, a real-time implementation (e.g., LiveNVS \cite{livenvs}) or even an immersive multi-spectral Gaussian splatting system \cite{franke2025vrsplatting} could make this use case far more accessible. While our approach is able to render in real-time, this was not the focus of our work. As such, performance can be increased with e.g. selected MLP evaluations based on frustum and/or occlusion culling.
\vspace{1cm}

Moreover, incorporating hierarchical partitioning methods~\cite{kerbl2024hierarchical} may enable larger unrestricted captures. While our current setup manages memory conservatively, this extension would be beneficial with the advent of drones with enhanced battery capabilities for long capture sequences.

Finally, generating artificial dataset~\cite{PEGASUS} using 3D Gaussian Splatting could benefit from the extension to multi-spectral data and the improved RGB rendering quality.

\section{Conclusion}

To conclude, we present \textsc{MS-Splatting}, a novel pipeline for view synthesis on multi-spectral images. At its core lies a unified \textit{neural color representation} that jointly encodes all spectral channels into a shared feature embedding. A tiny MLP then decodes this embedding to accurately recover each band’s appearance. Our extensive experiments show that this collaboration through a joint feature space leverages both spatial and spectral coherence, yielding consistent cross-spectral improvements across every evaluation metric.
Furthermore, we introduced a multi-spectral outdoor dataset spanning RGB, green, red, red-edge, and \nir~bands, which together with our code will be available openly.

\section*{Acknowledgments}
We thank \textbf{Martin Hundhausen} for supporting the scene recordings.
Lukas Meyer was funded by the 5G innovation program of the German Federal Ministry for Digital and Transport under the funding code 165GU103C.
Maximilian Weiherer was funded by the German Federal Ministry of Education and Research (BMBF), FKZ: 01IS22082 (IRRW). 
Linus Franke was supported in part by the ERC Advanced Grant NERPHYS (101141721, {https://project.inria.fr/nerphys}).
The authors are responsible for the content of this publication.
The authors gratefully acknowledge the scientific support and HPC resources provided by the Erlangen National High Performance Computing Center (NHR@FAU) of the Friedrich-Alexander-Universität Erlangen-Nürnberg (FAU) under the NHR project \textit{b162dc}. NHR funding is provided by federal and Bavarian state authorities. NHR@FAU hardware is partially funded by the German Research Foundation (DFG) – 440719683.

%%%%%%%%% REFERENCES
% Bibliography
\bibliographystyle{eg-alpha-doi}
\bibliography{egbib}

\clearpage

\FloatBarrier 
\newpage
\beginsupplement

\setcounter{section}{0}

\section*{Supplemental Material}

In this supplementary material, we provide extra ablation studies
and details on the dataset, experiments and implementation.

\section{Ablation on \textit{MS-Splatting}}

We extended the ablation studies described in Sec.~\ref{ssec:ablation} by an evaluation on the feature normalization (Sec.~\ref{sssec:sup_feature_norm}) and the positional encoding (Sec.~\ref{sssec:supp_positional_encoding}). 
Additionally, we provide the full tables of all channels in Tab.~\ref{tab:supp_densification_strategy},  Tab.~\ref{tab:supp_feature_emb}, Tab.~\ref{tab:supp_network_size}, Tab.~\ref{tab:supp_pos_encoding} .

\subsection{Feature Normalization}
\label{sssec:sup_feature_norm}

We also evaluated the impact of the feature normalization on our neural color representation. In Tab.~\ref{tab:supp_norm_ablation} both runs with and without normalization are shown. This leads to only a minor improvement on the PSNR. Besides this we sometimes encountered numerical instability with no feature normalization.

\begin{table}[h]
\caption[]{Evaluation of feature normalization. NoNorm has feature normalization deactivated}
\footnotesize
\centering
\begin{tabular}{ l | c c c  }
\toprule
\# Features & PSNR $(\uparrow)$ &SSIM $(\uparrow)$ & LPIPS $(\downarrow)$\\
\midrule
NoNorm & \cy 24.80& \cg 0.748  & \cg 0.310  \\
Norm    & \cg 25.02&  \cg 0.748 & \cg  0.310 \\
\bottomrule
\end{tabular}

\label{tab:supp_norm_ablation}
\end{table}

\subsection{Positional Encoding}
\label{sssec:supp_positional_encoding}

FeatSplat~\cite{martins2024featuresplattingbetternovel} proposed to input the viewing direction as an additional parameter to the MLP. Here, we tried whether the \textit{normalized} viewing direction or the normalized splat position can improve the results. Additionally, we tested positional encoding on both arguments with 0, 5, and 10 frequencies. In the case of 0, only the viewing direction with no encoding is meant. The results are shown in Tab.~\ref{tab:pe_ablation}. It is clearly visible that positional encoding on the viewing direction improves the result, while the positional encoding on the splats position degrades the evaluation results. From a rendering perspective, adding the viewing direction to the MLP seems to be similar to the view-dependent representation of spherical harmonics.

\begin{table}[h]
\caption{Comparison of positional encoding on the viewing direction (DE) and the splat position (PE). }
\footnotesize
\centering
\begin{tabular}{ l | c c c }
\toprule
Encoding & PSNR $(\uparrow)$ & SSIM $(\uparrow)$ & LPIPS $(\downarrow)$\\
\midrule
None          & 25.02 & 0.748 & 0.310 \\
PE[0]         & 24.48 & 0.739 & 0.326 \\
PE[5]         & 24.62 & 0.743 & 0.320 \\
PE[10]        & 24.57 & 0.743 & 0.322 \\
DE[0]         & \cg 25.29 & \cg 0.754 & \cg 0.304 \\
DE[5]         & 25.23 & 0.746 & 0.313 \\
DE[10]        & \cg  25.29 & 0.745 & 0.315 \\
PE[0]-DE[0]   & 25.09 & 0.753 & 0.307 \\
PE[5]-DE[5]   & 25.12 & 0.747 & 0.314 \\
PE[10]-DE[10] & 25.17 & 0.746 & 0.316 \\
\bottomrule
\end{tabular}
\label{tab:pe_ablation}
\end{table}

\section{Dataset Setup}

In the scope of this work we also propose the \textit{MS-Splatting} datset. This multi-spectral dataset consists out of seven outdoor scenes comprising forward-facing, bird-view and 360$^\circ$ captures.  

\subsection{Hardware}

\begin{figure}[h]
    \centering
        \vspace{-0.2cm}
    \includegraphics[width=0.8\linewidth]{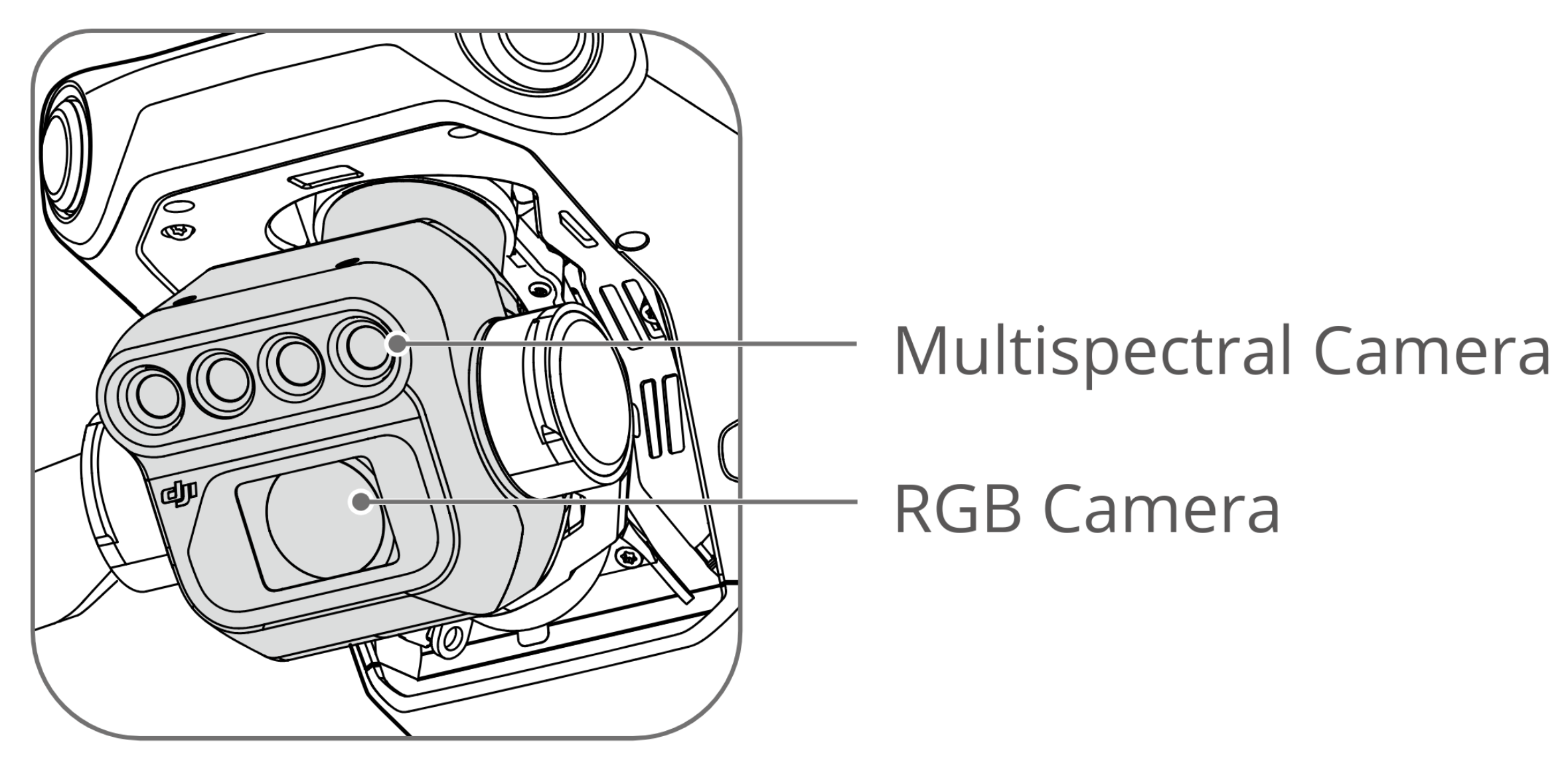}
    \caption{DJI Mavic 3M spatial layout of the cameras. Image taken from \cite{DJI2024_Mavic3M_UserManual}.}
    \label{fig:dji3m_spatial_layout}
    \vspace{-0.3cm}
\end{figure}

For our data acquisition, we employed the DJI Mavic 3M drone \cite{dji_mavic_3m}, which is equipped with a dual-camera system: one RGB sensor and four multi-spectral cameras covering the red (R), green (G), red-edge (RE), and near-infrared (NIR) bands. The RGB camera provides an image size of $5280 \times 3956$ pixels with a focal length of 24\,mm. Each multi-spectral camera is a single-band sensor with a resolution of $2592 \times 1944$ pixels and an equivalent focal length of 25\,mm. The spatial arrangement of the DJI Mavic 3M’s cameras is illustrated in Fig.~\ref{fig:dji3m_spatial_layout}. The minimum capture interval is 0.7\,s for RGB images and 2\,s when recording RGB and multi-spectral images simultaneously. Since the RGB and multi-spectral data are captured sequentially, spatial misalignment may occur, particularly in dynamic scenes or due to motion compensation by the drone’s stabilization system.

The four multi-spectral cameras are designed to capture both visible and invisible radiance. The Green (MS-G, 560$\pm$16\,nm) and Red (MS-R, 650$\pm$16\,nm) bands overlap with the RGB spectrum (G: 530--575\,nm, R: 590--630\,nm), but their narrower bandwidth provides higher spectral precision. The non-visible spectrum is covered by the Red-Edge (RE, 730$\pm$16\,nm) and Near-Infrared (NIR, 860$\pm$26\,nm) bands. The spectral response curves of these sensors were measured by Atherton \emph{et al.}~\cite{Atherton2024} 
and are shown in Fig.~\ref{fig:dji3m_spectral_response}.
\begin{figure}[h]
    \centering
    \resizebox{1.0\linewidth}{!}{
    \input{images/spectral_overlay_pgfplots_snippet}
    }
    \caption{DJI Mavic 3M spectral response curve. Data taken from \cite{Atherton2024, Kuurasuo2025}}
    \label{fig:dji3m_spectral_response}
\end{figure}

\subsection{Image Registration}

\begin{figure}[h]
    \centering
    \includegraphics[width=0.7\linewidth]{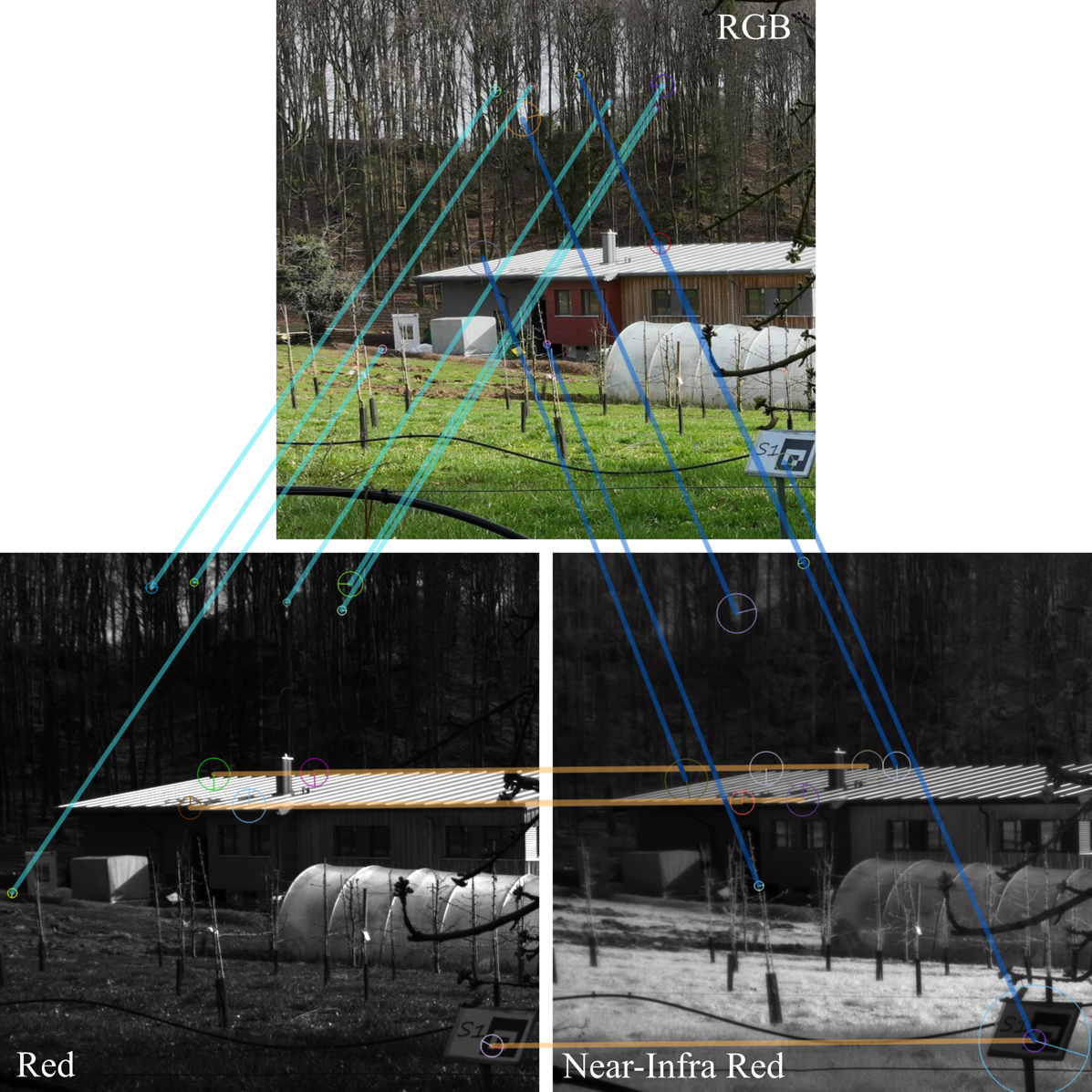}
    \caption{Exemplary SIFT~\cite{lowe1999object} (as used by COLMAPs SfM~\cite{schonberger2016structure}) feature matching between RGB, red and \nir~ channel. For visualization purpose only the best 6 matches are depicted.%\bernhard{SIFT und COLMAP tauchen im paper nicht auf. vermutlich wird irgendwo implizit gesagt, dass das genutzt wird, durch die angabe des frameworks, vielleicht kann man da noch expliziter sein}
    }
    \label{fig:sift_feature_matching}
\end{figure}

\textcolor{\modifycolor}{Our method relies on a standard SfM pipeline by Schönberger~\cite{schonberger2016structure}. We model each spectral band as an individual camera, using a per-band camera model with shared intrinsics across the entire scene.}
\textcolor{\modifycolor}{For the registration step, we employ SIFT features. Alignment across modalities such as NIR and RGB relies not only on texture but also on geometric structures, including object edges. Examples include roof lines and tree crowns, as visualized in Fig.~\ref{fig:sift_feature_matching}. These geometric cues enable reliable matching despite spectral differences and appearance variations caused by differing camera poses.}

\section{Multi-Spectral Metrics}

\paragraph*{Spectral Angular Mapper.}
The Spectral Angle Mapper (SAM)~\cite{Kruse_SAM} is a widely used method for spectral classification and comparison. It quantifies the similarity between two spectra by computing the angle between a $B$-dimensional reference feature vector $\vect{r}$ and a target vector $\vect{t}$. These feature vectors are constructed by stacking the reflectance values of all spectral bands into $B$-dimensional vectors at the pixel level. The resulting angle provides a measure of spectral similarity: smaller angles indicate a closer match between the spectra. Since the metric depends only on the angle between vectors, SAM is illumination invariant, as it does not account for vector magnitude~\cite{Kruse_SAM}. SAM is closely related to cosine similarity but yields an angle in radians instead of a normalized similarity score. It is computed as:

\begin{equation}
\text{SAM}(r, t) =
\arccos \left(
  \frac{\displaystyle \sum_{i=1}^{n} r_i \, t_i}
       {\sqrt{\displaystyle \sum_{i=1}^{n} r_i^2} \;
        \sqrt{\displaystyle \sum_{i=1}^{n} t_i^2}}
\right)
\end{equation}

where $\vect{r} \in \mathbb{R}^B$ is the reference feature vector, $\vect{t} \in \mathbb{R}^B$ is the target feature vector, and $B$ is the number of spectral bands.

\paragraph*{Spectral Correlation Mapper.}
The Spectral Correlation Mapper (SCM)~\cite{carvalho_SCM} is derived from the Pearson Correlation Coefficient and is closely related to SAM. In contrast to SAM, SCM centers the data around its mean, making it robust not only to scaling but also to offsets. The metric ranges from $-1$ (no correlation) to $1$ (perfect correlation). SCM is computed as:

\begin{equation}
\text{SCM}(r, t) =
\frac{\displaystyle \sum_{i=1}^{B} \left(r_i - \bar{r}\right)\left(t_i - \bar{t}\right)}
{\sqrt{\displaystyle \sum_{i=1}^{B} \left(r_i - \bar{r}\right)^2} \;
 \sqrt{\displaystyle \sum_{i=1}^{B} \left(t_i - \bar{t}\right)^2}}
\end{equation}

where $\vect{r} \in \mathbb{R}^B$ is the reference feature vector, $\vect{t} \in \mathbb{R}^B$ is the target feature vector, and $B$ is the number of spectral bands. The mean values are defined as $\bar{r} = \tfrac{1}{B} \sum_{i=1}^{B} r_i$ and $\bar{t} = \tfrac{1}{B} \sum_{i=1}^{B} t_i$.

\paragraph*{Spectral Information Divergence.}
The Spectral Information Divergence (SID)~\cite{Chang_SID} is an information-theoretic measure of spectral similarity.  
Instead of directly comparing reflectance values as vectors, SID interprets each spectrum as a probability distribution over the $B$ spectral bands.  
This is achieved by normalizing the reflectance values of spectra $\vect{r}$ and $\vect{t}$ into probability distributions $p$ and $q$, defined as $p_i = \tfrac{r_i}{\sum_{j=1}^{B} r_j}$ and $q_i = \tfrac{t_i}{\sum_{j=1}^{B} t_j}$.  
The divergence between the two distributions is then quantified using the Kullback–Leibler (KL) divergence in both directions to ensure symmetry.  
Small SID values indicate high similarity between spectral signatures, while larger values reflect greater divergence in spectral shape.  
Since SID compares relative distributions across bands rather than absolute reflectance magnitudes, it is invariant to scaling effects such as illumination changes.  
SID is computed as:

\begin{equation}
\text{SID}(r, t) =
\sum_{i=1}^{B} p_i \, \log \left(\frac{p_i}{q_i}\right)
+ \sum_{i=1}^{B} q_i \, \log \left(\frac{q_i}{p_i}\right)
\end{equation}

where $\vect{r}, \vect{t} \in \mathbb{R}^B$ are the reference and target feature vectors, and $p, q \in \mathbb{R}^B$ are the corresponding normalized probability distributions over $B$ spectral bands.

\section{Thermal Gaussian Re-implementation}
\label{sec:sup_tg_reimplementation}

\begin{table*}
\caption[]{Comparison between original ThermalGaussian (TG) on our warped dataset (W), our re-implementation (TG-Ours) on the warped dataset, and our re-implementation on the original dataset (O). Averaged across all 7 scenes.}
\scriptsize
\resizebox{1.0\textwidth}{!}{

\begin{tabular}{ l | c c c c c c | c c c c c c | c c c c c c | }
\toprule
& \multicolumn{6}{c|}{PSNR $(\uparrow)$} & \multicolumn{6}{c|}{SSIM $(\uparrow)$} & \multicolumn{6}{c|}{LPIPS $(\downarrow)$}\\
\midrule
& All & RGB & G & R & RE & NIR & All & RGB & G & R & RE & NIR & All & RGB & G & R & RE & NIR\\
\midrule
TG (W) & \cy 23.54 & \cy 19.81 & \cy 23.98 & \cy 25.66 & \cy 23.78 & \cy 24.49 & \cy 0.700 & \cy 0.565 & \cy 0.710 & \cy 0.763 & \cy 0.717 & 0.743 & \cy 0.342 & \cy 0.371 & \cy 0.336 & \cy 0.305 & \cy 0.336 & 0.362\\
TG-Ours (W) & 23.14 & 18.81 & 23.83 & 25.27 & 23.38 & 24.42 & 0.676 & 0.482 & 0.697 & 0.752 & 0.701 & \cy 0.749 & 0.378 & 0.483 & 0.367 & 0.331 & 0.357 & \cy 0.355\\
TG-Ours (O) & \cg 25.65 & \cg 21.17 & \cg 25.42 & \cg 27.79 & \cg 25.79 & \cg 28.13 & \cg 0.763 & \cg 0.633 & \cg 0.729 & \cg 0.827 & \cg 0.780 & \cg 0.849 & \cg 0.266 & \cg 0.306 & \cg 0.293 & \cg 0.240 & \cg 0.257 & \cg 0.232\\
\bottomrule
\end{tabular}}
\label{tab:tg_original_vs_tg_ours}
\end{table*}

To allow for proper comparison of methods on our datasets, we adapted the OMMG (One Multi-Modal Gaussian) approach introduced by ThermalGaussian \cite{thermalgaussian} for multi-spectral view-independent input data. For this, analog to their method, we model each Gaussian as containing separate spherical harmonics (SH) coefficients for every spectral band, e.g. RGB, MS-G, MS-R, MS-RE and NIR. 

To allow for independent camera poses across the multi-spectral dataset, during each training iteration, we select a single image from a given spectral band, and optimize only the SH coefficients corresponding to that band. This is different to the method of \cite{thermalgaussian}, since they render and optimize both the RGB and thermal channel in a single forward pass. Optimization of Gaussian positions, scales, rotations and opacities is unchanged and performed during all iterations. Standard Gaussian densification, as described by \cite{kerbl3Dgaussians}, is performed every 100 iterations.

Tab. \ref{tab:tg_original_vs_tg_ours} compares our re-implementation (TG-Ours) with the original ThermalGaussian implementation (TG). For TG, we utilized the warped version of our dataset and trained in four separate runs, each selecting the RGB images combined with one multi-spectral band, trained for 30,000 iterations. TG-Ours ran on the combined set of all 5 channels in a single model, once with warped images, and once with the original data. Since our approach splits the training of different bands into consecutive iterations, we chose 120,000 iterations. In the average over all 7 scenes, we observe that our re-implementation performs slightly worse than the original ThermalGaussian on warped datasets. However, TG-ours can leverage multi-spectral images with independent camera positions, which avoids errors introduced by image warping. We observe that TG-Ours on the original images consistently outperforms original ThermalGaussian. For this reason, we conclude that our re-implementation appropriately represents the method utilized by \cite{thermalgaussian} and we use it in the evaluation of our method.

\section{X-NeRF-Dataset Training Details} 
\label{sec:x_nerf_training}
For a proper comparison of the methods 3DGS~\cite{kerbl3Dgaussians}, \textit{ThermalGaussian}$ \dagger$, \textsc{TIMS}~\cite{gruen2025towards} and our method on the X-NeRF dataset~\cite{poggi2022xnerf} we adapted the training strategies to better fit the larger amount of training data (3 channels RGB, 10 channels multi-spectral (MS) and 1 channel infra-red (IR)).

For 3DGS, we trained 30.000 iterations of standard Gaussian Splatting separately on every modality (RGB, MS and IR), resulting in 12 models per scene. For \textit{ThermalGaussian}$ \dagger$, \textsc{TIMS}, and our method, all channels were combined in a single model. To account for the added spectral information, training was extended to 240,000 iterations for these three methods. For our method, the feature dimension was additionally increased to $f=16$. Detailed evaluation results can be found in Tabs. \ref{tab:x_nerf_evaluation_large} and \ref{tab:x_nerf_evaluation_large_spectral}. For comparability and reproducibility, we provide the set of test image indices used in every X-NeRF scene in Tab. \ref{tab:xnerf_dataset_indices}.

\section{Thermal Evaluation}

We evaluated each scene using the original ThermalGaussian~\cite{thermalgaussian}, \textit{ThermoNeRF}~\cite{hassan2024thermonerf}, and our implementation. We additionally added the smoothness loss from ThermalGaussian. It encourage spatial smoothness, and is defined through:
\begin{equation}
    \mathcal{L}_{\text{smooth}}
= \frac{1}{4M} \sum_{i,j}
\Bigl(\bigl|T_{i\pm1,j}-T_{i,j}\bigr|
      +\bigl|T_{i,j\pm1}-T_{i,j}\bigr|\Bigr)\,.
\end{equation}
where $T_{i,j}$ denotes the rendered thermal pixel at image coordinates $(i,j)$, and $M$ is the total number of rendered pixels.  While RGB images often exhibit sharp changes at object boundaries, thermal images typically vary smoothly in a local neighborhood.  Therefore, this loss yields smoother thermal reconstructions.

\begin{table}[b]
\caption[]{Comparison of the approaches \textit{ThermoNeRF}, ThermalGaussian~\cite{thermalgaussian} and our approach one six thermal scenes, combining RGB and thermal (T) images . Three scenes are from \textit{ThermoNeRF} \cite{mert} (FACE, LION, PAN) and three scenes from \textit{ThermoNeRF}~\cite{hassan2024thermonerf} (BUILD-A  (buildingA\_spring), ROBOT (double\_robot), KETTLE (heater\_water\_kettle))}
\scriptsize
%\backslashbox{{\tiny Method}}{{\tiny Modality}}
\resizebox{1.0\linewidth}{!}{

\begin{tabular}{ c | l | c c | c c | c c | }
\toprule
             & Metric & \multicolumn{2}{c|}{PSNR $(\uparrow)$} & \multicolumn{2}{c|}{SSIM $(\uparrow)$} & \multicolumn{2}{c|}{LPIPS $(\downarrow)$}\\ \midrule
 Dataset     &      & RGB   & T  & RGB    & T &  RGB   & T \\\midrule
\multirow{3}{*}{FACE}             & Thermal Gaus.  & 22.05 &  \cg 27.65   & 0.816  & 0.824   & 0.187  & \cg 0.217 \\
            & ThermoNeRF        & 9.76 & 10.43  & 0.254 & 0.395  & 0.712 & 0.657 \\
            & Ours               & \cg  26.74 & 28.73 & \cg  0.904 & \cg  0.874 & \cg  0.132 &  0.235 \\
            & Ours + $\mathcal{L}_{\text{smooth}}$         & 24.62 & 26.50 & 0.884 & 0.864 & 0.148 & 0.228 \\
            \midrule
\multirow{3}{*}{LION}   & Thermal Gaus.  & \cg 31.42 & \cg 30.21    & 0.936 & \cg 0.934    & 0.121 & \cg 0.122 \\
            & ThermoNeRF                 & 20.12 & 23.88  & 0.723 & 0.735  & 0.180 & 0.422 \\
            & Ours                        & 29.21 & 28.30 & 0.923 & 0.913 & 0.140 & 0.150 \\
            & Ours + $\mathcal{L}_{\text{smooth}}$        & 31.41 & 29.90 & \cg  0.948 & 0.931 & \cg  0.115 & 0.141 \\
            \midrule
\multirow{3}{*}{PAN}            & Thermal Gaus. & 29.23 & \cg 36.03    & 0.932 & \cg 0.939    & 0.123 & \cg 0.028 \\
            & ThermoNeRF        & 15.96 & 22.75  & 0.608 & 0.774  & 0.353 & 0.136 \\
            & Ours              & \cg 29.75 & 34.59 & \cg 0.946 & 0.929 &\cg 0.106 & 0.052 \\
            & Ours + $\mathcal{L}_{\text{smooth}}$        & 29.65 & 35.30 & 0.945 & 0.930 & 0.107 & 0.050 \\
            \midrule
\multirow{3}{*}{BUILD-A}            & Thermal Gaus.  & 22.70 & 28.79 & 0.812 & 0.975 & 0.253 & 0.121 \\
            & ThermoNeRF                             & 19.74 & 26.27  & 0.635 & 0.931  & 0.250 & 0.185 \\
            & Ours                                   & 23.24 & 28.60 & 0.838 & 0.973 & 0.193 & 0.098 \\
            & Ours + $\mathcal{L}_{\text{smooth}}$   & \cg 24.89 & \cg 28.85 & \cg 0.877 & \cg 0.976 & \cg 0.157 & \cg 0.077 \\
            \midrule
\multirow{3}{*}{ROBOT}           & Thermal Gaus.       & 19.90 & 30.62 & 0.727 & 0.944   & 0.353 & 0.118 \\
            & ThermoNeRF                               & 17.35 & 27.56  & 0.615 & 0.888  & 0.300 & 0.151 \\
            & Ours                                     & 22.64 & 33.00 & 0.836 & \cg 0.959 & 0.237 & 0.059 \\
            & Ours + $\mathcal{L}_{\text{smooth}}$     & \cg 23.15 & \cg 33.13 & \cg 0.845 & \cg 0.959 & \cg 0.226 & \cg 0.046 \\
            \midrule
\multirow{3}{*}{KETTLE}     & Thermal Gaus.  & 24.15 & \cg  34.73 & 0.719 & \cg  0.968 & 0.319 &  0.035 \\
           & ThermoNeRF              & 16.30 & 23.83  & 0.438 & 0.865  & 0.458 & 0.221 \\
           & Ours                    & \cg 26.27 & 33.89 & 0.848 & 0.955 & \cg 0.187 & \cg 0.030 \\
           & Ours + $\mathcal{L}_{\text{smooth}}$          & 26.24 & 33.72 & \cg 0.850 & 0.950 & \cg 0.187 & 0.031 \\

\bottomrule
\end{tabular}}

\label{tab:thermal_evaluation}
\end{table}

\begin{figure*}
     \centering
     \includegraphics[width=\linewidth]{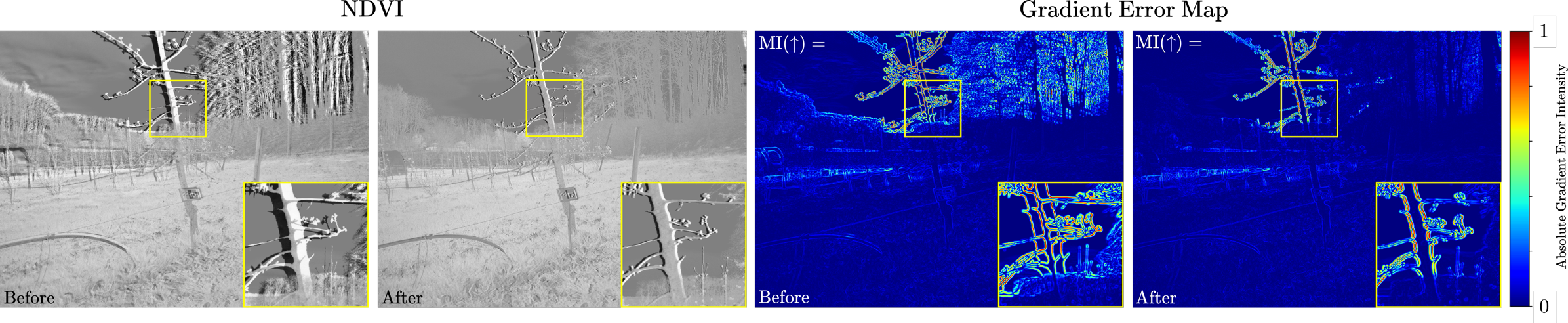}
      \caption{Image registration based on mutual information (MI). The top row visualizes an overlay of the \nir (NIR) and red edge (RE) channel with (right) and without registration (left). The bottom row shows the edge-based error map with (right) and without registration (left). The registration reduces the visual artifacts but artifacts at the top of the tree still remain.}
      \label{fig:motivation_image2}
\end{figure*}

We trained \textit{ThermalGaussian}~\cite{thermalgaussian} and \textit{ThermoNeRF}~\cite{hassan2024thermonerf} for 30,000 iterations, whereas our model was trained for 60,000 iterations.  We set the feature dimension to $d=8$, and use a single hidden layer with 32 neurons.  In addition, we concatenate the viewing direction $\vect{d}$ to the feature embedding $\vect{f}$.  We adapt our multi-spectral Gaussian strategy, as we first train on RGB only for 5,000 iterations, then continue with joint RGB–thermal optimization.  This allows the gaussian model to learn coarse geometry from RGB before incorporating the less spatially detailed thermal modality.
See Tab.~\ref{tab:thermal_evaluation} for quantitative results and Fig.~\ref{fig:eval_thermal_imags} for a visual comparison.

\section{Multi-Spectral Image Registration for Vegetation Indices}

As a baseline for our novel view synthesis vegetation index computation we need a method to estimate a transformation $T$ to map our target image $A$ to a reference image $B$. 
Therefore we utilize an image registration method which is based on mutual information (MI). This method is mainly used for the registration of multi-modal data such as computed tomography and magnetic resonance data \cite{ms-registration-mi, ms-registration, mm-registration-mi}.
The concept states that MI is maximized if two images are geometrically aligned and thus makes no assumption regarding the intensity relations between multi-modal data.
This is beneficial for multi-spectral channels such as NIR as it has less textural information such as red and green channels.
%\begin{figure}[b]
%     \centering
%     \includegraphics[width=\linewidth]{images/mi-baseline.png}
%      \caption{Image registration based on mutual information (MI). The top row visualizes an overlay of the \nir (NIR) and red edge (RE) channel with (right) and without registration (left). The bottom row shows the edge-based error map with (right) and without registration (left). The registration reduces the visual artifacts but artifacts at the top of the tree still remain.}
%      \label{fig:motivation_image2}
%\end{figure}

Mutual information \cite{mm-registration-mi} measures the degree of dependence between $A$ and $B$ by measuring the similarity between the joint probability distribution $p(a, b)$ and the product of the marginal distributions $p(a), p(b)$. In the case of complete independence MI converges towards $0$ and for dependent distributions MI maximizes with increasing similarity between $p(a)$ and $p(b)$. 
MI is defined by 
\begin{equation}
I(A;B) = \sum_{a \in A} \sum_{b \in B} p(a, b) \log \left( \frac{p(a, b)}{p(a) p(b)} \right),
\end{equation}
whereas $A$ and $B$ are random variables, $p(a, b)$ is defined as the joint probability distribution and $p(a), p(b)$ are the marginal distributions. If we consider $a$ and $b$ to be the images intensity values of corresponding pixel pairs in both images we can estimate the joint and marginal distributions by obtaining the joint and marginal histograms of the image pair \cite{mm-registration-mi}.

By maximizing the mutual information trough the application of a rigid transformation $T(\mathbf{x}) = R\mathbf{x} + \mathbf{t}$ on the target image $B$ we can determine a feasible set of registration parameters trough,
\begin{equation}
T^* = \operatorname*{argmax}_T I(A;T \circ B).
\end{equation}
By using Powell's multidimensional method \cite{powell} to maximize the mutual information $I(A;B)$ and defining a limited search space for the translation $t_x$ and $t_y$ and the rotation $\phi$ we can find a feasible transformation matrix to align our multi-spectral data. 

In Fig.~\ref{fig:motivation_image2} we demonstrate on the left the computation of the NDVI without and with MI image registration. To quantify the structural discrepancies between two images, we compute a gradient-based error map by measuring the absolute differences between their gradient magnitudes (Sobel), which is visualized on the right in Fig.~\ref{fig:motivation_image2}.

A result of an image registration of a NIR and red channel image is depicted in  Fig.~\ref{fig:motivation_image2}. In the top row an overlay of both images are visualized before and after registration. On the left image the blurry artifacts are visible and on the right a reduced intensity of errors is visible. This is also shown in the bottom row of Fig.~\ref{fig:motivation_image2}. It visualizes the edge/gradient error map of the registered images and emphasizes the error minimization due to the registration process. Despite image registration, the error remains present, albeit less significant.

\section{Vegetation Index Evaluation}\label{sec:vi_eval_long_baseline}

To ensure that a vegetation index such as NDVI is properly represented with \textit{MS-Splatting}, we designed a small dataset consisting out of 8 static scenes with only one image per band each.  
Due to the camera baseline $b$ of the NIR and R multi-spectral camera, it is necessary to record scenes with a depth such that $d \gg b$ to guarantee that parallax remains in the sub-pixel range.  
During capture, the drone was stationary and placed on a planar surface.  

To correct internal camera misalignment and account for the varying focal lengths of RGB and multi-spectral images, we warped the multi-spectral images into the RGB plane.  
First, we applied a factory-calibrated homography~\cite{DJI2023Mavic3MImageProcessingGuide} for coarse registration, followed by a second homography computed via feature matching for fine tuning.  
All images were then adjusted to the same resolution.  
For training, we used the five images, which where recorded from the exact same location, and followed the standard training strategy.  
We obtained an overall PSNR of 46\,dB and an SSIM of 0.991 averaged over all bands and datasets. After training, we computed NDVI for both the ground-truth and the predicted images. The predicted NDVI had an PSNR of  43\,dB and an SSIM of 0.977. 
Although this simplified evaluation setup yields artificially high scores, it demonstrates that our algorithm has the representational capacity for not only multiple spectral bands but also to capture vegetation indices such as NDVI.

\section{Multi-Spectral Clustering}
\label{ssec:segmentation_application}

Another valuable downstream application of \textit{MS-Splatting} is material-aware surface clustering. By grouping splats whose neural feature embeddings share similar reflectance properties, we can -- for example -- count fruits, distinguish healthy from diseased vegetation, or segment objects by material type. A nice side effect of our approach is that all of this information is already captured in the learned feature embedding.

\begin{figure}[h]
\centering
        \includegraphics[width=\linewidth]{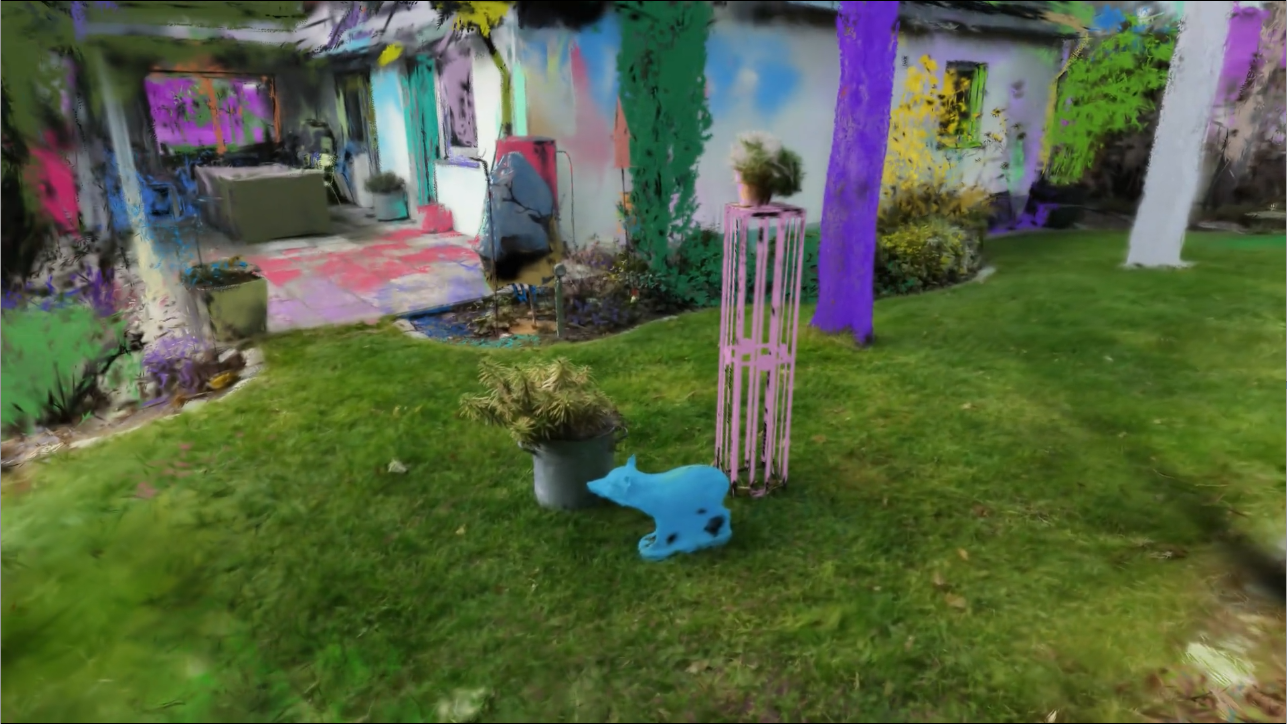}
\caption{Spectral-feature clustering based on the learned per-splat embeddings, with clusters reflecting similar spectral properties} 
\label{fig:application_figure}
\vspace{-0.2cm}
\end{figure}

To further encourage local consistency, we introduce a cosine-similarity loss between each feature and its spatial neighbors:
\begin{equation}
     \mathcal{L}_{\text{cos}}(\vect{f}_c, \mathcal{F}_s) = \frac{1}{|\mathcal{F}_s|} \sum_{\vect{f}_c \in \mathcal{F}_s}  \Bigl( 1 - \frac{\langle \vect{f}_c, \vect{f}_s\rangle}{\lVert \vect{f}_c \rVert  \lVert \vect{f}_s\rVert}  \Bigr),
\end{equation}
where $\vect{f}_c$ is the center and $\mathcal{F}_s$ is the set of the $k$ nearest neighbors (in Euclidean space). During training, we rebuild the $k$-NN graph ($k=16$) after each densification, split, or prune operation.
Once training is complete, we extract for each Gaussian splat its 3D position and corresponding feature vector, and then cluster these points using HDBSCAN~\cite{hdbscan}. Each resulting cluster is visualized to reveal material-consistent regions. An example of this clustering in the Garden scene is shown in Fig.~\ref{fig:application_figure}.

\begin{table}[h]
\caption{\label{tab:xnerf_dataset_indices}List of image indices in the test split of the respective scene from the X-NeRF dataset~\cite{poggi2022xnerf}. Same indices used across all channels. The first image has index 1.}
\centering
\small
\begin{tabular}{c | c  @{\hskip 1cm}  c | c}
\textbf{dataset} & \textbf{indices} & \textbf{dataset} & \textbf{indices} \\
\midrule
black     & 1, 8, 15, 22, 30 & hall      & 1, 8, 16, 23, 31 \\
bluechair & 1, 8, 15, 22, 30 & hall2     & 1, 7, 13, 20, 26, 33 \\
cvlab     & 1, 8, 15, 22, 30 & hall3     & 1, 8, 16, 23, 31 \\
fruits    & 1, 8, 15, 22, 30 & hall4     & 1, 8, 16, 23, 31 \\
green     & 1, 8, 15, 22, 30 & orange    & 1, 8, 15, 22, 30 \\
penguin   & 1, 8, 15, 22, 30 & penguin2  & 1, 8, 15, 22, 30 \\
dino      & 1, 8, 15, 22, 30 & puppets   & 1, 8, 15, 22, 30 \\
bins      & 1, 8, 15, 22, 30 & tech      & 1, 8, 15, 22, 30 \\
\end{tabular}
\end{table}

\begin{table*}[h]
\caption[]{Evaluation on our densification strategy with all spectral bands. Densification interval with \textit{Default} and \textit{MaxAverage} (Ours)Strategy. 3DGS default is listed as \textit{Default} - 100. The colors indicate the \colorbox{green!40}{best} and \colorbox{red!40}{worst} results.}
\scriptsize
\resizebox{1.0\textwidth}{!}{

\begin{tabular}{ l | c c c c c c | c c c c c c | c c c c c c | }
\toprule
& \multicolumn{6}{c|}{PSNR $(\uparrow)$} & \multicolumn{6}{c|}{SSIM $(\uparrow)$} & \multicolumn{6}{c|}{LPIPS $(\downarrow)$}\\
\midrule
& All & RGB & G & R & RE & NIR & All & RGB & G & R & RE & NIR & All & RGB & G & R & RE & NIR\\
\midrule

\textit{Default} - 100 & 25.02 & 20.67 & 24.83 & 27.10 & 25.12 & 27.43 & 0.748 & 0.600 & 0.717 & 0.815 & \cred 0.767 & 0.841 & 0.310 & 0.351 & 0.340 & 0.283 & 0.304 & 0.272\\
\textit{Default} - 200 & 25.02 & 20.73 & 24.82 & 27.12 & 25.12 & 27.38 & 0.748 & 0.599 & 0.717 & 0.815 & \cred 0.767 & 0.841 & 0.317 & 0.360 & 0.346 & 0.289 & 0.310 & 0.278\\
\textit{Default} - 300 & 25.08 & 20.67 & 24.95 & 27.23 & 25.18 & 27.44 & 0.748 & 0.599 & 0.718 & 0.815 & 0.768 & 0.843 & 0.319 & 0.363 & 0.349 & 0.291 & 0.311 & 0.279\\
\textit{Default} - 400 & 24.93 & 20.71 & 24.70 & 26.96 & 24.97 & 27.36 & 0.749 & 0.599 & 0.718 & 0.816 & 0.768 & 0.844 & 0.320 & 0.365 & 0.350 & 0.293 & 0.313 & 0.280\\
\textit{Default} - 500 & 24.86 & \cred 20.55 & 24.65 & \cred 26.83 & 24.97 & 27.35 & \cred 0.747 & \cred 0.596 & 0.717 & \cred 0.814 & \cred 0.767 & 0.843 & \cred 0.323 & \cred 0.368 & \cred 0.353 & \cred 0.295 & \cred 0.315 & \cred 0.282\\

Ours - 100 & 24.93 & 20.57 & 24.75 & 27.17 & 24.97 & 27.24 & 0.750 & 0.613 & \cred 0.716 & 0.816 & \cred 0.767 & \cred 0.839 & \cg 0.267 & \cg 0.305 & \cg 0.292 & \cg 0.244 & \cg 0.260 & \cg 0.235\\
Ours - 200 & 25.08 & 20.85 & 24.87 & 27.22 & 25.07 & 27.46 & 0.754 & 0.618 & 0.720 & 0.819 & 0.771 & 0.842 & 0.271 & 0.307 & 0.298 & 0.249 & 0.263 & 0.238\\
Ours - 300 & \cg 25.16 & \cg 20.91 & \cg 24.97 & \cg 27.38 & \cg 25.19 & 27.40 & \cg 0.758 & \cg 0.623 & \cg 0.723 & \cg 0.822 & \cg 0.775 & 0.845 & 0.271 & 0.308 & 0.298 & 0.248 & 0.262 & 0.237\\
Ours - 400 & 25.09 & 20.78 & 24.91 & 27.21 & 25.09 & \cg 27.52 & 0.757 & 0.620 & \cg 0.723 & \cg 0.822 & \cg 0.775 & \cg 0.846 & 0.275 & 0.314 & 0.303 & 0.252 & 0.267 & 0.241\\
Ours- 500 & \cred 24.85 & 20.59 & \cred 24.61 & 26.99 & \cred 24.94 & \cred 27.20 & 0.756 & 0.617 & \cg 0.723 & 0.820 & 0.774 & \cg 0.846 & 0.281 & 0.319 & 0.309 & 0.258 & 0.271 & 0.247\\

\bottomrule
\end{tabular}
}
\label{tab:supp_densification_strategy}

\end{table*}

\begin{table*}[h]
\caption[]{Comparison of variation in the neural feature embedding dimension $d$. Additionally to the main thesis we evaluated a deep MLP of four hidden layers ($D=4$) and 32 neurons per layer ($W=32$). The shallow MLP has a one hidden layer (D=1) and 32 neurons per layer (L=32). For the deep and shallow network, feature normalization is activated by default. Thus for the shallow MLP we deactivate feature normalization (NoNorm). The colors indicate the \colorbox{green!40}{best} and \colorbox{red!40}{worst} results.}

\scriptsize
\resizebox{1.0\textwidth}{!}{

\begin{tabular}{ l | c c c c c c | c c c c c c | c c c c c c | }
\toprule
& \multicolumn{6}{c|}{PSNR $(\uparrow)$} & \multicolumn{6}{c|}{SSIM $(\uparrow)$} & \multicolumn{6}{c|}{LPIPS $(\downarrow)$}\\
\midrule
& All & RGB & G & R & RE & NIR & All & RGB & G & R & RE & NIR & All & RGB & G & R & RE & NIR\\
\midrule
d = 2-NoNorm & 24.74 & 20.42 & 24.62 & 26.71 & 24.92 & 27.10 & 0.743 & 0.594 & 0.712 & 0.809 & 0.765 & 0.837 & 0.320 & 0.371 & 0.349 & 0.290 & 0.309 & 0.278\\
d = 4-NoNorm & 24.66 & 20.47 & 24.54 & 26.84 & 24.66 & 26.86 & 0.746 & 0.598 & 0.716 & 0.813 & 0.766 & 0.840 & 0.313 & 0.358 & 0.341 & 0.284 & 0.305 & 0.274\\
d = 8-NoNorm & 24.98 & 20.73 & 24.79 & \cg 27.14 & 24.98 & 27.31 & \cg 0.749 & \cg 0.603 & \cg 0.718 & \cg 0.816 & \cg 0.768 & \cg 0.842 & \cg 0.308 & \cg 0.349 & \cg 0.338 & \cg 0.281 & \cg 0.300 & \cg 0.270\\
d = 16-NoNorm & 24.80 & 20.61 & 24.65 & 26.94 & 24.84 & 27.02 & 0.748 & 0.601 & 0.717 & 0.814 & 0.767 & 0.840 & 0.310 & \cg 0.349 & 0.340 & 0.284 & 0.303 & 0.273\\
d = 32-NoNorm & 24.95 & 20.69 & 24.76 & 27.08 & 24.97 & 27.27 & 0.748 & 0.602 & 0.717 & 0.815 & \cg 0.768 & \cg 0.842 & 0.310 & 0.351 & 0.340 & 0.284 & 0.304 & 0.272\\
d = 64-NoNorm & 24.56 & 20.43 & 24.34 & 26.75 & 24.55 & 26.76 & 0.742 & 0.592 & 0.710 & 0.809 & 0.761 & 0.837 & 0.323 & 0.364 & 0.354 & 0.297 & 0.315 & 0.283\\
d = 2-Deep & \cred 24.40 & \cred 20.11 & \cred 24.31 & \cred 26.48 & 24.50 & 26.66 & \cred 0.738 & \cred 0.585 & \cred 0.708 & \cred 0.806 & \cred 0.760 & \cred 0.835 & \cred 0.333 & \cred 0.385 & \cred 0.361 & \cred 0.304 & \cred 0.323 & \cred 0.292\\
d = 4-Deep & 24.59 & 20.45 & 24.48 & 26.71 & 24.60 & 26.77 & 0.744 & 0.594 & 0.714 & 0.810 & 0.765 & 0.837 & 0.320 & 0.368 & 0.349 & 0.292 & 0.311 & 0.280\\
d = 8-Deep & 24.63 & 20.49 & 24.58 & 26.84 & 24.61 & 26.70 & 0.745 & 0.596 & 0.716 & 0.812 & 0.764 & 0.837 & 0.320 & 0.366 & 0.348 & 0.291 & 0.312 & 0.282\\
d = 16-Deep & 24.63 & 20.33 & 24.54 & 26.73 & 24.70 & 26.92 & 0.744 & 0.594 & 0.714 & 0.809 & 0.765 & 0.839 & 0.319 & 0.366 & 0.347 & 0.292 & 0.310 & 0.278\\
d = 32-Deep & 24.41 & 20.26 & 24.38 & 26.53 & \cred 24.42 & \cred 26.51 & 0.741 & 0.588 & 0.712 & 0.807 & 0.761 & 0.836 & 0.328 & 0.374 & 0.356 & 0.300 & 0.320 & 0.288\\
d = 64-Deep & 24.70 & 20.41 & 24.61 & 26.79 & 24.80 & 26.93 & 0.743 & 0.593 & 0.713 & 0.811 & 0.763 & \cred 0.835 & 0.322 & 0.368 & 0.351 & 0.294 & 0.313 & 0.282\\
d = 2 & 24.50 & \cred 20.11 & 24.47 & 26.59 & 24.67 & 26.70 & 0.739 & \cred 0.585 & 0.709 & \cred 0.806 & 0.761 & \cred 0.835 & 0.329 & 0.384 & 0.357 & 0.302 & 0.317 & 0.286\\
d = 3 & 24.93 & 20.54 & 24.81 & \cg 27.14 & 25.05 & 27.17 & 0.745 & 0.596 & 0.714 & 0.812 & 0.767 & 0.838 & 0.319 & 0.367 & 0.349 & 0.290 & 0.309 & 0.278\\
d = 4 & 24.55 & 20.39 & 24.45 & 26.67 & 24.56 & 26.74 & 0.744 & 0.593 & 0.714 & 0.811 & 0.764 & 0.838 & 0.317 & 0.364 & 0.346 & 0.290 & 0.308 & 0.277\\
d = 5 & 24.56 & 20.34 & 24.51 & 26.70 & 24.52 & 26.76 & 0.742 & 0.593 & 0.712 & 0.809 & 0.761 & 0.836 & 0.319 & 0.364 & 0.347 & 0.293 & 0.312 & 0.281\\
d = 6 & 24.85 & 20.64 & 24.68 & 27.05 & 24.86 & 27.06 & 0.747 & 0.600 & 0.716 & 0.814 & 0.765 & 0.840 & 0.312 & 0.356 & 0.341 & 0.284 & 0.305 & 0.274\\
d = 7 & 24.73 & 20.49 & 24.57 & 26.85 & 24.76 & 27.02 & 0.745 & 0.600 & 0.715 & 0.812 & 0.764 & 0.839 & 0.314 & 0.357 & 0.342 & 0.287 & 0.307 & 0.275\\
d = 8 & \cg 25.06 & \cg 20.75 & \cg 24.88 & \cg 27.14 & 25.11 & \cg 27.51 & \cg 0.749 & 0.602 & \cg 0.718 & 0.815 & \cg 0.768 & \cg 0.842 & \cg 0.308 & \cg 0.349 & \cg 0.338 & 0.282 & 0.301 & \cg 0.270\\
d = 16 & 25.02 & 20.67 & 24.83 & 27.10 & \cg 25.12 & 27.43 & 0.748 & 0.600 & 0.717 & 0.815 & 0.767 & 0.841 & 0.310 & 0.351 & 0.340 & 0.283 & 0.304 & 0.272\\
d = 32 & 24.92 & 20.59 & 24.77 & 27.09 & 24.97 & 27.22 & 0.747 & 0.599 & \cg 0.718 & 0.814 & 0.767 & 0.840 & 0.311 & 0.353 & 0.340 & 0.283 & 0.303 & 0.273\\
d = 64 & 24.85 & 20.54 & 24.67 & 26.98 & 24.91 & 27.18 & 0.747 & 0.598 & \cg 0.718 & 0.814 & 0.767 & 0.841 & 0.311 & 0.355 & 0.341 & 0.284 & 0.304 & 0.272\\
\bottomrule
\end{tabular}}
\label{tab:supp_feature_emb}

\end{table*}

\begin{table*}[h]
\caption[]{Variation on the spectral feature decoding network. We sweep over different layer depths $L$ and layer widths $W$. The colors indicate the \colorbox{green!40}{best} and \colorbox{red!40}{worst} results.}

\scriptsize
\resizebox{1.0\textwidth}{!}{

\begin{tabular}{ l | c c c c c c | c c c c c c | c c c c c c | }
\toprule
& \multicolumn{6}{c|}{PSNR $(\uparrow)$} & \multicolumn{6}{c|}{SSIM $(\uparrow)$} & \multicolumn{6}{c|}{LPIPS $(\downarrow)$}\\
\midrule
& All & RGB & G & R & RE & NIR & All & RGB & G & R & RE & NIR & All & RGB & G & R & RE & NIR\\
\midrule

$L=0$, $W=16 $& 24.89 & 20.57 & 24.73 & 27.00 & 24.97 & 27.24 & 0.747 & 0.598 & 0.717 & 0.814 & 0.766 & 0.840 & 0.312 & 0.352 & 0.341 & 0.285 & 0.306 & 0.275\\
$L=0$, $W=32 $& 24.91 & 20.58 & 24.80 & 27.06 & 24.98 & 27.19 & 0.746 & 0.596 & 0.716 & 0.813 & 0.765 & 0.840 & 0.314 & 0.356 & 0.343 & 0.287 & 0.308 & 0.276\\
$L=0$, $W=64 $& 24.96 & 20.64 & 24.87 & 27.02 & 25.02 & 27.33 & 0.747 & 0.599 & 0.717 & 0.813 & 0.766 & 0.841 & 0.312 & 0.353 & 0.340 & 0.285 & 0.305 & 0.274\\
$L=0$, $W=128$ & 24.97 & 20.65 & 24.75 & 27.11 & 25.03 & 27.39 & 0.747 & 0.599 & 0.717 & 0.815 & 0.767 & 0.841 & 0.311 & 0.353 & 0.340 & 0.284 & \cg 0.304 & 0.273\\
$L=0$, $W=192$ & \cg 25.04 & 20.61 & 24.91 & 27.21 & \cg 25.14 & 27.37 & 0.748 & 0.600 & 0.718 & 0.815 & \cg 0.768 & \cg 0.842 & 0.311 & 0.353 & 0.340 & 0.283 & 0.305 & 0.273\\
$L=0$, $W=256$ & 24.80 & 20.68 & 24.63 & 26.94 & 24.78 & 27.02 & 0.747 & 0.599 & 0.717 & 0.814 & 0.766 & 0.839 & 0.312 & 0.353 & 0.340 & 0.284 & 0.306 & 0.275\\

$L=1$, $W=16 $& 24.59 & 20.29 & 24.33 & 26.66 & 24.70 & 27.02 & 0.742 & 0.593 & 0.711 & 0.808 & 0.763 & 0.838 & 0.317 & 0.361 & 0.345 & 0.290 & 0.309 & 0.277\\
$L=1$, $W=32 $& 25.02 & 20.67 & 24.83 & 27.10 & 25.12 & \cg 27.43 & 0.748 & 0.600 & 0.717 & 0.815 & 0.767 & 0.841 & \cg 0.310 & \cg 0.351 & 0.340 & 0.283 & \cg 0.304 & \cg 0.272\\
$L=1$, $W=64 $& 25.03 & \cg 20.73 & 24.85 & 27.13 & 25.13 & 27.39 & 0.748 & \cg 0.602 & 0.717 & \cg 0.816 & 0.766 & 0.840 & 0.311 & 0.353 & 0.340 & \cg 0.282 & \cg 0.304 & 0.273\\
$L=1$, $W=128$ & \cg 25.04 & 20.68 & \cg 24.93 & \cg 27.22 & 25.10 & 27.36 & \cg 0.749 & \cg 0.602 & \cg 0.719 & \cg 0.816 & \cg 0.768 & 0.841 & \cg 0.310 & 0.353 & \cg 0.339 & 0.283 & \cg 0.304 & 0.273\\
$L=1$, $W=192$ & 24.72 & 20.41 & 24.63 & 26.88 & 24.68 & 27.04 & 0.743 & 0.594 & 0.714 & 0.809 & 0.762 & 0.837 & 0.325 & 0.371 & 0.351 & 0.296 & 0.319 & 0.288\\
$L=1$, $W=256$ & 24.34 & 20.33 & 24.28 & 26.40 & 24.45 & 26.27 & 0.732 & 0.581 & 0.706 & 0.801 & 0.752 & 0.824 & 0.351 & 0.405 & 0.371 & 0.318 & 0.344 & 0.316\\

$L=2$, $W=16 $& 24.97 & 20.65 & 24.79 & 27.08 & 25.05 & 27.33 & 0.747 & 0.599 & 0.717 & 0.815 & 0.767 & 0.841 & 0.312 & 0.356 & 0.340 & 0.284 & \cg 0.304 & 0.273\\
$L=2$, $W=32 $& 24.74 & 20.60 & 24.52 & 26.89 & 24.71 & 27.01 & 0.746 & 0.599 & 0.717 & 0.813 & 0.765 & 0.840 & 0.314 & 0.356 & 0.342 & 0.287 & 0.309 & 0.275\\
$L=2$, $W=64 $& 24.81 & 20.62 & 24.66 & 26.96 & 24.87 & 27.02 & 0.746 & 0.598 & 0.716 & 0.813 & 0.766 & 0.839 & 0.315 & 0.359 & 0.343 & 0.287 & 0.306 & 0.276\\
$L=2$, $W=128$ & 23.85 & 19.82 & 23.95 & 25.96 & 23.99 & 25.54 & 0.721 & 0.558 & 0.697 & 0.792 & 0.744 & 0.816 & 0.373 & 0.435 & 0.396 & 0.337 & 0.362 & 0.332\\
$L=2$, $W=192$ & 20.17 & 15.80 & 20.58 & 22.58 & 20.54 & 21.36 & 0.611 & 0.443 & 0.591 & 0.694 & 0.627 & 0.700 & 0.502 & 0.582 & 0.520 & 0.461 & 0.496 & 0.454\\
$L=2$, $W=256$ & \cred 16.25 & \cred 12.79 & \cred 16.38 & \cred 18.14 & \cred 15.73 & \cred 18.26 & \cred 0.480 & \cred 0.313 & \cred 0.446 & \cred 0.497 & \cred 0.519 & \cred 0.627 & \cred 0.689 & \cred 0.798 & \cred 0.725 & \cred 0.651 & \cred 0.669 & \cred 0.600\\

\bottomrule
\end{tabular}
}

\label{tab:supp_network_size}
\end{table*}

\begin{table*}
\caption[]{Combination of different positional encoding. Either encoding on the splat position (PE) or the normalized viewing direction (DE). The available frequencies are either [5] or [10] and [0] indicates the plain vector with no encoding. The colors indicate the \colorbox{green!40}{best} and \colorbox{red!40}{worst} results.}

\scriptsize
\resizebox{1.0\textwidth}{!}{

\begin{tabular}{ l | c c c c c c | c c c c c c | c c c c c c | }
\toprule
& \multicolumn{6}{c|}{PSNR $(\uparrow)$} & \multicolumn{6}{c|}{SSIM $(\uparrow)$} & \multicolumn{6}{c|}{LPIPS $(\downarrow)$}\\
\midrule
& All & RGB & G & R & RE & NIR & All & RGB & G & R & RE & NIR & All & RGB & G & R & RE & NIR\\
\midrule

None & 25.02 & 20.67 & 24.83 & 27.10 & 25.12 & 27.43 & 0.748 & 0.600 & 0.717 & 0.815 & 0.767 & 0.841 & 0.310 & 0.351 & 0.340 & 0.283 & 0.304 & 0.272\\
PE[0] & \cred 24.48 & \cred 20.04 & \cred 24.45 & \cred 26.64 & \cred 24.55 & \cred 26.79 & \cred 0.739 & \cred 0.584 & \cred 0.710 & \cred 0.807 & \cred 0.760 & \cred 0.836 & \cred 0.326 & \cred 0.375 & \cred 0.355 & \cred 0.297 & \cred 0.317 & \cred 0.284\\
PE[5] & 24.62 & 20.28 & 24.50 & 26.67 & 24.68 & 27.04 & 0.743 & 0.593 & 0.713 & 0.809 & 0.763 & 0.838 & 0.320 & 0.366 & 0.349 & 0.291 & 0.312 & 0.281\\
PE[10] & 24.57 & 20.14 & \cred 24.45 & 26.70 & 24.74 & 26.89 & 0.743 & 0.590 & 0.713 & 0.810 & 0.763 & 0.837 & 0.322 & 0.367 & 0.351 & 0.293 & 0.316 & \cred 0.284\\

DE[0] & \cg 25.29 & \cg 20.86 & \cg 25.12 & 27.29 & \cg 25.45 & \cg 27.82 & \cg 0.754 & \cg 0.612 & \cg 0.724 & \cg 0.820 & 0.771 & \cg 0.843 & \cg 0.304 & \cg 0.347 & \cg 0.333 & \cg 0.276 & \cg 0.296 & \cg 0.265\\
DE[5] & 25.23 & 20.65 & 25.01 & \cg 27.41 & 25.34 & 27.80 & 0.746 & 0.599 & 0.716 & 0.814 & 0.765 & 0.840 & 0.313 & 0.359 & 0.343 & 0.284 & 0.304 & 0.272\\
DE[10] & \cg 25.29 & 20.83 & 25.06 & 27.40 & \cg 25.45 & 27.78 & 0.745 & 0.597 & 0.715 & 0.813 & 0.763 & 0.838 & 0.315 & 0.363 & 0.344 & 0.286 & 0.307 & 0.275\\

PE[0]-DE[0] & 25.09 & 20.83 & 24.90 & 27.20 & 25.15 & 27.44 & 0.753 & 0.610 & \cg 0.724 & \cg 0.820 & \cg 0.772 & \cg 0.843 & 0.307 & 0.353 & 0.336 & 0.278 & 0.299 & 0.268\\
PE[5]-DE[5] & 25.12 & 20.65 & 24.95 & 27.16 & 25.27 & 27.65 & 0.747 & 0.599 & 0.718 & 0.814 & 0.765 & 0.840 & 0.314 & 0.360 & 0.344 & 0.285 & 0.306 & 0.274\\
PE[10]-DE[10] & 25.17 & 20.63 & 24.99 & 27.27 & 25.28 & 27.73 & 0.746 & 0.597 & 0.716 & 0.813 & 0.764 & 0.838 & 0.316 & 0.365 & 0.345 & 0.287 & 0.307 & 0.276\\

\bottomrule
\end{tabular}
}
\label{tab:supp_pos_encoding}
\end{table*}

\begin{table*}

\caption[]{Comparison on GARDEN scene. The colors indicate the \colorbox{green!40}{best} and \colorbox{yellow!40}{second best} results.}

\scriptsize
\centering
\setlength{\tabcolsep}{3pt}

\resizebox{\textwidth}{!}{
\begin{tabular}{ l | c | c c c c c c | c c c c c c | c c c c c c | }
\toprule
 & & \multicolumn{6}{c|}{PSNR $(\uparrow)$} & \multicolumn{6}{c|}{SSIM $(\uparrow)$} & \multicolumn{6}{c|}{LPIPS $(\downarrow)$}\\
\midrule
 & IC & All & RGB & G & R & RE & NIR & All & RGB & G & R & RE & NIR & All & RGB & G & R & RE & NIR\\
\midrule
ThermalGaus. & $\times$ & 26.10 & 23.45 & 26.30 & 25.06 & 28.95 & 26.72 & 0.739 & 0.690 & 0.732 & 0.717 & 0.795 & 0.762 & 0.315 & 0.290 & 0.309 & 0.294 & 0.329 & 0.351\\
ThermoNeRF & $\times$ & 22.91 & 18.72 & 23.51 & 21.27 & 26.93 & 24.10 & 0.589 & 0.371 & 0.615 & 0.555 & 0.704 & 0.700 & 0.438 & 0.457 & 0.433 & 0.436 & 0.439 & 0.426\\
3DGS & \checkmark & 26.91 & \cy 23.47 & 26.85 & 26.19 & 29.38 & 28.69 & 0.777 & \cy 0.692 & 0.747 & 0.775 & 0.825 & 0.843 & 0.270 & \cy 0.217 & 0.281 & 0.235 & 0.313 & 0.303\\
ThermalGaus.$ \dagger$ & \checkmark & 27.14 & 23.14 & \cy 27.19 & 26.13 & 30.05 & 29.20 & 0.761 & 0.646 & 0.742 & 0.759 & 0.824 & 0.833 & 0.315 & 0.305 & 0.326 & 0.299 & 0.326 & 0.316\\
TIMS & \checkmark & \cg 27.77 & \cg 24.14 & \cg 27.53 & \cg 26.95 & \cg 30.50 & \cg 29.74 & \cg 0.796 & \cg 0.718 & \cg 0.765 & \cg 0.797 & \cg 0.845 & \cg 0.855 & \cg 0.239 & \cg 0.213 & \cy 0.253 & \cy 0.226 & \cy 0.256 & \cg 0.247\\
MultiSpec-FeatSplat & \checkmark & 26.80 & 22.85 & \cy 27.19 & \cy 26.85 & 29.39 & 27.74 & \cy 0.787 & 0.691 & \cy 0.756 & \cy 0.795 & \cy 0.842 & \cy 0.850 & \cy 0.243 & 0.249 & \cg 0.249 & \cg 0.219 & \cg 0.250 & \cy 0.249\\
Ours & \checkmark & \cy 27.44 & 23.11 & \cg 27.53 & 26.64 & \cy 30.36 & \cy 29.56 & 0.779 & 0.676 & 0.750 & 0.782 & 0.837 & 0.849 & 0.255 & 0.251 & 0.266 & 0.241 & 0.264 & 0.254\\
\bottomrule
\end{tabular}
}

\end{table*}

\begin{table*}

\caption[]{Comparison on SINGLE TREE scene. The colors indicate the \colorbox{green!40}{best} and \colorbox{yellow!40}{second best} results.}

\scriptsize
\centering
\setlength{\tabcolsep}{3pt}

\resizebox{\textwidth}{!}{
\begin{tabular}{ l | c | c c c c c c | c c c c c c | c c c c c c | }
\toprule
 & & \multicolumn{6}{c|}{PSNR $(\uparrow)$} & \multicolumn{6}{c|}{SSIM $(\uparrow)$} & \multicolumn{6}{c|}{LPIPS $(\downarrow)$}\\
\midrule
 & IC & All & RGB & G & R & RE & NIR & All & RGB & G & R & RE & NIR & All & RGB & G & R & RE & NIR\\
\midrule
ThermalGaus. & $\times$ & 22.49 & 18.32 & 23.36 & 24.20 & 23.06 & 23.51 & 0.667 & 0.428 & \cg 0.678 & 0.746 & 0.715 & 0.767 & 0.395 & 0.411 & 0.380 & 0.340 & 0.381 & 0.460\\
ThermoNeRF & $\times$ & 17.25 & 14.03 & 17.66 & 18.00 & 17.22 & 19.35 & 0.477 & 0.144 & 0.522 & 0.573 & 0.511 & 0.633 & 0.695 & 0.793 & 0.689 & 0.673 & 0.671 & 0.648\\
3DGS & \checkmark & 22.53 & 19.38 & 22.07 & 23.44 & 22.36 & 25.41 & 0.677 & 0.490 & 0.598 & 0.731 & 0.712 & 0.853 & 0.359 & \cg 0.263 & 0.378 & 0.372 & 0.357 & 0.427\\
ThermalGaus.$ \dagger$ & \checkmark & 22.28 & 17.38 & 21.96 & 23.27 & 22.72 & 26.06 & 0.644 & 0.375 & 0.602 & 0.722 & 0.691 & 0.831 & 0.405 & 0.435 & 0.446 & 0.385 & 0.385 & 0.372\\
TIMS & \checkmark & \cy 25.27 & \cg 20.39 & \cy 24.48 & \cy 26.26 & \cy 25.86 & \cy 29.33 & \cy 0.730 & \cg 0.538 & 0.662 & \cy 0.798 & \cy 0.775 & 0.875 & \cg 0.260 & \cy 0.273 & \cg 0.303 & \cg 0.247 & \cg 0.241 & \cg 0.237\\
MultiSpec-FeatSplat & \checkmark & 24.68 & 19.36 & 24.14 & 25.66 & 25.17 & 29.05 & 0.726 & 0.515 & 0.663 & 0.795 & \cy 0.775 & \cy 0.880 & 0.275 & 0.304 & 0.319 & 0.258 & 0.250 & 0.245\\
Ours & \checkmark & \cg 25.56 & \cy 20.13 & \cg 24.81 & \cg 26.67 & \cg 26.31 & \cg 29.87 & \cg 0.731 & \cy 0.521 & \cy 0.664 & \cg 0.804 & \cg 0.781 & \cg 0.884 & \cy 0.271 & 0.297 & \cy 0.315 & \cy 0.256 & \cy 0.245 & \cy 0.242\\
\bottomrule
\end{tabular}
}

\end{table*}

\begin{table*}

\caption[]{Comparison on LAKE scene. The colors indicate the \colorbox{green!40}{best} and \colorbox{yellow!40}{second best} results.}

\scriptsize
\centering
\setlength{\tabcolsep}{3pt}

\resizebox{\textwidth}{!}{
\begin{tabular}{ l | c | c c c c c c | c c c c c c | c c c c c c | }
\toprule
 & & \multicolumn{6}{c|}{PSNR $(\uparrow)$} & \multicolumn{6}{c|}{SSIM $(\uparrow)$} & \multicolumn{6}{c|}{LPIPS $(\downarrow)$}\\
\midrule
 & IC & All & RGB & G & R & RE & NIR & All & RGB & G & R & RE & NIR & All & RGB & G & R & RE & NIR\\
\midrule
ThermalGaus. & $\times$ & 21.03 & 20.02 & 21.44 & 24.19 & 19.63 & 19.86 & 0.698 & 0.567 & \cg 0.720 & 0.801 & \cy 0.676 & 0.725 & 0.393 & \cy 0.475 & \cy 0.411 & \cy 0.343 & \cy 0.380 & \cy 0.359\\
ThermoNeRF & $\times$ & 16.62 & 16.13 & 17.37 & 20.38 & 14.51 & 14.70 & 0.554 & 0.440 & 0.609 & 0.721 & 0.498 & 0.504 & 0.671 & 0.792 & 0.664 & 0.591 & 0.660 & 0.646\\
3DGS & \checkmark & 21.16 & 20.76 & 20.88 & 24.63 & 19.54 & 20.01 & 0.658 & 0.587 & 0.616 & 0.764 & 0.627 & 0.696 & 0.459 & 0.500 & 0.518 & 0.467 & 0.407 & 0.402\\
ThermalGaus.$ \dagger$ & \checkmark & 21.77 & 20.99 & 22.23 & 25.44 & 19.69 & 20.50 & 0.677 & 0.610 & 0.667 & 0.782 & 0.624 & 0.701 & 0.435 & 0.492 & 0.462 & 0.395 & 0.435 & 0.389\\
TIMS & \checkmark & 20.13 & 20.38 & 20.23 & 21.95 & 18.68 & 19.42 & 0.651 & 0.603 & 0.641 & 0.729 & 0.604 & 0.676 & 0.445 & 0.493 & 0.470 & 0.418 & 0.443 & 0.402\\
MultiSpec-FeatSplat & \checkmark & \cy 22.47 & \cy 21.16 & \cy 22.89 & \cg 26.20 & \cg 20.72 & \cg 21.38 & \cg 0.720 & \cg 0.648 & \cy 0.712 & \cg 0.816 & \cg 0.683 & \cy 0.742 & \cg 0.382 & \cg 0.457 & \cg 0.407 & \cg 0.336 & \cg 0.377 & \cg 0.331\\
Ours & \checkmark & \cg 22.53 & \cg 21.58 & \cg 23.14 & \cy 26.13 & \cy 20.46 & \cy 21.34 & \cy 0.711 & \cy 0.636 & 0.695 & \cy 0.806 & 0.672 & \cg 0.744 & \cy 0.386 & \cg 0.457 & 0.416 & 0.346 & 0.382 & \cg 0.331\\
\bottomrule
\end{tabular}
}

\end{table*}

\begin{table*}

\caption[]{Comparison on FRUIT TREES scene. The colors indicate the \colorbox{green!40}{best} and \colorbox{yellow!40}{second best} results.}

\scriptsize
\centering
\setlength{\tabcolsep}{3pt}

\resizebox{\textwidth}{!}{
\begin{tabular}{ l | c | c c c c c c | c c c c c c | c c c c c c | }
\toprule
 & & \multicolumn{6}{c|}{PSNR $(\uparrow)$} & \multicolumn{6}{c|}{SSIM $(\uparrow)$} & \multicolumn{6}{c|}{LPIPS $(\downarrow)$}\\
\midrule
 & IC & All & RGB & G & R & RE & NIR & All & RGB & G & R & RE & NIR & All & RGB & G & R & RE & NIR\\
\midrule
ThermalGaus. & $\times$ & 20.85 & 16.12 & 21.46 & 23.78 & 20.08 & 22.81 & 0.560 & 0.425 & 0.535 & 0.645 & 0.551 & 0.646 & 0.409 & 0.418 & 0.421 & 0.386 & 0.412 & 0.406\\
ThermoNeRF & $\times$ & 19.38 & 15.02 & 20.74 & 22.43 & 17.59 & 21.13 & 0.479 & 0.243 & 0.497 & 0.608 & 0.447 & 0.599 & 0.506 & 0.549 & 0.519 & 0.473 & 0.518 & 0.474\\
3DGS & \checkmark & 20.92 & 16.69 & 21.02 & 22.89 & 20.37 & 23.61 & 0.603 & 0.487 & 0.535 & 0.654 & 0.603 & 0.739 & 0.323 & \cg 0.280 & 0.354 & 0.351 & 0.311 & 0.321\\
ThermalGaus.$ \dagger$ & \checkmark & 21.66 & 16.42 & 22.14 & 24.50 & 21.02 & 24.21 & 0.625 & 0.443 & 0.590 & 0.711 & 0.629 & 0.753 & 0.327 & 0.359 & 0.357 & 0.312 & 0.317 & 0.289\\
TIMS & \checkmark & 21.92 & 17.11 & 22.25 & 24.65 & 21.25 & 24.35 & 0.660 & \cy 0.511 & 0.611 & 0.737 & 0.664 & \cy 0.774 & \cy 0.268 & \cy 0.286 & \cy 0.298 & \cy 0.254 & 0.260 & 0.239\\
MultiSpec-FeatSplat & \checkmark & \cy 22.72 & \cy 17.22 & \cy 23.15 & \cy 25.70 & \cy 22.12 & \cy 25.44 & \cg 0.672 & \cg 0.515 & \cg 0.626 & \cg 0.753 & \cg 0.679 & \cg 0.785 & \cg 0.258 & 0.291 & \cg 0.286 & \cg 0.241 & \cg 0.246 & \cg 0.224\\
Ours & \checkmark & \cg 22.90 & \cg 17.42 & \cg 23.16 & \cg 26.07 & \cg 22.21 & \cg 25.64 & \cy 0.668 & 0.506 & \cy 0.624 & \cy 0.752 & \cy 0.675 & \cg 0.785 & 0.269 & 0.301 & 0.301 & \cy 0.254 & \cy 0.257 & \cy 0.231\\
\bottomrule
\end{tabular}
}

\end{table*}

\begin{table*}

\caption[]{Comparison on GOLF scene. The colors indicate the \colorbox{green!40}{best} and \colorbox{yellow!40}{second best} results.}

\scriptsize
\centering
\setlength{\tabcolsep}{3pt}

\resizebox{\textwidth}{!}{
\begin{tabular}{ l | c | c c c c c c | c c c c c c | c c c c c c | }
\toprule
 & & \multicolumn{6}{c|}{PSNR $(\uparrow)$} & \multicolumn{6}{c|}{SSIM $(\uparrow)$} & \multicolumn{6}{c|}{LPIPS $(\downarrow)$}\\
\midrule
 & IC & All & RGB & G & R & RE & NIR & All & RGB & G & R & RE & NIR & All & RGB & G & R & RE & NIR\\
\midrule
ThermalGaus. & $\times$ & 28.70 & 22.21 & 29.39 & 31.87 & 29.56 & 30.46 & 0.780 & 0.551 & 0.833 & 0.885 & 0.819 & 0.812 & 0.355 & 0.448 & 0.314 & 0.278 & 0.340 & 0.394\\
ThermoNeRF & $\times$ & 24.04 & 18.06 & 24.85 & 25.05 & 24.24 & 28.02 & 0.602 & 0.270 & 0.706 & 0.750 & 0.652 & 0.633 & 0.711 & 0.994 & 0.652 & 0.575 & 0.701 & 0.634\\
3DGS & \checkmark & 27.62 & 24.98 & 26.04 & 28.89 & 26.43 & 31.77 & 0.812 & 0.710 & 0.782 & 0.856 & 0.807 & 0.904 & 0.431 & \cg 0.303 & 0.497 & 0.478 & 0.455 & 0.424\\
ThermalGaus.$ \dagger$ & \checkmark & 29.72 & 24.02 & 29.59 & 31.10 & 29.76 & 34.60 & 0.806 & 0.591 & 0.826 & 0.885 & 0.820 & 0.921 & 0.425 & 0.553 & 0.419 & 0.353 & 0.440 & 0.350\\
TIMS & \checkmark & \cy 31.58 & \cy 25.06 & \cy 31.75 & \cy 33.48 & \cy 31.65 & \cg 36.46 & 0.854 & \cg 0.719 & 0.854 & 0.905 & \cy 0.859 & 0.942 & \cg 0.235 & \cy 0.313 & \cg 0.247 & \cy 0.218 & \cg 0.215 & \cg 0.175\\
MultiSpec-FeatSplat & \checkmark & 30.79 & 23.92 & 31.35 & 33.08 & 30.57 & 35.51 & \cg 0.857 & \cy 0.711 & \cy 0.861 & \cy 0.906 & \cg 0.871 & \cg 0.947 & \cy 0.245 & 0.340 & \cy 0.253 & \cg 0.214 & \cy 0.226 & \cy 0.183\\
Ours & \checkmark & \cg 31.96 & \cg 25.24 & \cg 32.29 & \cg 33.87 & \cg 32.77 & \cy 36.04 & \cy 0.856 & 0.699 & \cg 0.863 & \cg 0.911 & \cg 0.871 & \cy 0.945 & 0.267 & 0.360 & 0.277 & 0.224 & 0.257 & 0.213\\
\bottomrule
\end{tabular}
}

\end{table*}

\begin{table*}

\caption[]{Comparison on BUD SWELLING scene. The colors indicate the \colorbox{green!40}{best} and \colorbox{yellow!40}{second best} results.}

\scriptsize
\centering
\setlength{\tabcolsep}{3pt}

\resizebox{\textwidth}{!}{
\begin{tabular}{ l | c | c c c c c c | c c c c c c | c c c c c c | }
\toprule
 & & \multicolumn{6}{c|}{PSNR $(\uparrow)$} & \multicolumn{6}{c|}{SSIM $(\uparrow)$} & \multicolumn{6}{c|}{LPIPS $(\downarrow)$}\\
\midrule
 & IC & All & RGB & G & R & RE & NIR & All & RGB & G & R & RE & NIR & All & RGB & G & R & RE & NIR\\
\midrule
ThermalGaus. & $\times$ & 23.24 & 18.25 & 23.23 & 25.90 & 24.00 & 24.82 & 0.699 & 0.586 & 0.686 & 0.770 & 0.731 & 0.724 & 0.298 & 0.290 & 0.307 & 0.293 & 0.271 & 0.327\\
ThermoNeRF & $\times$ & 17.16 & 12.81 & 17.83 & 20.59 & 16.26 & 18.33 & 0.397 & 0.124 & 0.433 & 0.599 & 0.388 & 0.443 & 0.698 & 0.800 & 0.699 & 0.669 & 0.668 & 0.652\\
3DGS & \checkmark & 22.95 & 19.59 & 21.10 & 24.60 & 23.10 & 26.35 & 0.709 & 0.658 & 0.601 & 0.743 & 0.727 & 0.818 & 0.263 & \cy 0.182 & 0.322 & 0.334 & 0.235 & 0.241\\
ThermalGaus.$ \dagger$ & \checkmark & 23.17 & 18.55 & 22.18 & 25.89 & 23.08 & 26.14 & 0.710 & 0.577 & 0.645 & 0.796 & 0.725 & 0.808 & 0.266 & 0.271 & 0.319 & 0.269 & 0.245 & 0.228\\
TIMS & \checkmark & 24.29 & \cg 20.26 & 23.01 & 26.62 & 24.35 & 27.19 & \cy 0.769 & \cg 0.686 & 0.692 & 0.838 & \cg 0.782 & 0.850 & \cg 0.195 & \cg 0.180 & \cy 0.248 & \cy 0.199 & \cg 0.179 & \cy 0.167\\
MultiSpec-FeatSplat & \checkmark & \cg 25.19 & 19.92 & \cg 23.91 & \cg 28.19 & \cy 24.87 & \cg 29.07 & \cg 0.771 & \cy 0.666 & \cg 0.701 & \cg 0.849 & 0.780 & \cg 0.861 & \cy 0.197 & 0.205 & \cg 0.244 & \cg 0.195 & \cg 0.179 & \cg 0.161\\
Ours & \checkmark & \cy 25.11 & \cy 19.93 & \cy 23.77 & \cy 27.75 & \cg 25.13 & \cy 28.95 & 0.766 & 0.654 & \cy 0.696 & \cy 0.843 & \cy 0.781 & \cy 0.856 & 0.205 & 0.213 & 0.253 & 0.206 & \cy 0.183 & 0.168\\
\bottomrule
\end{tabular}
}

\end{table*}

\begin{table*}

\caption[]{Comparison on SOLAR scene. The colors indicate the \colorbox{green!40}{best} and \colorbox{yellow!40}{second best} results.}

\scriptsize
\centering
\setlength{\tabcolsep}{3pt}

\resizebox{\textwidth}{!}{
\begin{tabular}{ l | c | c c c c c c | c c c c c c | c c c c c c | }
\toprule
 & & \multicolumn{6}{c|}{PSNR $(\uparrow)$} & \multicolumn{6}{c|}{SSIM $(\uparrow)$} & \multicolumn{6}{c|}{LPIPS $(\downarrow)$}\\
\midrule
 & IC & All & RGB & G & R & RE & NIR & All & RGB & G & R & RE & NIR & All & RGB & G & R & RE & NIR\\
\midrule
ThermalGaus. & $\times$ & 22.42 & 20.33 & 22.72 & 24.62 & 21.19 & 23.25 & 0.754 & 0.705 & 0.784 & 0.780 & 0.732 & 0.766 & 0.230 & 0.264 & \cg 0.212 & 0.200 & 0.237 & 0.238\\
ThermoNeRF & $\times$ & 14.98 & 12.94 & 14.41 & 16.65 & 14.72 & 16.16 & 0.404 & 0.196 & 0.409 & 0.507 & 0.400 & 0.509 & 0.520 & 0.671 & 0.483 & 0.458 & 0.493 & 0.497\\
3DGS & \checkmark & 23.24 & \cy 21.17 & 22.45 & 25.99 & 22.17 & 24.42 & 0.809 & 0.735 & 0.793 & 0.857 & 0.808 & 0.852 & 0.231 & \cy 0.234 & 0.248 & 0.204 & 0.237 & 0.234\\
ThermalGaus.$ \dagger$ & \checkmark & 23.49 & 20.67 & 22.97 & 26.31 & 22.69 & 24.82 & 0.804 & 0.707 & 0.791 & 0.859 & 0.810 & 0.853 & 0.236 & 0.292 & 0.251 & 0.189 & 0.238 & 0.212\\
TIMS & \checkmark & \cy 23.95 & \cg 21.66 & 23.13 & 26.36 & \cy 23.18 & \cy 25.43 & 0.824 & \cg 0.752 & 0.805 & 0.867 & 0.830 & 0.865 & \cg 0.194 & \cg 0.225 & \cy 0.213 & \cy 0.158 & \cg 0.194 & \cg 0.178\\
MultiSpec-FeatSplat & \checkmark & 23.73 & 20.80 & \cy 23.19 & \cy 26.76 & 23.07 & 24.82 & \cy 0.829 & \cy 0.738 & \cy 0.811 & \cy 0.884 & \cy 0.839 & \cy 0.872 & 0.212 & 0.264 & 0.228 & 0.167 & 0.209 & 0.190\\
Ours & \checkmark & \cg 24.05 & 20.82 & \cg 23.23 & \cg 27.41 & \cg 23.30 & \cg 25.49 & \cg 0.831 & 0.737 & \cg 0.812 & \cg 0.893 & \cg 0.841 & \cg 0.875 & \cy 0.206 & 0.260 & 0.224 & \cg 0.156 & \cy 0.207 & \cy 0.184\\
\bottomrule
\end{tabular}
}

\end{table*}

\begin{table*}
\centering
\caption[]{Spectral similarity evaluation. Each block is (SAM $\downarrow$, SCM $\uparrow$, SID $\downarrow$). Colors: \colorbox{green!40}{best}, \colorbox{yellow!40}{second best}.}
\label{tab:spectral_sim_dataset}
\resizebox{1.0\textwidth}{!}{%
\setlength{\tabcolsep}{5pt}

\begin{tabular}{ l | c c c | c c c | c c c | c c c | c c c | c c c | c c c || c c c }
\toprule
& \multicolumn{3}{c|}{SOLAR} & \multicolumn{3}{c|}{FRUIT TREES} & \multicolumn{3}{c|}{SINGLE TREE} & \multicolumn{3}{c|}{BUD SWELLING} & \multicolumn{3}{c|}{GOLF} & \multicolumn{3}{c|}{LAKE} & \multicolumn{3}{c|}{GARDEN} & \multicolumn{3}{c}{Average} \\
\cmidrule(lr){2-4}\cmidrule(lr){5-7}\cmidrule(lr){8-10}\cmidrule(lr){11-13}\cmidrule(lr){14-16}\cmidrule(lr){17-19}\cmidrule(lr){20-22}\cmidrule(lr){23-25}
& SAM $(\downarrow)$ & SCM $(\uparrow)$ & SID $(\downarrow)$ & SAM $(\downarrow)$ & SCM $(\uparrow)$ & SID $(\downarrow)$ & SAM $(\downarrow)$ & SCM $(\uparrow)$ & SID $(\downarrow)$ & SAM $(\downarrow)$ & SCM $(\uparrow)$ & SID $(\downarrow)$ & SAM $(\downarrow)$ & SCM $(\uparrow)$ & SID $(\downarrow)$ & SAM $(\downarrow)$ & SCM $(\uparrow)$ & SID $(\downarrow)$ & SAM $(\downarrow)$ & SCM $(\uparrow)$ & SID $(\downarrow)$ & SAM $(\downarrow)$ & SCM $(\uparrow)$ & SID $(\downarrow)$ \\
\midrule
3DGS & 0.122 & 0.861 & 0.037 & 0.207 & 0.696 & 0.075 & 0.168 & 0.742 & 0.050 & 0.125 & 0.912 & 0.042 & 0.119 & 0.816 & 0.024 & 0.195 & 0.711 & 0.064 & 0.083 & 0.943 & 0.014 & \textbf{0.146} & \textbf{0.812} & \textbf{0.044} \\
ThermalGaus.$ \dagger$ & 0.114 & 0.855 & 0.031 & 0.171 & 0.762 & 0.054 & 0.154 & 0.743 & 0.047 & 0.130 & 0.908 & 0.047 & 0.073 & 0.915 & 0.010 & 0.165 & 0.769 & 0.046 & 0.080 & 0.951 & \colorbox{yellow!40}{0.012} & \textbf{0.127} & \textbf{0.843} & \textbf{0.035} \\
TIMS & \colorbox{green!40}{0.108} & 0.863 & \colorbox{yellow!40}{0.028} & 0.156 & 0.797 & \colorbox{yellow!40}{0.044} & \colorbox{yellow!40}{0.114} & \colorbox{yellow!40}{0.847} & 0.027 & \colorbox{green!40}{0.111} & 0.927 & 0.034 & \colorbox{yellow!40}{0.065} & \colorbox{yellow!40}{0.931} & \colorbox{yellow!40}{0.008} & 0.189 & 0.694 & 0.065 & \colorbox{green!40}{0.072} & \colorbox{green!40}{0.957} & \colorbox{green!40}{0.011} & \textbf{0.116} & \textbf{0.860} & \textbf{0.031} \\
MultiSpec-FeatSplat & 0.114 & \colorbox{yellow!40}{0.875} & 0.030 & \colorbox{green!40}{0.152} & \colorbox{green!40}{0.808} & \colorbox{green!40}{0.040} & 0.119 & 0.837 & \colorbox{yellow!40}{0.026} & 0.116 & \colorbox{yellow!40}{0.929} & \colorbox{yellow!40}{0.033} & 0.071 & 0.926 & 0.009 & \colorbox{yellow!40}{0.141} & \colorbox{yellow!40}{0.831} & \colorbox{yellow!40}{0.031} & 0.083 & 0.948 & \colorbox{yellow!40}{0.012} & \colorbox{yellow!40}{\textbf{0.114}} & \colorbox{yellow!40}{\textbf{0.879}} & \colorbox{yellow!40}{\textbf{0.026}} \\
Ours & \colorbox{yellow!40}{0.110} & \colorbox{green!40}{0.884} & \colorbox{green!40}{0.026} & \colorbox{yellow!40}{0.153} & \colorbox{yellow!40}{0.807} & \colorbox{green!40}{0.040} & \colorbox{green!40}{0.110} & \colorbox{green!40}{0.859} & \colorbox{green!40}{0.023} & \colorbox{yellow!40}{0.114} & \colorbox{green!40}{0.930} & \colorbox{green!40}{0.032} & \colorbox{green!40}{0.059} & \colorbox{green!40}{0.948} & \colorbox{green!40}{0.006} & \colorbox{green!40}{0.137} & \colorbox{green!40}{0.834} & \colorbox{green!40}{0.030} & \colorbox{yellow!40}{0.078} & \colorbox{yellow!40}{0.952} & \colorbox{green!40}{0.011} & \colorbox{green!40}{\textbf{0.109}} & \colorbox{green!40}{\textbf{0.888}} & \colorbox{green!40}{\textbf{0.024}} \\
\bottomrule
\end{tabular}

}%
\end{table*}

\begin{table*}

\caption[]{X-NeRF dataset, individual metrics. Average across all X-NerF scenes.}
\label{tab:x_nerf_evaluation_large}

\scriptsize
\centering
\setlength{\tabcolsep}{3pt}

\resizebox{\textwidth}{!}{
\begin{tabular}{ l | c c c c c c c c c c c c c | c c c c c c c c c c c c c | c | }
\toprule
& \multicolumn{13}{c|}{PSNR $(\uparrow)$} & \multicolumn{13}{c|}{SSIM $(\uparrow)$} & \multicolumn{1}{c|}{LPIPS $(\downarrow)$}\\
\midrule
 & All & RGB & MS0 & MS1 & MS2 & MS3 & MS4 & MS5 & MS6 & MS7 & MS8 & MS9 & IR & All & RGB & MS0 & MS1 & MS2 & MS3 & MS4 & MS5 & MS6 & MS7 & MS8 & MS9 & IR & RGB\\
\midrule
3DGS & 30.99 & \cg 35.36 & 30.63 & 30.04 & 30.90 & 30.39 & 30.74 & 30.05 & 30.87 & 30.57 & 31.48 & 29.77 & \cg 31.14 & 0.896 & \cg 0.948 & 0.898 & 0.888 & 0.902 & 0.886 & 0.900 & 0.861 & 0.894 & 0.878 & 0.896 & 0.879 & \cy 0.922 & \cg 0.243\\
ThermalGaus.$ \dagger$ & \cy 31.47 & 32.87 & \cy 31.95 & \cy 31.65 & \cy 31.60 & \cy 31.48 & \cy 31.78 & \cy 31.22 & \cy 31.64 & \cy 31.41 & \cy 31.90 & \cg 30.49 & 29.63 & \cy 0.916 & 0.929 & \cy 0.935 & \cy 0.927 & \cy 0.926 & \cy 0.917 & \cy 0.920 & \cy 0.897 & \cy 0.912 & \cy 0.899 & \cy 0.912 & \cy 0.899 & 0.918 & 0.311\\
TIMS & 29.07 & 31.45 & 29.41 & 29.23 & 29.39 & 29.24 & 29.36 & 28.85 & 29.05 & 28.99 & 29.14 & 27.54 & 27.20 & 0.889 & 0.932 & 0.906 & 0.898 & 0.897 & 0.888 & 0.891 & 0.866 & 0.881 & 0.868 & 0.879 & 0.867 & 0.895 & 0.288\\
Ours & \cg 32.02 & \cy 32.88 & \cg 32.37 & \cg 32.52 & \cg 32.54 & \cg 32.25 & \cg 32.44 & \cg 32.26 & \cg 32.45 & \cg 32.10 & \cg 32.14 & \cy 30.41 & \cy 29.93 & \cg 0.930 & \cy 0.940 & \cg 0.947 & \cg 0.941 & \cg 0.940 & \cg 0.934 & \cg 0.934 & \cg 0.914 & \cg 0.926 & \cg 0.916 & \cg 0.926 & \cg 0.912 & \cg 0.926 & \cy 0.280\\
\bottomrule
\end{tabular}
}

\end{table*}

\begin{table*}
\centering
\caption[]{Spectral similarity evaluation of the X-NeRF dataset. Each block is (SAM $\downarrow$, SCM $\uparrow$, SID $\downarrow$). Colors: \colorbox{green!40}{best}, \colorbox{yellow!40}{second best}.}
\label{tab:x_nerf_evaluation_large_spectral}

\setlength{\tabcolsep}{5pt}

\resizebox{1.0\textwidth}{!}{%
\setlength{\tabcolsep}{5pt}

\begin{tabular}{ l | c c c | c c c | c c c | c c c | c c c | c c c | c c c | c c c | c c c || c c c }
\toprule
& \multicolumn{3}{c|}{bins} & \multicolumn{3}{c|}{black} & \multicolumn{3}{c|}{bluechair} & \multicolumn{3}{c|}{cvlab} & \multicolumn{3}{c|}{dino} & \multicolumn{3}{c|}{fruits} & \multicolumn{3}{c|}{green} & \multicolumn{3}{c|}{hall} & \multicolumn{3}{c|}{hall2} \\
\cmidrule(lr){2-4}\cmidrule(lr){5-7}\cmidrule(lr){8-10}\cmidrule(lr){11-13}\cmidrule(lr){14-16}\cmidrule(lr){17-19}\cmidrule(lr){20-22}\cmidrule(lr){23-25}\cmidrule(lr){26-28}
& SAM $(\downarrow)$ & SCM $(\uparrow)$ & SID $(\downarrow)$ 
& SAM $(\downarrow)$ & SCM $(\uparrow)$ & SID $(\downarrow)$ 
& SAM $(\downarrow)$ & SCM $(\uparrow)$ & SID $(\downarrow)$ 
& SAM $(\downarrow)$ & SCM $(\uparrow)$ & SID $(\downarrow)$ 
& SAM $(\downarrow)$ & SCM $(\uparrow)$ & SID $(\downarrow)$ 
& SAM $(\downarrow)$ & SCM $(\uparrow)$ & SID $(\downarrow)$ 
& SAM $(\downarrow)$ & SCM $(\uparrow)$ & SID $(\downarrow)$ 
& SAM $(\downarrow)$ & SCM $(\uparrow)$ & SID $(\downarrow)$ 
& SAM $(\downarrow)$ & SCM $(\uparrow)$ & SID $(\downarrow)$ \\
\midrule
3DGS & 0.173 & 0.286 & 0.052 & 0.103 & 0.708 & 0.054 & 0.090 & 0.786 & 0.057 & 0.064 & 0.905 & 0.012 & 0.077 & 0.806 & 0.037 & 0.044 & 0.945 & \colorbox{yellow!40}{0.005} & 0.063 & 0.924 & 0.018 & 0.067 & 0.941 & 0.049 & 0.065 & 0.959 & 0.025 \\
TG$ \dagger$ & \colorbox{green!40}{0.117} & \colorbox{yellow!40}{0.446} & \colorbox{green!40}{0.020} & 0.091 & \colorbox{yellow!40}{0.714} & 0.028 & \colorbox{green!40}{0.074} & \colorbox{green!40}{0.808} & 0.046 & \colorbox{yellow!40}{0.057} & 0.914 & \colorbox{yellow!40}{0.008} & \colorbox{yellow!40}{0.062} & \colorbox{yellow!40}{0.822} & 0.022 & \colorbox{green!40}{0.042} & \colorbox{yellow!40}{0.950} & \colorbox{green!40}{0.004} & \colorbox{green!40}{0.047} & \colorbox{green!40}{0.959} & 0.012 & 0.056 & 0.956 & \colorbox{yellow!40}{0.026} & \colorbox{yellow!40}{0.058} & \colorbox{yellow!40}{0.970} & \colorbox{yellow!40}{0.024} \\
TIMS & 0.132 & 0.418 & \colorbox{yellow!40}{0.025} & \colorbox{yellow!40}{0.088} & 0.698 & \colorbox{yellow!40}{0.027} & \colorbox{yellow!40}{0.075} & 0.800 & \colorbox{yellow!40}{0.041} & 0.058 & \colorbox{yellow!40}{0.916} & \colorbox{yellow!40}{0.008} & 0.064 & 0.820 & \colorbox{yellow!40}{0.018} & \colorbox{green!40}{0.042} & 0.949 & \colorbox{green!40}{0.004} & \colorbox{green!40}{0.047} & \colorbox{yellow!40}{0.958} & \colorbox{yellow!40}{0.010} & \colorbox{yellow!40}{0.055} & \colorbox{yellow!40}{0.957} & 0.028 & 0.081 & 0.940 & 0.041 \\
Ours & \colorbox{yellow!40}{0.129} & \colorbox{green!40}{0.482} & 0.026 & \colorbox{green!40}{0.075} & \colorbox{green!40}{0.759} & \colorbox{green!40}{0.021} & \colorbox{green!40}{0.074} & \colorbox{yellow!40}{0.802} & \colorbox{green!40}{0.030} & \colorbox{green!40}{0.052} & \colorbox{green!40}{0.921} & \colorbox{green!40}{0.006} & \colorbox{green!40}{0.057} & \colorbox{green!40}{0.847} & \colorbox{green!40}{0.014} & \colorbox{yellow!40}{0.043} & \colorbox{green!40}{0.953} & \colorbox{green!40}{0.004} & \colorbox{yellow!40}{0.048} & 0.953 & \colorbox{green!40}{0.005} & \colorbox{green!40}{0.054} & \colorbox{green!40}{0.964} & \colorbox{green!40}{0.020} & \colorbox{green!40}{0.056} & \colorbox{green!40}{0.974} & \colorbox{green!40}{0.014} \\
\bottomrule
\end{tabular}
}

\resizebox{1.0\textwidth}{!}{%
\setlength{\tabcolsep}{5pt}

\begin{tabular}{ l | c c c | c c c | c c c | c c c | c c c | c c c | c c c || c c c }
\toprule
& \multicolumn{3}{c|}{hall3} & \multicolumn{3}{c|}{hall4} & \multicolumn{3}{c|}{orange} & \multicolumn{3}{c|}{penguin} & \multicolumn{3}{c|}{penguin2} & \multicolumn{3}{c|}{puppets} & \multicolumn{3}{c|}{tech} & \multicolumn{3}{c}{Average} \\
\cmidrule(lr){2-4}\cmidrule(lr){5-7}\cmidrule(lr){8-10}\cmidrule(lr){11-13}\cmidrule(lr){14-16}\cmidrule(lr){17-19}\cmidrule(lr){20-22}\cmidrule(lr){23-25}
& SAM $(\downarrow)$ & SCM $(\uparrow)$ & SID $(\downarrow)$ 
& SAM $(\downarrow)$ & SCM $(\uparrow)$ & SID $(\downarrow)$ 
& SAM $(\downarrow)$ & SCM $(\uparrow)$ & SID $(\downarrow)$ 
& SAM $(\downarrow)$ & SCM $(\uparrow)$ & SID $(\downarrow)$ 
& SAM $(\downarrow)$ & SCM $(\uparrow)$ & SID $(\downarrow)$ 
& SAM $(\downarrow)$ & SCM $(\uparrow)$ & SID $(\downarrow)$ 
& SAM $(\downarrow)$ & SCM $(\uparrow)$ & SID $(\downarrow)$ 
& SAM $(\downarrow)$ & SCM $(\uparrow)$ & SID $(\downarrow)$ \\
\midrule
3DGS & 0.089 & 0.942 & 0.034 & 0.099 & 0.793 & 0.027 & 0.050 & 0.898 & 0.013 & 0.266 & 0.488 & 0.240 & 0.055 & 0.935 & 0.022 & 0.080 & 0.895 & 0.116 & 0.075 & 0.850 & 0.024 & \textbf{0.091} & \textbf{0.816} & \textbf{0.049} \\
TG$ \dagger$ & 0.078 & \colorbox{yellow!40}{0.959} & \colorbox{yellow!40}{0.023} & \colorbox{yellow!40}{0.072} & \colorbox{yellow!40}{0.886} & \colorbox{yellow!40}{0.013} & \colorbox{green!40}{0.043} & \colorbox{green!40}{0.912} & \colorbox{yellow!40}{0.009} & \colorbox{yellow!40}{0.087} & \colorbox{green!40}{0.855} & \colorbox{yellow!40}{0.035} & \colorbox{yellow!40}{0.051} & \colorbox{yellow!40}{0.941} & \colorbox{yellow!40}{0.018} & \colorbox{yellow!40}{0.046} & \colorbox{yellow!40}{0.945} & 0.031 & \colorbox{yellow!40}{0.062} & \colorbox{yellow!40}{0.874} & \colorbox{yellow!40}{0.012} & \colorbox{yellow!40}{\textbf{0.065}} & \colorbox{yellow!40}{\textbf{0.869}} & \colorbox{yellow!40}{\textbf{0.021}} \\
TIMS & 0.081 & 0.950 & 0.027 & 0.080 & 0.847 & 0.015 & \colorbox{yellow!40}{0.046} & 0.907 & \colorbox{yellow!40}{0.009} & \colorbox{yellow!40}{0.087} & 0.847 & \colorbox{yellow!40}{0.035} & \colorbox{green!40}{0.050} & \colorbox{green!40}{0.943} & \colorbox{green!40}{0.016} & 0.049 & 0.936 & \colorbox{yellow!40}{0.022} & 0.073 & 0.842 & 0.015 & \textbf{0.069} & \textbf{0.858} & \colorbox{yellow!40}{\textbf{0.021}} \\
Ours & \colorbox{yellow!40}{0.079} & \colorbox{green!40}{0.962} & \colorbox{green!40}{0.022} & \colorbox{green!40}{0.068} & \colorbox{green!40}{0.905} & \colorbox{green!40}{0.011} & \colorbox{green!40}{0.043} & \colorbox{yellow!40}{0.910} & \colorbox{green!40}{0.008} & \colorbox{green!40}{0.072} & \colorbox{yellow!40}{0.851} & \colorbox{green!40}{0.034} & 0.055 & 0.936 & \colorbox{yellow!40}{0.018} & \colorbox{green!40}{0.045} & \colorbox{green!40}{0.950} & \colorbox{green!40}{0.015} & \colorbox{green!40}{0.058} & \colorbox{green!40}{0.894} & \colorbox{green!40}{0.010} & \colorbox{green!40}{\textbf{0.063}} & \colorbox{green!40}{\textbf{0.879}} & \colorbox{green!40}{\textbf{0.016}} \\
\bottomrule
\end{tabular}
}

\end{table*}

\begin{table*}[h]
\caption{\textsc{Lake} and \textsc{Garden} scenes: memory (MB), number of splats, and training time}
\label{tab:memory_splats_lake_garden}
\tiny
\centering
\setlength{\tabcolsep}{4pt}
\renewcommand{\arraystretch}{1.1}
\resizebox{\linewidth}{!}{
\begin{tabular}{l | c c c c c | c | c c c c c | c || c c c c c | c | c c c c c | c }
\toprule
& \multicolumn{12}{c|}{\textsc{Lake} } & \multicolumn{12}{c|}{\textsc{Garden}} \\
& \multicolumn{6}{c|}{\textbf{Lake Memory (MB)} $\downarrow$} 
& \multicolumn{5}{c|}{\textbf{Lake \# Splats} $\downarrow$} 
& \textbf{Train} 
& \multicolumn{6}{c|}{\textbf{Garden Memory (MB)} $\downarrow$} 
& \multicolumn{5}{c|}{\textbf{Garden \# Splats} $\downarrow$} 
& \textbf{Train} \\
\cmidrule(lr){2-7}\cmidrule(lr){8-12}\cmidrule(lr){14-19}\cmidrule(lr){20-24}
& RGB & G & R & RE & NIR & Total & RGB & G & R & RE & NIR & (h:m:s) 
& RGB & G & R & RE & NIR & Total & RGB & G & R & RE & NIR & (h:m:s) \\
\midrule
3DGS~\cite{kerbl3Dgaussians} & 595.81 & 260.91 & 144.55 & 263.70 & 191.76 & 1456.74 
     & 2.613.207 & 2.609.129 & 1.445.470 & 2.637.019 & 1.917.627 & 01:41:03
     & 262.15 & 107.93 & 120.93 & 75.04 & 71.49 & 637.54 
     & 1.149.797 & 1.079.264 & 1.209.323 & 750.358 & 714.917 & 1:26:18 \\
TIMS~\cite{gruen2025towards}
     & \multicolumn{5}{c|}{------------------~~~~2689.60~~~~------------------} & 2689.60 
     & \multicolumn{5}{c|}{------------------~~~~5.557.032~~~~------------------} & 04:49:11
     & \multicolumn{5}{c|}{------------------~~~~927.37~~~~~------------------} & 927.37 
     & \multicolumn{5}{c|}{------------------~~~~1.916.049~~~~~------------------} & 02:36:09 \\
TG (Ours) 
     & \multicolumn{5}{c|}{------------------~~~~1634.38~~~~------------------} & 1634.38
     & \multicolumn{5}{c|}{------------------~~~~3.376.816~~~~------------------} & 03:27:18
     & \multicolumn{5}{c|}{------------------~~~~532.00~~~~~------------------} & 532.00 
     & \multicolumn{5}{c|}{------------------~~~~1.099.171~~~~~------------------} & 01:50:36 \\
Ours 
     & \multicolumn{5}{c|}{------------------~~~~326.14~~~~------------------} & 326.14 
     & \multicolumn{5}{c|}{------------------~~~~4.796.045~~~~------------------} & 02:34:50
     & \multicolumn{5}{c|}{------------------~~~~135.01~~~~~------------------} & 135.01 
     & \multicolumn{5}{c|}{------------------~~~~1.985.252~~~~~------------------} & 02:02:44 \\
\bottomrule
\end{tabular}
}
\end{table*}

\begin{table*}
\caption[]{\label{tab:less_spectral_images}Our method only using the indicated percentage of non-RGB spectral images.}

\scriptsize
\centering
\begin{tabular}{ l | c c c c c c | c c c c c c | c }
\toprule
& \multicolumn{6}{c|}{PSNR $(\uparrow)$} & \multicolumn{6}{c|}{SSIM $(\uparrow)$} & \multicolumn{1}{c}{LPIPS $(\downarrow)$}\\
\midrule
& All & RGB & G & R & RE & NIR & All & RGB & G & R & RE & NIR & RGB \\
\midrule
100\% & \cg 25.65 & \cg 21.17 & \cg 25.42 & \cg 27.79 & \cg 25.79 & \cg 28.13 & \cg 0.763 & \cg 0.633 & \cg 0.729 & \cg 0.827 & \cg 0.780 & \cg 0.849 & 0.306\\
40\% & 24.31 & 20.91 & 23.88 & 26.27 & 24.16 & 26.38 & 0.726 & 0.621 & 0.676 & 0.790 & 0.733 & 0.809 & 0.310 \\
20\% & 23.01 & 20.72 & 22.40 & 24.71 & 22.53 & 24.72 & 0.683 & 0.609 & 0.620 & 0.745 & 0.680 & 0.763 & 0.310 \\
10\% & 22.00 &   20.55 & 21.37 & 23.78 & 21.18 & 23.17 & 0.640 &   0.605 & 0.566 & 0.698 & 0.624 & 0.707  & 0.304  \\
5\% &   21.40 & 20.61 &   20.83 &   22.84 &   20.57 &   22.19 &   0.610 & 0.610 &   0.535 &   0.658 &   0.582 &   0.667 &   \cg 0.295 \\
\bottomrule
\end{tabular}
\end{table*}

\begin{figure*}
\resizebox{1.0\textwidth}{!}{
\input{value_spider.tex}
}

    \caption{Spectral spider plot. A random image from the evaluation dataset is taken and we selected pixel x=100 and y=100. For every spectral band we plot the amplitude of the ground truth pixel and the amplitude of the methods 3DGS, \textit{ThermalGaussian}$ \dagger$ (TG), TIMS, MultiSpec-FeatSplat and ours.}
    \label{fig:value_spectral_spider}

\end{figure*}

\begin{figure*}[h]
    \centering

\resizebox{1.0\textwidth}{!}{
\input{error_spider}
}
    
    \caption{Error spider plot. For this plot we computed the D-SSIM averaged over the respective dataset. GT is selected as a reference and the plot shows the average error of individual spectral bands. We evaluated 3DGS~\cite{kerbl3Dgaussians}, \textit{ThermalGaussian}$ \dagger$ (TG), TIMS~\cite{gruen2025towards}, MultiSpec-FeatSplat and our method \textit{MS-Splatting}.}
    \label{fig:error_spectral_spider}
\end{figure*}

\clearpage
\newpage

\begin{figure*}[h]
    \centering
    \includegraphics[width=0.98\textwidth]{images/appendix/ms-solar.pdf}

    \caption{Visual comparison on the SOLAR scene with all available spectral-bands. The used \textit{ThermalGaussian}$ \dagger$ method in this comparison is our multi-spectral re-implementation.}

    \label{fig:comp_images_ms_solar}
\end{figure*}

\begin{figure*}[h]
    \centering
   \includegraphics[width=0.98\textwidth]{images/appendix/ms-bauwagen.pdf}

    \caption{Visual comparison on the LAKE scene with all available spectral-bands. The used \textit{ThermalGaussian}$ \dagger$ method in this comparison is our multi-spectral re-implementation.}
    \label{fig:comp_images_ms_bauwagen}
\end{figure*}

\begin{figure*}[h]
    \centering
   \includegraphics[width=0.98\textwidth]{images/appendix/ms-LWG-vegetation.pdf}
    \caption{Visual comparison on the BUD SWELLING scene with all available spectral-bands. The used \textit{ThermalGaussian}$ \dagger$ method in this comparison is our multi-spectral re-implementation.}

    \label{fig:comp_images_ms_LWG_vegetation}
\end{figure*}

\begin{figure*}[h]
    \centering
   \includegraphics[width=0.98\textwidth]{images/appendix/ms-single-tree-1.pdf}
    \caption{Visual comparison on the SINGLE TREE scene with all available spectral-bands. The used \textit{ThermalGaussian}$ \dagger$ method in this comparison is our multi-spectral re-implementation.}
    \label{fig:comp_images_ms_single_tree_1}
\end{figure*}

\begin{figure*}[h]
    \centering
    \includegraphics[width=0.98\textwidth]{images/appendix/ms-golf.pdf}
    \caption{Visual comparison on the GOLF scene with all available spectral-bands. The used \textit{ThermalGaussian}$ \dagger$ method in this comparison is our multi-spectral re-implementation.}
    \label{fig:comp_images_ms_golf}
\end{figure*}

\begin{figure*}[htbp]
    \centering
    \includegraphics[width=0.98\textwidth]{images/appendix/ms-garden.pdf}

    \caption{Visual comparison on the GARDEN scene with all available spectral-bands. The used \textit{ThermalGaussian}$ \dagger$ method in this comparison is our multi-spectral re-implementation.}
    \label{fig:comp_images_ms_garden}
\end{figure*}

\end{document}